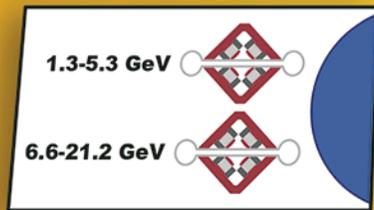
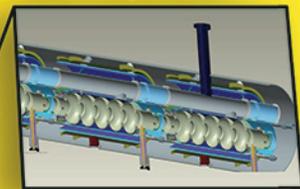
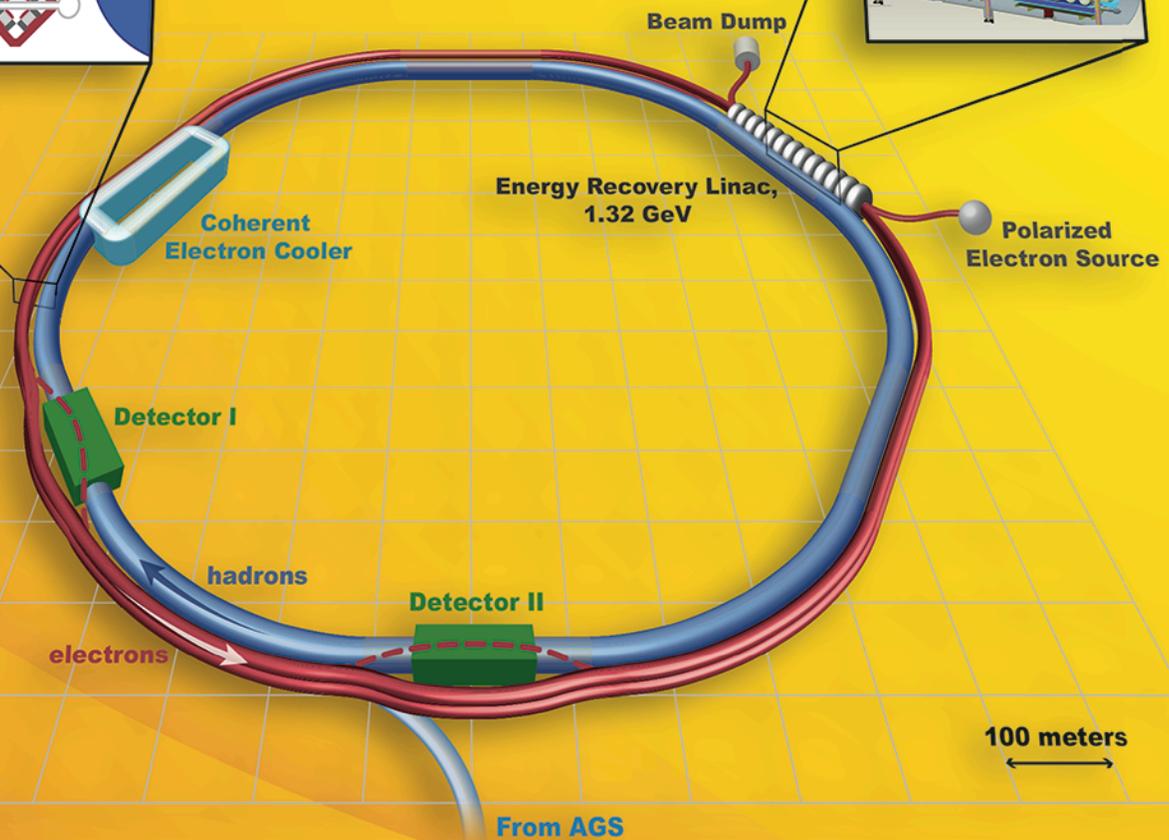

# eRHIC Design Study
## An Electron-Ion Collider at BNL

**DECEMBER 2014**

# eRHIC Design Study
## An Electron-Ion Collider at BNL

# Authors

*(Unless noted otherwise, all are members of the BNL staff)*

E.C. Aschenauer, M.D. Baker, A. Bazilevsky, K. Boyle (RBRC), S. Belomestnykh, I. Ben-Zvi,
S. Brooks, C. Brutus, T. Burton, S. Fazio, A. Fedotov, D. Gassner, Y. Hao, Y. Jing, D. Kayran,
A. Kiselev, M.A.C. Lamont, J.-H. Lee, V. N. Litvinenko, C. Liu, T. Ludlam, G. Mahler, G. McIntyre,
W. Meng, F. Meot, T. Miller, M. Minty, B.Parker, R. Petti, I. Pinayev, V. Ptitsyn, T. Roser, M. Stratmann
(Uni.-Tuebingen), E. Sichtermann (LBNL), J. Skaritka, O. Tchoubar, P. Thieberger, T. Toll, D. Trbojevic,
N. Tsoupas, J. Tuozzolo, T. Ullrich, E. Wang, G. Wang, Q. Wu, W. Xu, L. Zheng

# Acknowledgements

*The authors are indebted to the following colleagues for important discussions, comments, and other contributions in the preparation of this report:*

J.L. Albacete (Universidad de Granada), S. Berg, M. Blaskiewicz, W. Brooks (Univ. Technica Valparaiso), K. Brown, A. Deshpande (Stony Brook Univ.), M. Diehl (DESY), V. Guzey (PNPI), H. Hahn, W. Jackson, Y. Kovchegov (Ohio State), R. Lambiase, T. Lappi (Univ. Jyvaskyla), D. Lissauer, Y. Luo, B. Martin, C. Marquet (CNRS), R. Michnoff, K. Mirabella, A. Morgan, D. Morrison, D. Mueller (Univ. Bochum), H. Paukkunen (Univ. Jyvaskyla), A. Pendzick, J.-W. Qiu, J. Sandberg, R. Sassot (Univ. Buenos Aires), H. Spiesberger (Univ. Mainz), R. Than, R. Venugopalan, B.-W. Xiao (CCNU), Zh. Xu, A. Zaltsman

# 1 INTRODUCTION AND OVERVIEW

This document presents BNL's plan for an electron-ion collider, eRHIC, a major new research tool that builds on the existing RHIC facility to advance the long-term vision for Nuclear Physics to discover and understand the emergent phenomena of Quantum Chromodynamics (QCD), the fundamental theory of the strong interaction that binds the atomic nucleus.

We describe the scientific requirements for such a facility, following up on the community-wide 2012 white paper, "Electron-Ion Collider: the Next QCD Frontier" [1], and present a design concept that incorporates new, innovative accelerator techniques to provide a cost-effective upgrade of RHIC with polarized electron beams colliding with the full array of RHIC hadron beams. The new facility will deliver electron-nucleon luminosity of $10^{33}$-$10^{34}$ cm$^{-2}$sec$^{-1}$ for collisions of 15.9 GeV polarized electrons on either 250 GeV polarized protons or 100 GeV/u heavy ion beams. The facility will also be capable of providing an electron beam energy of 21.2 GeV, at reduced luminosity. We discuss the on-going R&D effort to realize the project, and present key detector requirements and design ideas for an experimental program capable of making the "golden measurements" called for in Ref. [1]. We outline Brookhaven's plan to complete the scientific program of RHIC and make a smooth transition to the first eRHIC experiments by mid-to-late 2020s.

## 1.1 The Need for an Electron Ion Collider

In the four decades since the discovery of quarks, experiments have revealed an unexpected richness of nature as described by QCD. The substructure of the nucleon is not a simple system of three quarks, but a complex interaction of valence quarks and gluons, the force carriers of the strong interaction, along with virtual quarks and antiquarks. A full understanding of the relationship of this dynamic substructure to the observed spectrum of hadrons remains a challenge for theory and experiment. Each of these constituents carries its own intrinsic spin and orbital angular momentum, and a global program of precision measurements with high-energy polarized beams has begun to quantify how each contributes to the overall spin of the nucleon. The mechanism by which this complex dynamical system of quarks and gluons results in the characteristic spin-1/2 of the nucleon is not yet understood.

Neutrons and protons bound inside atomic nuclei exhibit collective behavior that, under extreme conditions, reveals its "QCD substructure". We now know through laboratory experiments, with high-energy heavy ion collisions at RHIC and the CERN LHC, that at temperatures and densities similar to those of the nascent universe moments after the Big Bang, nuclear matter is transformed to a state in which the relevant degrees of freedom are the quarks and gluons, rather than neutrons and protons. This quark gluon plasma takes the form of a strongly coupled fluid whose transport properties include a shear viscosity-to entropy density ratio consistent with the "perfect liquid", a conjectured quantum lower bound derived from string theory techniques. These results have brought widespread interest to the study of condensed matter of the strong force, and the understanding that the formation and evolution of this extreme phase of QCD matter is dominated by the properties of gluons at high density [2].

The most energetic nuclear collisions, including electron-proton collisions at HERA, point to the dominance of gluons in the structure of nuclear matter when probed at high energies (small Bjorken $x$). This arises from the property that the gluon carries a non-zero color charge. Thus gluons, unlike their electromagnetic analogue (the electrically neutral photon), can interact directly with each other.

The energy of self-interaction among gluons accounts for a significant fraction of the nucleon mass. In collisions at higher and higher energy, the density of gluons seen in the nucleon increases rapidly and without apparent limit, implying a correspondingly rapid and unchecked rise in the total nucleon-nucleon interaction. At high enough energy, this increase would violate unitarity, and therefore the growth of the gluon density must slow and saturate at some point. While this saturation has not yet been clearly observed, the mechanism for such an effect, and its consequences, are the subject of much theoretical activity. It is widely conjectured that such a saturated gluonic state, a "color glass condensate", may have universal properties and form the initial state for the quark gluon plasma produced in heavy ion collisions. The observation and quantitative study of this remarkable state of matter is predicted to be within reach of collisions of electrons with heavy ions in the Electron Ion Collider, eRHIC, being proposed here.

The exploration of nucleon structure and nuclear interactions at high energies in recent decades has brought many discoveries. It has opened surprising new avenues for the study of fundamental properties of strongly interacting matter and the role of QCD in the formation and structure of our natural world. A broad consensus now exists that new discoveries await, and that the next level of research, made possible by current accelerator technology, calls for a new facility colliding high-energy beams of electrons with beams of nucleons and heavy ions. Such a facility should have the capability to explore the structure of QCD matter with the precision of electromagnetic probes at high enough energies and with sufficient intensity to access the gluon-dominated regime with unprecedented statistical precision, and with polarized beams to enable a complete picture of the spin structure of the nucleon. The anticipated physics reach and a defining set of key measurements for such an Electron Ion Collider are given in Ref. [1]. The specific capabilities of eRHIC for this research are discussed in Section 1 of this document.

## 1.2 The eRHIC Design Concept

The eRHIC accelerator is designed to provide timely and cost-effective realization of the Electron Ion Collider (EIC) physics program, taking full advantage of recent advances in accelerator technology. This design adds a high current, multi-pass Energy Recovery Linac (ERL) and electron recirculation rings to the existing RHIC hadron facility to provide a polarized electron beam with energy 15.9 GeV colliding with ion species ranging from polarized protons with a top energy of 250 GeV to fully stripped Uranium ions with energies up to 100 GeV/u, and e-nucleon luminosity of $10^{33}$-$10^{34}$ cm$^{-2}$sec$^{-2}$.

As described in Section 3, the current eRHIC design uses just two Fixed Field Alternating Gradient (FFAG) magnet rings to carry the recirculating electrons. Recent studies have shown that the FFAG configuration provides a very simple and robust transport line, with large acceptance. The two FFAG rings transport 12 electron beam passes through the main ERL linac, operating at 1.32 GeV with a beam current of 10 mA, to produce a 15.9 GeV final electron beam. The FFAG rings can support 16 passes, producing a 21.2 GeV beam, but with luminosity reduced by a factor of 2-3 due to limits on beam power loss through synchrotron radiation.

To achieve the very **high** luminosity without requiring an unacceptably large electron beam current, the emittance of the hadron beam has to be very small – about 10 times smaller than presently available in hadron beams. This is also a requirement for certain small-angle physics measurements required of an EIC. Such small emittance requires a level of beam cooling that can only be achieved using Coherent electron Cooling (CeC), a novel form of beam cooling that promises to cool ion and proton beams by a factor of 10, both transversely and longitudinally, in less than 30 minutes. CeC will be tested in 2015-2016 in a proof-of-principle experiment at RHIC by a collaboration of scientists from BNL, JLab, and TechX. R&D is also under way at BNL on a high beam intensity ERL, and, at BNL, JLab, and MIT on source techniques for producing a 50 mA polarized electron beam.

A high-luminosity IR configuration for eRHIC has been designed, and is described in Section 3. In it, the beams cross at a 10 mrad angle, with zero magnetic field along the electron beam trajectory, and hence minimal synchrotron radiation in the +/-4.5m space reserved for the detector. For such a scheme to work, the finite-length bunches must be rotated so that they pass through each other at an effective angle of zero degrees as they cross. This "crab crossing" requires highly specialized RF cavities that are currently in the development stage.

The conceptual design for eRHIC is well advanced. As noted, in order to meet science-driven performance goals within realistic cost constraints the design incorporates several challenging technology developments. These are being addressed by intensive R&D efforts at BNL and elsewhere. We view it as realistic that the technical issues can be settled within the next five years and a final design readied to begin construction by the end of this decade.

There is not yet a detailed cost estimate for eRHIC, but the entire design process is aiming at a highly cost effective facility. The target is to provide the day-one eRHIC machine with one high-luminosity intersection region equipped with crab crossing cavities and a second interaction region upgradable to low beta star and available for a second detector. eRHIC detectors will also be very cost effective, as they can take full advantage of the existing infrastructure in the STAR and PHENIX experimental halls. A discussion of detector requirements, and available technology choices, is given in Section 4.

# 2 THE SCIENCE OF eRHIC

Quantum Chromodynamics (QCD), the theory of strong interactions, is a cornerstone of the Standard Model of modern physics. It explains all strongly interacting matter in terms of point-like quarks interacting by the exchange of gauge bosons, known as gluons. This strongly interacting matter is responsible for 99% of the visible mass in the universe. This mass derives from emergent phenomena of QCD that are not evident from its Lagrangian. Other phenomena include chiral symmetry breaking and confinement, which are fundamental features of the strong interactions. Lattice gauge theory and effective field theories have taught us that the rich and complex structure of QCD arises primarily from the dynamics of gluons with contributions from the quark sea. Unlike photons, the carriers of the electromagnetic force, gluons interact with each other. The underlying non-linear dynamics of this self-interaction is key in understanding QCD, but is hard to put under theoretical control. Despite the central role of gluons, their properties and dynamics remain largely unexplored. Despite the many successes in our understanding of QCD, some profound mysteries remain and leave our knowledge incomplete.

Experimental results suggest that both nucleons (see Figure 2-1) and nuclei, when viewed at high energies, appear as dense systems of gluons, dominating not only the hadronic structure but also creating fields whose intensity may be the strongest known in nature. The quest to probe this universal gluonic regime drives the development of eRHIC.

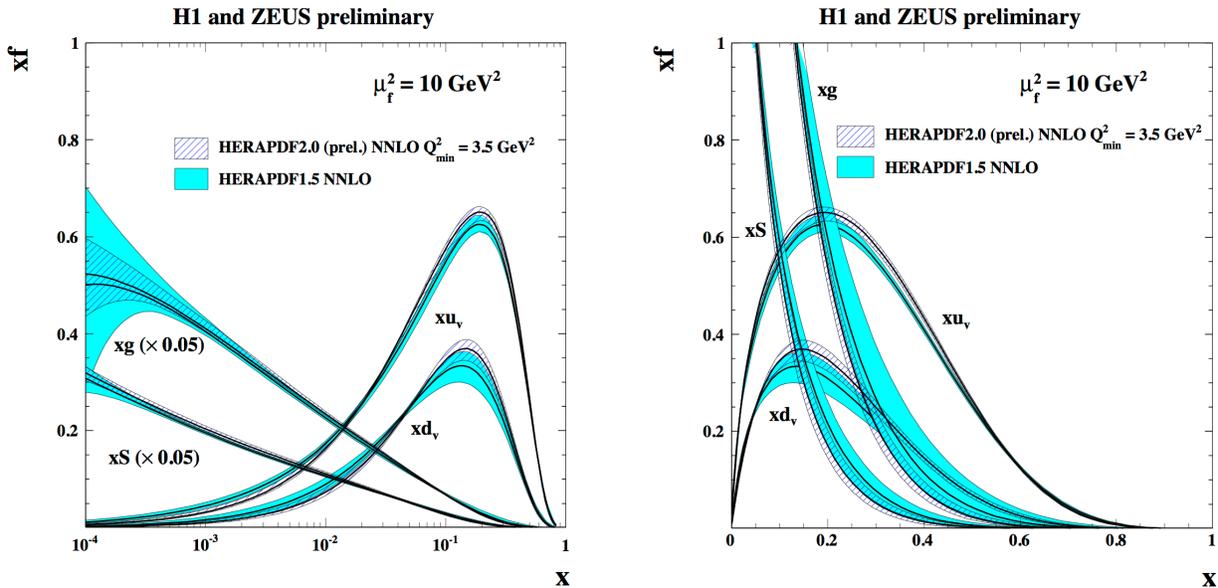

Figure 2-1: Proton Parton Distribution Functions (PDFs) of gluons, valence and sea quarks plotted as functions of Bjorken $x$. Already at a parton momentum $x \sim 0.3$ gluons dominate the nucleon structure.

In the 2007 Nuclear Physics Long Range Plan [3] a set of overarching questions has been defined for the subfield of Nuclear/Hadronic Physics. The goal was to guide the community in breaking the next nuclear science frontier. The questions are:
- What are the phases of strongly interacting matter, and what roles do they play in the cosmos?
- What is the internal landscape of the nucleons?

- What does QCD predict for the properties of strongly interacting matter?
- What governs the transition of quarks and gluons into pions and nucleons?
- What is the role of gluons and gluon self-interactions in nucleons and nuclei?
- What determines the key features of QCD, and what is their relation to the nature of gravity and space-time?

Answers to all but the first of these pressing questions make the realization of an EIC indispensable. Such a facility will address directly and with high precision questions that relate to our fundamental understanding of QCD. In the EIC White Paper [1] the questions have been further detailed:

- **How are the sea quarks and gluons, and their spins, distributed in space and momentum inside the nucleon?** How are these quark and gluon distributions correlated with overall nucleon properties, such as spin direction? What is the role of the orbital motion of sea quarks and gluons in building up the nucleon spin?

- **Where does the saturation of gluon densities set in?** Is there a simple boundary that separates this region from that of more dilute quark-gluon matter? If so, how do the distributions of quarks and gluons change as one crosses the boundary? Does this saturation produce matter of universal properties in the nucleon and all nuclei viewed at nearly the speed of light?

- **How does the nuclear environment affect the distribution of quarks and gluons and their interactions in nuclei?** How does the transverse spatial distribution of gluons compare to that in the nucleon? How does nuclear matter respond to a fast moving color charge passing through it? Is this response different for light and heavy quarks?

The parameters for an EIC can be derived directly from the above questions and were used to guide the design of eRHIC. A high-energy collider is needed to reach well into the gluon-dominated regime. As one increases the energy of the electron-nucleon collisions, one can access regions of progressively higher gluon density ($\sqrt{s} \sim x^{-1/2}$) as illustrated in Figure 2-1. **Electron beams** are needed to bring to bear the unmatched precision of electromagnetic interactions as a probe. Deep inelastic scattering (DIS) of electrons with hadrons is dominated by one photon exchange preserving the properties of partons in the hadronic wave functions because there is no direct color interaction between the exchanged photon and the partons. This is in contrast to hadron-hadron scattering where the parton scattering occurs dominantly through color exchange. Electron beams also allow for the precise determination of the indispensable kinematic variables $x_{Bj}, Q^2$ from the scattered lepton, which provide a clean access to the parton kinematics. **Polarized nucleon and electron beams** are needed to determine the correlations of sea quark and gluon distributions with the nucleon spin. **Heavy ion beams** are required to provide precocious access to the regime of saturated gluon densities. The scale that defines this novel regime, the saturation scale $Q_s$, increases with increasing ion mass making the nucleus an efficient **amplifier** of the physics of high gluon densities. A **wide range of ion beams from light to heaviest mass** offers a precise dial in the study of propagation-length for color charges in nuclear matter. **High luminosity** is required to unravel the multidimensional dependencies of the different physics processes on the kinematic variables $x, Q^2, p_T, z,$ and $\Phi$, representing respectively, the momentum fraction of the parton on which the photon scatters, the squared momentum transfer to the lepton, the transverse momentum of the final state hadron with respect to the virtual photon in the center-of-mass of the virtual photon and the nucleon, the momentum fraction of the final state hadron with respect to the virtual photon and the azimuthal angle of the final state hadron with respect to the lepton plane.

While past or existing DIS experiments were and are very successful in determining the polarized quark structure of the proton and of some light and intermediate-size nuclei, none matches the unique capabilities of an eRHIC. Fixed target experiments such as HERMES, COMPASS, and those at JLab are not suitable to reach deep into the gluon-dominated region since they do not provide the necessary kinematic reach in $x$ and $Q^2$. HERA, till to-day the only high-energy e+p collider provided the best measurement of the gluonic structure of the proton but neither provided nuclear beams nor polarized nucleon beams. The unique kinematic reach of eRHIC for polarized electron-proton and electron-ion collisions is illustrated in Figure 2-2.

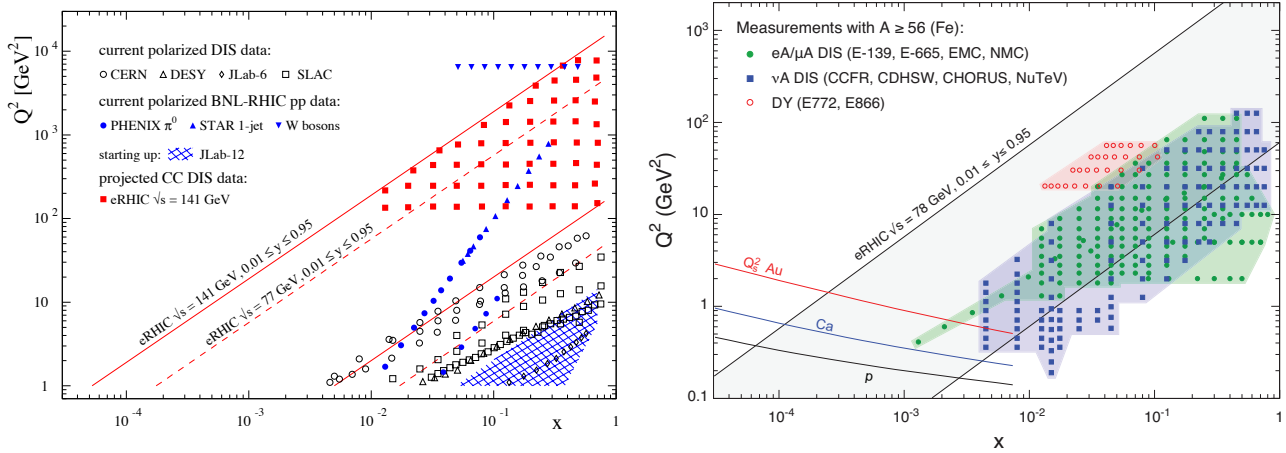

Figure 2-2: *Left*: The range in parton momentum fraction *x* vs. the square of the momentum transferred by the electron to the proton $Q^2$ accessible with eRHIC in polarized e+p collisions from 15 GeV on 100 GeV to 20 GeV on 250 GeV compared to past (CERN, DESY, SLAC) and existing (JLAB, COMPASS) facilities as well to polarized p+p collisions at RHIC. *Right*: The kinematic acceptance in $x$ and $Q^2$ of completed lepton-nucleus (DIS) and Drell-Yan (DY) experiments (all fixed target) compared to the eRHIC acceptance at nominal beam energies of $E_e = 15$ GeV and $E_A = 100$ GeV ($\sqrt{s} = 78$ GeV). The acceptance bands for eRHIC are defined by $Q^2=xsy$ with $0.01 \leq y \leq 0.95$. The red, blue and black curves indicate the predicted saturation scale for Au, Ca, and protons, respectively.

eRHIC will distinguish itself from all past, current, and contemplated facilities around the world by addressing the above questions with the highest, unprecedented precision for the first time and at one facility. In Table 2-1 the unique key measurements and their underlying physics goals that will be made possible at an eRHIC are summarized, and several of them are outlined in more detail in the following sections. In Sec. 1.1 we describe studies in polarized e+p collisions related to the spin and the 2+1-dimensional momentum and spatial structure of the nucleon. Sec. 1.1 discusses measurements in e+A with focus on the high gluon density regime as well studies of hadronization and energy-loss. Chapter 4 describes in detail how the requirements to perform all the different outlined physics measurement with high precision and low systematic uncertainties have been integrated in potential detector designs.

| Deliverables | Observables | What we learn | |
|---|---|---|---|
| **Proton Spin** | | | |
| Polarized gluon distribution $\Delta g$ | Scaling violation in inclusive DIS | Gluon contribution to proton spin | Sec. 2.1.1<br>Figure 2-3 to<br>Figure 2-6 |
| polarized quark and anti-quark densities | Semi-incl. DIS for pions and kaons | quark contribution to proton spin; asym. $\Delta \bar{u} - \Delta \bar{d}$; $\Delta s$ | |
| Novel electroweak spin structure functions | Inclusive DIS at high $Q^2$ | Flavor separation at medium $x$ and high $Q^2$ | Figure 2-7 |
| **The motion of quarks and gluons in the proton** | | | |
| Sivers & unpolarized quark and gluon TMDs | Semi-incl. DIS with transverse polarization; di-hadron (di-jet) | Quantum Interference & Spin-Orbital correlations<br>3D Imaging of quark's motion: valence + sea<br>QCD dynamics in a unprecedented $Q^2$ ($P_T$) range | Sec. 2.1.2<br>Figure 2-9<br>Figure 2-10 |
| Chiral-odd functions; Transversity; Boer-Mulders | Semi-incl. DIS with transverse polarisation | 3$^{rd}$ basic quark PDFs, tensor charge<br>Novel spin-dependent hadronization effect<br>QCD dynamics in a chiral-odd sector with a wide $Q^2$ ($P_T$) coverage | Figure 2-11 |
| **The tomographic images of the proton** | | | |
| GPDs of sea quarks and gluons | DVCS and $J/\Psi$, $\rho^0$, $\phi$ production cross section and polarization asymmetries | Transverse spatial distrib. of sea quarks and gluons; total angular momentum and spin orbit correlations | Sec. 2.1.3<br>Figure 2-13 to<br>Figure 2-16 |
| GPDs of valence and sea quarks | Electro-production of $\pi^+$, $K$ and $\rho^+$, $K^*$ | Dependence on quark flavor and polarization | |
| **QCD matter at an extreme gluon density** | | | |
| Gluon momentum distribution $g_A(x,Q^2)$ | $F_2$, $F_L$ and $F_2^{charm}$ | Nuclear wave function;<br>$Q^2$ evolution: onset of DGLAP violation<br>Saturation<br>$A$-dependence of (anti-)shadowing | Sec. 2.2.1<br>Figure 2-20 to<br>Figure 2-23 |
| $k_T$-dependent gluon distributions $f(x,k_T)$; gluon correlations | Di-hadron correlations | Non-linear QCD evolution/universality;<br>Saturation scale $Q_s$ | Figure 2-25<br>Figure 2-26 |
| Spatial gluon distribution $f(x,b_T)$; gluon correlations | Diffractive dissociation $\sigma_{diff}/\sigma_{tot}$<br>$d\sigma/dt$ and $d\sigma/dQ^2$ for vector mesons & DVCS | Non-linear QCD small-$x$ evolution;<br>Saturation dynamics;<br>black disk limit | Figure 2-29 to<br>Figure 2-33 |
| **Quark hadronization** | | | |
| Transport coefficients in nuclear matter | Productions of light and heavy hadrons and jets in semi-incl. DIS | Color neutralization: mass dependence of hadronization;<br>Multiple scattering and mass dependence of energy loss;<br>Medium effect of heavy quarkonium production | Sec. 2.2.2<br>Figure 2-35<br>Figure 2-36 |
| Fluctuations of the nuclear density | $\phi$-modulation of light and heavy meson production in semi-incl. DIS | Color fluctuations - connection to heavy ion physics; | |

Table 2-1: Unique key measurements at eRHIC

## 2.1 Electron Proton Scattering

As described above, eRHIC will open up the unique opportunity to go far beyond our current largely one-dimensional picture of the nucleon. It will enable parton femtoscopy by correlating the information on the individual parton contribution to the nucleon spin with its transverse momentum and spatial distribution inside the nucleon. To understand how the constituents of the proton carry the proton's spin has been a defining question in hadron structure for several decades now, but remains unanswered. Unraveling the proton spin presents the formidable challenge of understanding an essential feature of how a complex strongly-interacting many-body system organizes itself to produce a simple result and goes directly to the heart of exploring and understanding the QCD dynamics of matter. To provide definitive answers in this area will be among the key tasks of eRHIC.

Partons can have a small momentum component transverse to the direction of their parent fast moving nucleon. Experimental evidence supports an average transverse momentum of a few hundred MeV. However, much of our understanding of nucleon structure today is in terms of integrated parton distributions that describe the distribution of longitudinal momentum within a fast-moving nucleon, with $k_T$ effects being integrated over. Transverse momentum distributions (TMDs) are an essential step toward a more comprehensive understanding of the parton structure of the nucleon in QCD. eRHIC will enable precise and detailed measurements of TMDs over a broad kinematic range. TMDs not only quantify the magnitude of the parton transverse momentum, but also the transverse momentum direction, yielding strikingly asymmetric azimuthal distributions. A golden measurement at eRHIC will be the Sivers asymmetry, a particular angular correlation between the target polarization and the direction of a produced final state hadron in polarized semi-inclusive DIS (SIDIS). At the parton level, the Sivers effect probes a spin-orbit coupling in QCD and is described by a TMD that quantifies how strongly the transverse momentum from orbital motion is coupled to spin. These partonic spin-orbit correlations are analogous to those observed in atomic systems such as hydrogen atoms, but in the strong coupling regime. . It was found that the Sivers functions are not universal in hard-scattering reactions. Depending on the process, the associated color Lorentz forces will act in different ways on the parton. In DIS, the final-state interaction between the struck parton and the nucleon remnant is attractive. In contrast, for the Drell-Yan process it becomes an initial-state interaction and is repulsive. As a result, the Sivers functions contribute with opposite signs to the single-spin asymmetries for these two processes. This is a fundamental prediction about the nature of QCD color interactions, directly rooted in the quantum nature of the interactions.

The high luminosity and large kinematic reach of eRHIC offers unique possibilities for exploring the spatial distribution of sea quarks and gluons in the nucleon and in nuclei. The transverse position of the quark or gluon on which the scattering took place is obtained by a Fourier transform from the momentum transfer of the scattered nucleon or nucleus. By choosing particular final states, measurements at eRHIC will be able to selectively probe the spatial distribution of sea quarks and gluons in a wide range of longitudinal momentum fractions $x$. Such "tomographic images" will provide essential insight into QCD dynamics inside hadrons, such as the interplay between sea quarks and gluons and the role of pion degrees of freedom at large transverse distances. The quantities that encode such tomographic information are Generalized Parton Distributions (GPDs). GPDs directly quantify, unlike TMDs, the quark and gluon orbital angular momenta, the other essential ingredients in understanding the spin of the proton apart from the contributions of the quark and gluon intrinsic spins.

Details about the required eRHIC performance to reach these scientific goals are given in the following.

## 2.1.1 The Longitudinal Spin of the Nucleon

Helicity-dependent parton densities encode the information to what extent quarks and gluons with a given momentum fraction *x* tend to have their spins aligned with the spin direction of a nucleon. The most precise knowledge about these non-perturbative quantities, along with estimates of their uncertainties, is gathered from comprehensive global QCD analyses [4] to all available data taken in spin-dependent proton-proton collisions and DIS, with and without additional identified hadrons in the final state.

Apart from being essential for a comprehensive understanding of the partonic structure of hadronic matter, helicity PDFs draw much their relevance from their relation to one of the most fundamental and basic but yet not satisfactorily answered questions in hadronic physics, namely how the spin of a nucleon is composed of the spins and orbital angular momenta of quarks and gluons. The integrals of helicity PDFs over all momentum fractions *x* (first moments) at a resolution scale $Q^2$, $\Delta f(Q^2) \equiv \int_0^1 \Delta f(x, Q^2) dx$, provide information about the contribution of a given parton flavor *f* to the spin of the nucleon. A precise determination of the polarized gluon $\Delta g(x,Q^2)$ and quark $\Delta q(x,Q^2)$ distribution functions in a broad kinematic regime is a primary goal of eRHIC. Current determinations of $\Delta g$ suffer from both a limited $x$-$Q^2$ coverage and fairly large theoretical scale ambiguities in polarized p+p collisions for inclusive (di)jet and pion production [5].

Several channels are sensitive to $\Delta g$ in e+p scattering at collider energies such as DIS jet or charm production, but QCD scaling violations in inclusive polarized DIS have been identified as the golden measurement. Scaling violations are a key prediction of QCD for PDFs and have been used successfully at HERA to determine the unpolarized gluon distribution with high precision. The inclusive DIS structure function $g_1(x,Q^2)$ is the most straightforward probe in spin physics and has been determined at various fixed-target experiments at medium-to-large values of *x* in the last two decades. It is also the best-understood quantity from a theoretical point of view.

A consistent framework up to beyond next-to-next-to-leading order accuracy will be in place by the time of first eRHIC operations and is required in order to match the size of residual theoretical scale uncertainties to the anticipated unprecedented level of precision for a polarized DIS experiment. To achieve the latter, systematic uncertainties need to be controlled extremely well, which imposes stringent requirements on the detector performance, acceptance, and the design of the interaction region.

For studying DIS scaling violations, i.e., $dg_1(x,Q^2)/dlogQ^2$, efficiently, it is not only essential to have good precision but also to cover the largest possible range in $Q^2$ for any given fixed value of *x*. The accessible range in $Q^2$ is again linked (via the inelasticity *y*) to the capabilities of detecting electrons in an as wide as possible range of momenta and scattering angles.

To estimate the impact of eRHIC on our understanding of helicity PDFs [6], sets of pseudo-data were generated with the PEPSI Monte Carlo generator for different c.m.s. energies within the typical DIS kinematics ($Q^2 \geq 1\text{GeV}^2$, $W^2 \geq 10$ GeV$^2$, and $0.01 < y < 0.95$). The range of y is further restricted from below by constraining the depolarization factor *D(y)* of the virtual photon to be larger than 0.1.

To ensure detection of the scattered lepton we require a minimum momentum of 0.5 GeV, and, in case of SIDIS, only hadrons with a momentum larger than 1 GeV and a fractional energy in the range *0.2<z<0.9* are accepted. Monte Carlo data for the ratio $g_1/F_1$ in DIS and SIDIS are generated in 4 [5] bins per decade in *log $Q^2$ [log x]*. We note that the typical size of the experimentally relevant double spin asymmetry
$A_{||}(x,Q^2)=D(y)g_1(x,Q^2)/F_1(x,Q^2)$ at the lowest *x* values accessible at eRHIC can be as small as a few times $10^{-4}$, depending on the yet unknown behavior of $\Delta g(x,Q^2)$ in this kinematic regime. This size sets the scale at which one needs to control systematic uncertainties due to detector performance or luminosity measurements. Most likely, the dominant source of systematic uncertainty will be the determination of the beam polarizations.

Figure 2-3 illustrates the simulated data sets for inclusive polarized DIS at eRHIC for the three different choices of c.m.s. energies. The error bars reflect the expected statistical accuracy for a integrated luminosity of 10 fb$^{-1}$ and assuming 70% beam polarizations, this corresponds to 6 month running time at $10^{33}$cm$^{-2}$s$^{-1}$ assuming 50% efficiency at

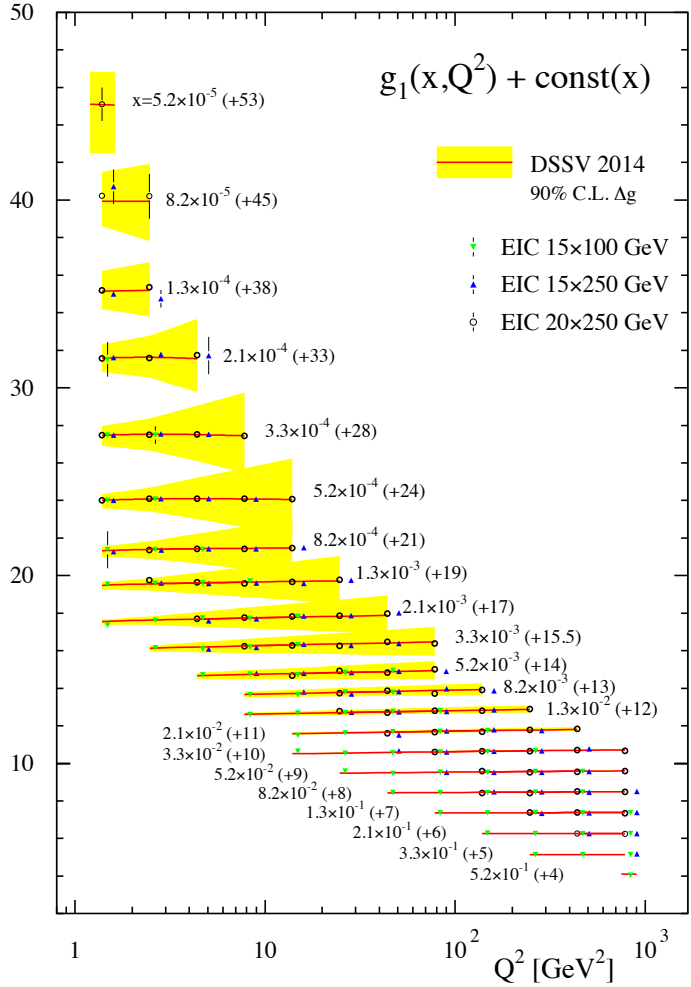

Figure 2-3: Projected eRHIC data for the structure function $g_1(x,Q^2)$ for different combinations of electron and proton beam energies (15 GeV on 100 GeV/250 GeV and 20 GeV on 250 GeV). Constants are added to $g_1$ to separate the different $x$ bins. The solid lines are the result of the DSSV 2014 best fit, and the shaded bands illustrate the current uncertainty estimate for 90% C.L. variations of $\Delta g$. DSSV 2014 includes all currently available (SI)DIS and polarized p+p data, for details see [6]. The statistical uncertainties correspond to an integrated luminosity of 10 fb$^{-1}$ and 70% beam polarizations.

15 GeV on 250 GeV. The existing fixed target DIS data (see Figure 2-2 (left)) populate only the lower left corner of the kinematic plane shown.

The simulated *inclusive* DIS-data are used in a pQCD-fit based on DSSV 2014 to study what precision can be achieved for the first moments of the flavor singlet combination $\Delta\Sigma$ and the gluon helicity density $\Delta g$, which both enter the proton spin rule $\frac{1}{2} = \frac{1}{2}\Delta\Sigma + \Delta g + \sum_q L_q^z + L_g^z$, with $L_{q,g}^z$ denoting the contribution from orbital angular momentum, which is not directly measurable in inclusive DIS. Figure 2-4 shows the running integral for the gluon contribution $\int_{x_{min}}^{1} dx\, \Delta g(x, Q^2)$ at $Q^2$=10 GeV$^2$ as function of the lower integration limit $x_{min}$ for the central value of DSSV 2014 (red line) and its 90% C.L. uncertainty band (cyan). The impact of the simulated inclusive DIS pseudo data for three different eRHIC center-of-mass energies is illustrated by the significantly reduced uncertainty bands (grey to marine blue). eRHIC will determine the integral of the helicity gluon distribution at an $x_{min}$ of 5×10$^{-5}$ to about +/- **10%**. Figure 2-5 shows the running integral $\int_{x_{min}}^{1} dx\, \Delta\Sigma(x, Q^2)$, which represents the contribution of all quark flavors to the spin of the proton. As for the gluons, the impact of the eRHIC data is unprecedented and allows determining the integral with an accuracy of about +/- **20%**. Comparing the convergence of both integrals depicted in Figure 2-4 and Figure 2-5 as function of $x_{min}$, it is obvious that the quark singlet $\Delta\Sigma(x, Q^2)$ converges less well. The poor convergence is imprinted on the current fits by imposing a constraint affecting the unmeasured small-$x$ behavior of $\Delta\Sigma(x, Q^2)$, in particular, of the strangeness helicity density. eRHIC will be able to verify the validity of this assumption, e.g., by determining $\Delta s(x)$ down to unprecedented values of $x$. It is expected that such measurements will considerably improve the convergence of $\Delta\Sigma(x, Q^2)$ shown in Figure 2-5 and to reduce the uncertainties to a similar level as for $\Delta g$.

Figure 2-6 demonstrates that combining the information on $\Delta\Sigma$ and $\Delta g$ from eRHIC will yield an excellent indirect constraint on the total amount of quark and gluon orbital angular momentum present in the proton. Would the eRHIC data confirm the functional forms of the polarized quark and gluon distributions of the DSSV 2014 best fit (as indicated by the red lines), only very little net angular momentum is expected at a scale of $Q^2$=10 GeV$^2$.

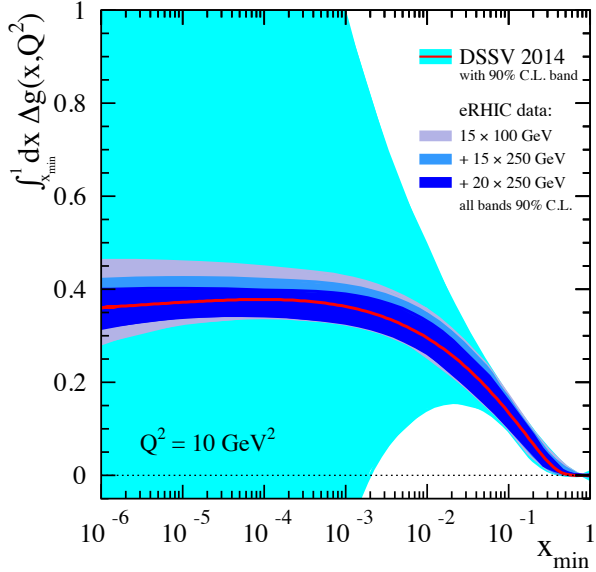

Figure 2-4: The running integral $\int_{x_{min}}^{1} dx \Delta g(x, Q^2)$ and 90% C.L uncertainty estimates based on DSSV 2014 with and without including various sets of simulated eRHIC data.

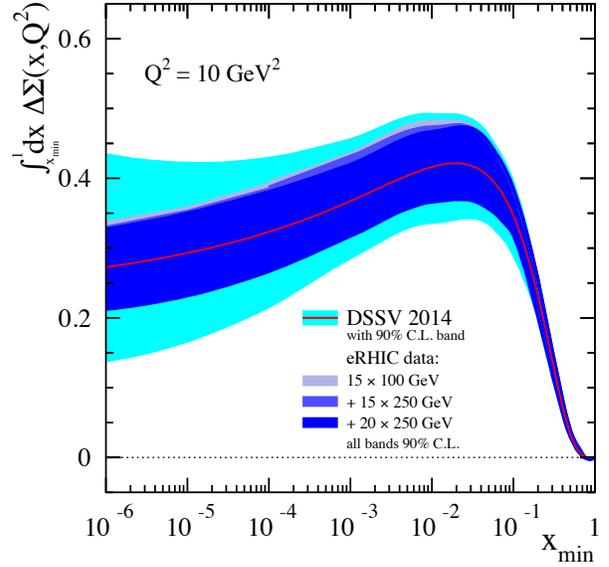

Figure 2-5: As in Figure 2-4 but now for $\int_{x_{min}}^{1} dx \Delta\Sigma(x, Q^2)$

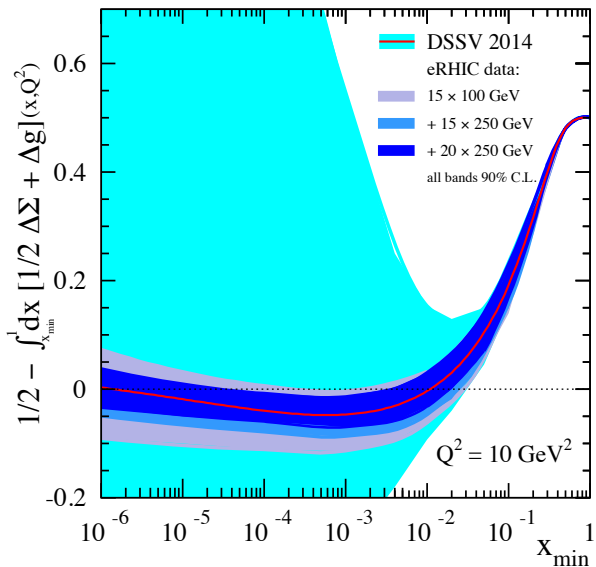

Figure 2-6: The expected net contribution from quark and gluon orbital angular momentum as a function of xmin calculated from the results for $\Delta g$ and $\Delta\Sigma$ shown in Figure 2-4 and Figure 2-5.

The use of charged leptons to probe the structure of nucleons through electroweak interactions has proven to be an invaluable tool in our exploration of the strong force. Experiments on deep inelastic scattering ep → eX dominantly proceed via the exchange of a virtual photon between the electron and the nucleon. However, at high enough momentum transfer $Q^2$, the exchange of massive Z and $W^\pm$ gauge bosons contributes as well to neutral and charged current DIS, respectively. Charged current (CC) interactions in DIS lepton scattering measurements have been performed at HERA in $e^\pm p$ collisions and at various neutrino scattering experiments.

**They are inaccessible at fixed target charged lepton beam facilities where $Q^2 \ll M^2_W$.**

eRHIC provides a number of essential advantages in the study of (un)polarized structure functions and parton distribution functions through electroweak interactions over previous and existing facilities:

- As the asymmetries and relative likelihood of $Z^0$ and $W^\pm$ exchange increase with $Q^2$ due to the large mass of the $Z^0$ and $W^\pm$, larger c.m.s. energies are more favorable for such measurements.

- Because of the maximum parity violating nature of the CC current interaction, it provides access to the flavor structure of the nucleons **without** the complication of tagging the struck quark flavor from some observed final-state hadron in SIDIS through the use of fragmentation functions.

- In addition, advances in accelerator and source technologies will provide luminosities up to $10^{33}$ cm$^{-2}$s$^{-1}$, two orders of magnitude higher than what was available at HERA, which will yield unprecedented precision in electroweak observables in DIS

- eRHIC will have the ability for bunch-by-bunch variations of the sign of the longitudinal polarization of the hadron beams. This will for the first time allow measuring polarized parton distributions through single-spin asymmetries in CC interactions.

- A broader $Q^2$ and $y$ acceptance than at fixed target facilities, and variable beam energy, also allow for a separation of the various structure functions entering the CC cross section.

Figure 2-7 shows the simulated single spin asymmetries for CC DIS off polarized proton and neutron beams assuming a c.m.s. energy of $\sqrt{s}$ = 141 GeV [7]. The top panel shows $A_L^{W-,p}$, which is positive and takes values ranging from a few percent at the smallest x value to more than 80% at $x \simeq 0.7$. $A_L^{W-,n}$ (bottom panel) is negative and somewhat smaller in size, reaching about −50% at $x \simeq 0.7$. The estimated errors reflect the statistical accuracy for an integrated luminosity of 10 fb$^{-1}$ after unfolding detector smearing and radiative effects. The event kinematic $(x,Q^2)$ is reconstructed from the hadronic final state.

At high x these asymmetries give direct access to the polarization values $\Delta q/q$ for u and d quarks. The current world data constrain the polarizations to approach 1 for $\Delta u/u$ and approximately −0.6 for $\Delta d/d$ in global pQCD fits. While $\Delta u/u$ at large x is pretty well constrained from existing fixed-target DIS data, there are theoretical expectations based on the concept of "helicity retention" [8] that $\Delta d/d$ should also approach 1 for $x \rightarrow 1$. Such a behavior would require a dramatic change in the trend seen in the present fixed-target data. Measurements of $A_L^{W-,n}$ would be particularly suited to study a possible sign change in $\Delta d/d$ at large values of x. To make this more quantitative, the dotted lines in Figure 2-7 are obtained with a special set from DSSV where $\Delta d/d \rightarrow 1$ is enforced. The resulting $A_L^{W-,n}$ for x > 0.2 are quite different from the standard DSSV predictions (green bands in Figure 2-7), these differences can be easily resolved with the statistical precision available at eRHIC.

At smaller values of x, where valence quark contributions are dying out, various combinations of light sea quark polarizations can be studied. Of special interest is the (un)polarized strange sea quark distributions, which still constitutes a big uncertainty in predictions for many beyond the Standard Model physics observables. It is noted again that CC observables allow one to tag the struck quark flavor without involving fragmentation functions, which yields sizable additional uncertainties on the quark distributions.

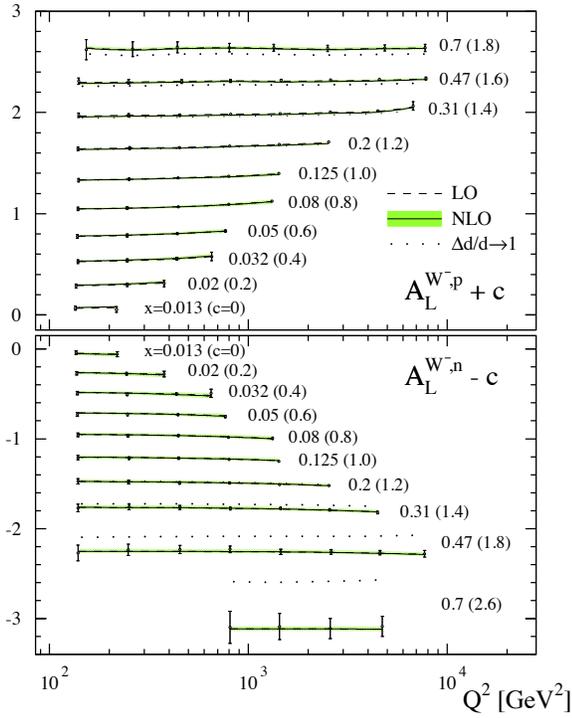

Figure 2-7: Projected single-spin asymmetries $A_L^{W^-,p}$ (top panel) and $A_L^{W^-,n}$ (bottom panel) for $\sqrt{s} \sim 141$ GeV (open circles) compared to LO and NLO calculations using the DSSV helicity densities. The dotted line shows an alternative DSSV set, which enforces $\Delta d/d \to 1$ as $x \to 1$ (see text). The shaded bands correspond to the DSSV uncertainty estimates. Note that a constant c is added to each bin as indicated.

In summary, the eRHIC longitudinal polarized e+p data will clarify without any doubt the intrinsic spin contributions from quarks and gluons to the spin of the proton. Utilizing the complementarity of inclusive and semi-inclusive as well as charge current measurements, eRHIC data will provide a full flavor separation and unprecedented constraints on the functional form of the polarized parton distributions as function of $x$ and $Q^2$. This information combined with the measurements eRHIC will provide to constrain GPDs will for the first time allow one to unravel the orbital angular momenta of quarks and gluons yielding a complete decomposition of the different contributions to the spin of the proton. Quark and gluon orbital angular momenta will be part of another suite of unique measurements at eRHIC aiming at the nucleons spatial structure (see section 2.1.3).

## 2.1.2 The Confined Motion of Partons in Nucleons: TMDs

The consolidated understanding of the nucleon structure from DIS experiments is till today basically one-dimensional. From inclusive DIS we learn about the longitudinal motion of partons in a fast moving nucleon, i.e. their light-cone momentum fraction x. In inclusive DIS the nucleon appears as a bunch of fast-moving quarks, antiquarks and gluons, whose transverse momenta are not resolved. A fast moving nucleon is Lorentz-contracted but its transverse size is still about 1 fm, which is a typical scale of non-pertubative interactions, where phenomena such as confinement are at work. Important questions in this context are:

- How are quarks spatially distributed inside the nucleon?
- How do they move in the transverse plane?
- Do they carry orbital angular momentum?
- Is there a correlation between orbital motion of quarks, their spin, and the spin of the nucleon?
- How can we access information on such spin-orbit correlations, and what will this tell us about the nucleon?

Recent theoretical progress has put many of these questions on a firm field-theoretical basis. We still lack quantitative answers to most questions, but we have now a much better idea on how to obtain them. The past decade has also witnessed tremendous experimental achievements, which lead to fascinating new phenomenological insights into the structure of the nucleon. The above questions address two complementary aspects of the nucleon structure: the distribution of quarks and gluons in the transverse plane in momentum space and in coordinate space. The field-theoretical tools adequate to describe the former are the Transverse Momentum Dependent Parton Distribution Functions (TMD PDFs, or, shortly, TMDs). The field-theoretical objects tailored to describe the spatial distributions of quarks in the transverse plane are the Generalized Parton Distributions (GPDs).

The focus of this section is on TMDs and their partonic interpretation. There will be also a short introduction about Transverse Momentum Dependent Fragmentation Functions (TMD FFs). The TMDs contain information on both the longitudinal and transverse (sometimes called intrinsic) motion of quarks and gluons inside a fast moving nucleon. When including spin degrees of freedom TMDs link information on the intrinsic spin of a parton ($s_{q,g}$) and their transverse motion ($k_\perp^{q,g}$) to the spin direction of the parent nucleon. At leading twist level the most general spin dependent TMD is usually denoted by $f_1^{q,g}(x, k_\perp^{q,g}; s_{q,g}, S)$. At leading order, there are eight such combinations, leading to eight independent TMDs, see Figure 2-8.

### Leading Twist TMDs

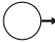

Figure 2-8: Leading twist TMDs classified according to the polarizations of the quark (f, g, h) and nucleon (U, L, T). For gluons a similar classification of TMDs exists.

A similar correlation between spin and transverse motion can occur in the fragmentation process of a transversely polarized quark/gluon, with spin vector $s_{q,g}$ and three-momentum $k_q$, into a hadron with longitudinal momentum fraction $z$ and transverse momentum $p_\perp$ (with respect to the direction of the fragmenting parton); such a mechanism is called the **Collins effect** and manifests itself in the fragmentation function via a $s_q \cdot (k_q \times p_\perp)$ correlation which leads to specific azimuthal modulations of the observed hadrons.

In the following, we use the **Sivers function** as an example for how well transverse momentum dependent distribution functions can be measured at eRHIC. The Sivers function $f_{1T}^{\perp a}(x, k_\perp)$ appears in the distribution of unpolarized partons inside a transversely polarized proton. It links the parton's intrinsic motion to the spin of the proton $f_1^a(x, k_\perp; S) = f_1^a(x, k_\perp) - \frac{1}{M} F_{1T}^{\perp a}(x, k_\perp) S \cdot (\hat{P} \times \widehat{k_\perp})$.

The Sivers function offers new information and plays a crucial role in our understanding of the nucleon structure: Its very origin is a clear indication of the existence of parton orbital motion in the proton wave function and its expected process dependence is related to fundamental QCD effects. Till today the Sivers function was **only** observed in the valence quark region at fixed target experiments.

The Sivers asymmetries for $\pi^+$ production at eRHIC were simulated using the transverse spin Monte Carlo generator gmc_trans [9]. Beam energies of 15 GeV for the electron and 100 GeV for the proton were used. The parameterization of [10] was used for the up and down quark Sivers distribution functions. The Sivers distributions of sea quarks are currently unknown, and, therefore, only the positivity limit, $f_1^a(x, k_\perp; S) \geq 0$, can be applied as an upper bound. As saturation of the positivity limit is already ruled out by existing data, a modest Sivers distribution of 10% of the positivity limit (i.e. 10% of $f_{\text{unpolarized}}$) was used for each sea quark flavor. Events were generated for $Q^2 > 1$ GeV$^2$ using the GRSV-2000 LO standard scenario PDFs and the DSS fragmentation functions. A cut of $0.01 < y < 0.95$ was also applied, and events in which the generated $\pi^+$ had a momentum fraction $z < 0.1$ were also rejected.

Events were binned four-dimensionally in $x$, $Q^2$, $z$ and hadron $p_T$ (w.r.t. the direction of the exchanged virtual photon) and the mean Sivers asymmetry per bin was calculated. Statistical uncertainties correspond to an integrated luminosity of 10 fb$^{-1}$. This corresponds to 6 month of running time at $10^{33}$cm$^{-2}$s$^{-1}$, assuming 50% efficiency at 15 GeV on 250 GeV.

The results of the simulation are shown in Figure 2-9. Two representative $z$ and $p_T$ bins are selected; the growth of asymmetries with both $z$ and $p_T$ can be seen. In each ($z$, $p_T$) bin, asymmetries are shown as a function of $x$ for four different $Q^2$ bins between 1 and 10 GeV$^2$. We note that there is no TMD evolution included in any of the available MC codes at present. Points for different $Q^2$ are re-scaled by a factor to separate them, while the error bars remain unchanged.

By construction, given the Sivers input of only 10% of the positivity bound for sea quarks, the generated asymmetries at low x are expected to be small. However, with a data set on the order of 10 fb$^{-1}$ even these modest asymmetries will be measurable simultaneously as a function of $x$, $Q^2$, $z$, and $p_T$ at eRHIC. Such a multi-dimensional analysis of the Sivers function (and other TMDs) is vital to truly ascertain its properties, and is a unique strength of eRHIC. Present and upcoming data are too limited in their kinematic reach and statistical precision to allow for such type of analyses to be performed. Only with eRHIC can we gather a full understanding of the physics of TMDs.

To separate pertubative and non-pertubative contributions to the observed $p_t$ of hadrons is experimentally a challenging task, especially the study of primordial transverse momentum ($k_T$) in the nucleon and nucleus has a long history (for examples of some of the earliest work see [11,12]). While a theoretically precise definition and separation can be challenging (see e.g. Reference [13] for a modern discussion), it was recognized in the 1980s that a distinction could be made between three different sources of transverse momentum:

- The non-pertubative transverse momentum, referred to as "intrinsic or primordial" $k_T$;
- The $p_T$ generated during the collision by either hard QCD processes, i.e. photon-gluon fusion (PGF), QCD Compton (QCDC), or parton showering;
- The $p_T$ acquired in the hadronization process.

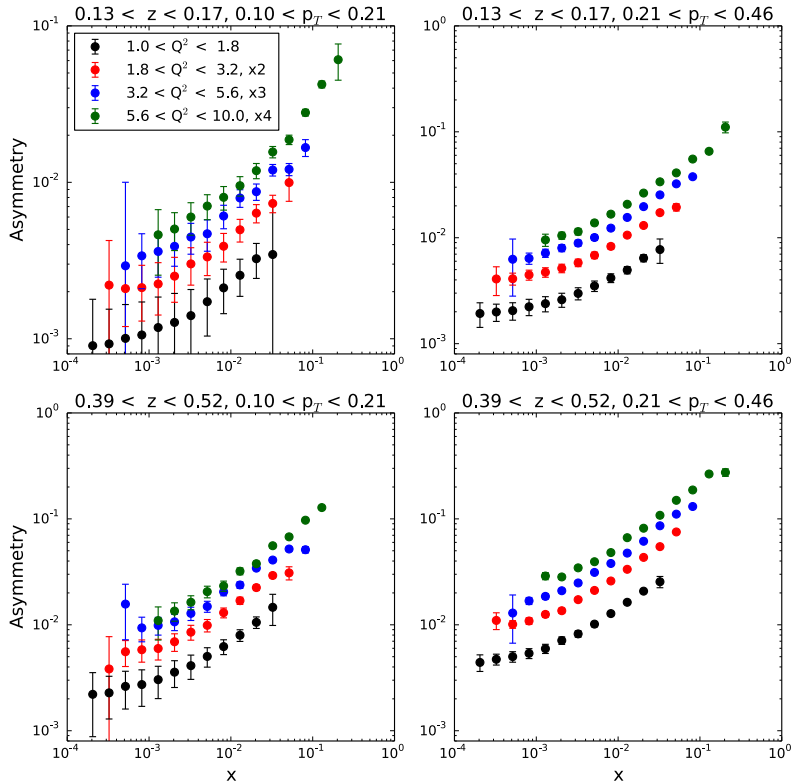

Figure 2-9: Simulated Sivers asymmetries as a function of $x$ for different bins in $Q^2$, $z$ and $p_T$. Values for different $Q^2$ bins are scaled by a factor to separate them, but error bars remain unscaled

Models were implemented in MC generators [14], and experimental collaborations such as EMC [15] were able to use experimentally distinct signatures of the sources of transverse momentum $p_T$ to tune the model parameters meaningfully. In particular, they found that these different mechanisms contributed to the $p_T$ in distinct regions of Feynman-x ($x_F = \frac{2p_\parallel^*}{W}$; here $p_\parallel^*$ is the longitudinal momentum of the particle in the virtual photon-proton CM frame with respect to the direction of the beam proton). Primordial $k_T$ contributes directly to the "current jet" or the $x_F>0$ region, and the hadron remnant recoil leads to an equal and opposite contribution at $x_F<0$. These effects are distributed proportional to $x_F$ and therefore contribute primarily at $|x_F|>0.2$. In contrast, both hard and soft QCD are essentially radiation due to the acceleration of the scattered parton and are concentrated in the current jet (positive $x_F$ region) with some contribution near $x_F=0$ and very little impact on the hadron remnant jet (negative $x_F$ region). Finally, the effects of hadronization, e.g., cluster fragmentation or string breaking effects, are largely independent of $x_F$ and tend to be subdominant in any case.

The EMC collaboration [15] measured the so called "seagull plot" showing $<p^2_T>$ vs. $x_F$, for produced charged hadrons. It was shown that the $x_F<-0.2$ particles were the ideal measures of primordial $k_T$, while the $x_F>0.2$ particles measure a combination of intrinsic and dynamic (QCD) effects.

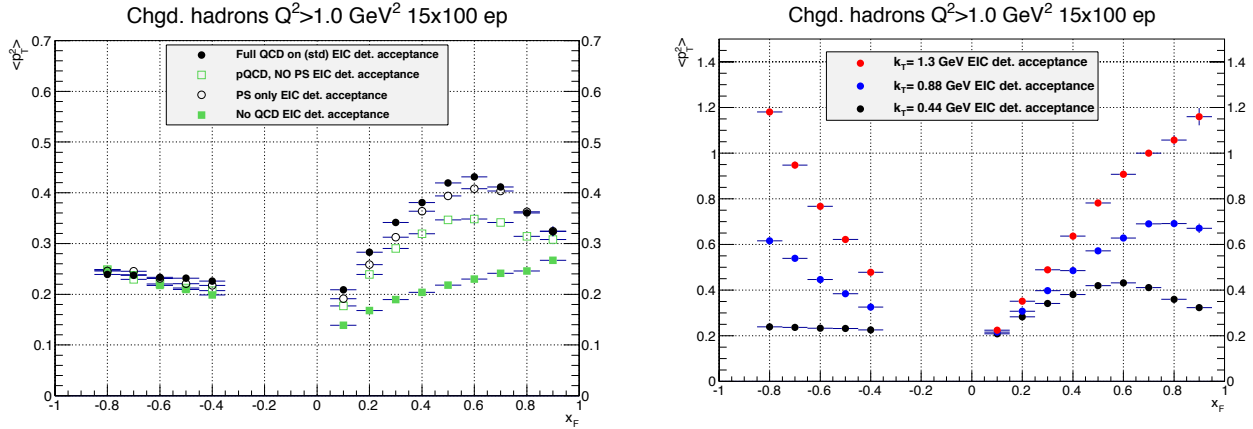

Figure 2-10: (left) The seagull plot ($<p^2_T>$ vs. $x_F$) for 15 GeV on 100 GeV e+p collisions as obtained with the LEPTO-PHI MC for the default rms-$k_T$ value: 0.44 GeV for four cases: all pQCD processes turned on (standard); standard pQCD but no parton shower; parton shower only; no pQCD. (right) The seagull plot ($<p^2_T>$ vs. $x_F$) for 15 GeV on 100 GeV e+p collisions for a variety of rms-$k_T$ values: 0.44 GeV (default), 0.88 GeV, and 1.3 GeV. These plots were simulated assuming the acceptance of the model eRHIC detector and just 0.4 pb$^{-1}$ of data.

In recent years, interest in primordial $k_T$ has been on the rise, mostly in the form of related topics such as unintegrated parton distributions [16], transverse-momentum dependent parton distributions [17], and saturation momentum scales in very low-x e+p or, more accessibly, in e+A collisions [2]. As described earlier all of these topics are key e+p and e+A measurements for eRHIC.

One of the first eRHIC measurements will be the W-dependence of the $<p^2_T>$ for both the current jet ($x_F>0.2$) and the hadron beam-remnant jet ($x_F<-0.2$), which will allow for the separation of the effects of non-pertubative $k_T$ from parton showering. This will provide an important e+p baseline for studies of possible saturation-based enhancement of primordial $k_T$ in e+A.

All currently studied eRHIC detector concepts will all have Roman Pots to measure protons in the far forward direction (with respect to the direction of the proton or heavy ion beam). These Roman Pots will have significant acceptance for positively charged particles (almost all protons in practice) from $-0.85 < x_F < -0.35$ allowing us to make these primordial $k_T$ measurements using seagull plots. Figure 2-10 (left) illustrates the point that the positive $x_F$ particles reflect a combination of primordial $k_T$ and of various QCD effects while the negative $x_F$ particles are best suited to study effects due to non-pertubative $k_T$. Figure 2-10 (right) illustrates the tremendous sensitivity of studying $<p^2_T>$ vs. $x_F$ to primordial $k_T$. These plots correspond to a tiny integrated luminosity (0.4 pb$^{-1}$). With the anticipated integrated annual luminosities for eRHIC it will be possible to measure the primordial $k_T$ for different quark flavors, including heavy quarks, as function of $x$ and $Q^2$.

In addition to its impact on the seagull plot, primordial transverse momenta also lead to an azimuthal asymmetry in the produced hadrons. This asymmetry, first pointed out by Cahn [11], occurs because the quark and incoming lepton have a higher cross-section if they are more head-on, leading to a preference for the scattered quark to be in the lepton-hadron scattering plane, but with the opposite orientation ($\varphi=\pi$). These effects, along with the $O(\alpha_S)$ analog in the Photon Gluon Fusion and QCD Compton processes [18,19] were implemented in old e+p MC generators by the E665 Collaboration [19]. Following the prescription from Reference [19] a new MC generator was created.

A simple Fourier decomposition of the produced hadrons yields significant effects.

Define:
$$\frac{d^2N}{dx_F d\varphi} = A(x_F) + B(x_F)\cos\varphi + C(x_F)\cos 2\varphi ,$$
where $\varphi$ is the azimuthal angle around the virtual photon direction with the scattered electron being defined as $\varphi=0$. It should be noted that some phenomenological approaches towards azimuthal asymmetry include a spectacular increase in the number of terms (see for instance equation 61 of Reference [20]), but for unpolarized (or polarization-averaged) collisions, over most of the kinematic space, the "B/A" Cahn effect tends to be dominant. The B/A parameter is a very strong function of $k_T$ and a weak function of the strength of all types of pQCD effects, as can be seen in Figure 2-11, which, again, represents a tiny integrated luminosity (0.4 pb$^{-1}$) of data. Similarly to the seagull plots, with the anticipated integrated luminosities for eRHIC it will be possible to measure the primordial $k_T$ for different quark flavors, including heavy quarks, as a function of $x$ and $Q^2$.

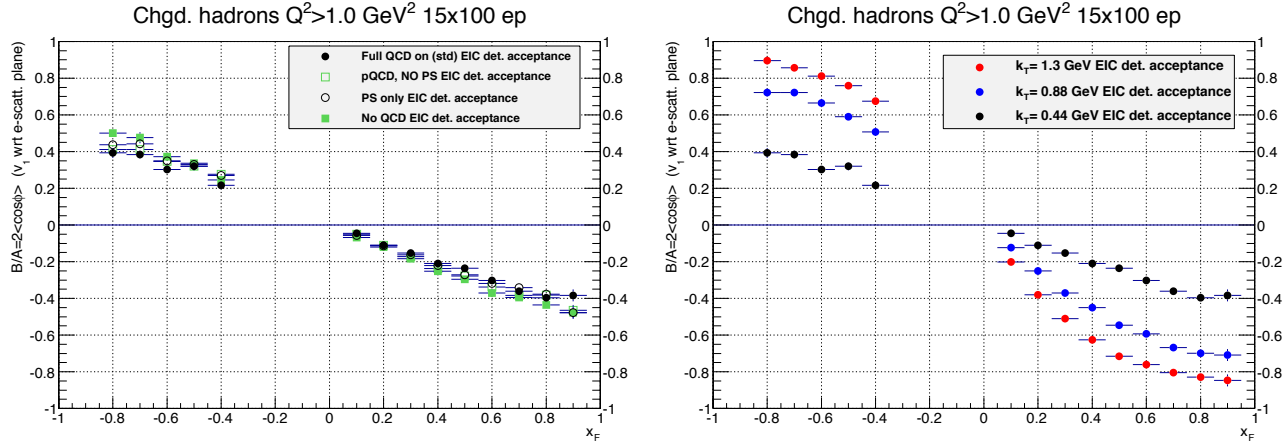

Figure 2-11: (left) The azimuthal asymmetry parameter B/A (or $2v_1$ w.r.t. the lepton scattering plane) vs. $x_F$ for different MC-models: four cases: all pQCD processes turned on (standard); standard pQCD but no parton shower; parton shower only; no pQCD, all with a rms-$k_T$ value of 0.44 GeV. (right) The azimuthal asymmetry parameter B/A vs. $x_F$ in LEPTO-PHI for a variety of rms-$k_T$ values: 0.44 GeV (default), 0.88 GeV, and 1.3 GeV. These plots were simulated assuming the acceptance of the model eRHIC detector and just 0.4 pb$^{-1}$ of data.

## 2.1.3 The Spatial Imaging of Quarks and Gluons

The internal landscape of the nucleon and nuclei in terms of the fundamental quarks and gluons can be studied in different hard processes and can be characterized by different quantities (distributions). Hard exclusive reactions such as deeply virtual Compton scattering (DVCS) and the exclusive production of mesons give access to the spatial distribution of partons in the transverse plane as encoded either in generalized parton distributions (GPDs) or, at small $x$, in dipole scattering amplitudes.

GPDs unify the information contained in the well-known form factors, and standard one-dimensional parton distributions and quantify various correlations/distributions of quarks and gluons in terms of their momentum fractions and positions in the transverse plane. Thus, GPDs provide a rigorous framework for studies of the three-dimensional parton structure of hadrons as well as many additional important aspects of the hadron structure such as the parton angular momentum contributing to the proton spin, the spin and flavor content, and the role of chiral symmetry.

At the moment, our knowledge about GPDs is mostly limited to valence quark GPDs (Hermes, Jefferson Lab 6 GeV, also Compass, and Jefferson Lab 12 GeV in the near future) and limited rather low precision data from HERA. A high-energy, high-luminosity eRHIC will be the ideal machine for detailed quantitative studies of hard exclusive reactions and the so far unexplored sea quark and gluon GPDs:

- One essential aspect of the GPD program is to obtain the transverse distribution of quarks and gluons in the nucleon/nucleus through precise measurements of the *t* dependence of cross sections for various exclusive processes, in particular, DVCS and the production of J/Ψ, Φ, π, K, etc. mesons. In the nucleon case, covering the interval $0 \approx |t| \leq 2$ GeV$^2$ will enable one to map out the parton distributions in the transverse plane down to an impact parameter *b* of about 0.1 fm.

- One area where eRHIC excels is the large range in $Q^2$ available in each *x* interval. QCD evolution equations of GPDs, similarly to the PDF case, allow one to globally fit the data using flexible parameterizations of GPDs and to extract accurate and model-independent information on GPDs. The large lever arm in $Q^2$ is also critical to establish details of the reaction mechanisms such as scaling properties or the relevance of higher twist effects.

- Another clear advantage of eRHIC compared to previous experiments is the availability of different polarizations for the lepton and proton beams that can be used to fully disentangle the various different GPDs from a large range of experimental observables. While DVCS is sensitive to singlet quark and gluon GPDs, other exclusive diffractive processes (electroproduction of ρ, J/Ψ, Φ, etc.) and non-diffractive processes (electroproduction of $\pi^+$, $K^+$, etc.) will allow one to access and disentangle the spin and flavor dependences of GPDs. Note that the non-diffractive processes push the requirements for high luminosity much further than DVCS or other diffractive processes.

- Exclusive processes with nuclei in a collider and, subsequently, the spatial imaging of sea quarks and gluons in nuclei will be studied for the first time. All the processes mentioned above will benefit from the high luminosity of eRHIC as well as excellent detection capabilities and particle identification guaranteeing exclusivity.

In conclusion, a high-energy high-luminosity eRHIC, studying various deep exclusive processes through cross sections and polarization observables, would uniquely extend and complement our currently very limited knowledge of the 2+1D partonic structure of the nucleon/nucleus to the region dominated by sea quarks and gluons. One of the motivations to measure these processes is the quest for an understanding of the decomposition of the nucleon spin in terms of quark and gluon total angular momenta [21]:

$$J^Q(Q^2) + J^G(Q^2) = 1/2 \text{ with } J^Q(Q^2) = \sum_{q=u,d,s} J^q(Q^2).$$

Comprehensive GPD studies (up to NNLO accuracy) of small-$x_B$ DVCS data measured by H1 and ZEUS collaborations have been performed. It was found that the functional form of the t-dependence cannot be pinned down, and an access to the polarization dependence is not feasible when having only unpolarized DVCS cross section and the lepton beam charge asymmetry measurements available [22]. The virtual Compton process contains twelve helicity amplitudes (or equivalently twelve complex Compton form factors (CFF) [23]. The measurement of CFFs should be considered a primary task, as important as the measurement of inclusive structure functions. The (partial) disentanglement of the various CFFs offers then a clean access to GPDs, labeled as *H* and *E*. A high-luminosity eRHIC experiment with transversely polarized protons certainly provides a unique opportunity for precise measurements of CFFs and to explore their partonic interpretation in the small-$x_B$ region, i.e., $x_B < 0.01$. The set of relevant CFFs at twist-two level is then reduced to $\mathcal{H}$ and $\mathcal{E}$ only. Hence, we can restrict ourselves to two observables, namely, the unpolarized DVCS cross-section and the single transverse proton-spin azimuthal asymmetry. In the partonic interpretation of DVCS data we are interested in the transverse distribution of sea quarks and gluons at small $x_B$ for an unpolarized and for a transversely polarized proton.

In the following the capabilities eRHIC will be discussed to constrain different GPDs through DVCS [24] and exclusive vector meson production. Figure 2-12 shows the $x$-$Q^2$ coverage for 20 GeV on 250 GeV e+p collisions for DVCS and exclusive J/Ψ production at eRHIC. The numbers of events per $x$-$Q^2$ bin correspond to an integrated luminosity of 10 fb$^{-1}$.

To study the potential of eRHIC pseudo data for the unpolarized DVCS cross section (see Figure 2-13) and the transverse proton spin azimuthal asymmetry compared to the current world data, both have been used in a global fit utilizing a flexible GPD model [24] to extract GPD $H$ and $E$.

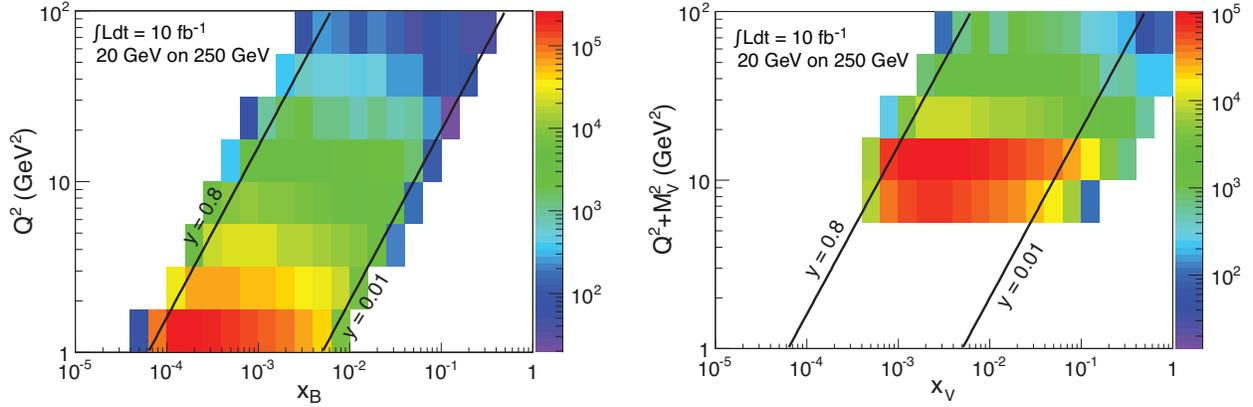

Figure 2-12: (left) Expected distribution of DVCS events in bins of $x$ and $Q^2$, i.e., the contribution of the Bethe-Heitler process to the process ep → epγ has been subtracted. (right) Expected number of events for exclusive J/Ψ production in bins of $x_V$ (defined as $(Q^2 + M_V^2)/(2p \cdot q)$) and $Q^2$.

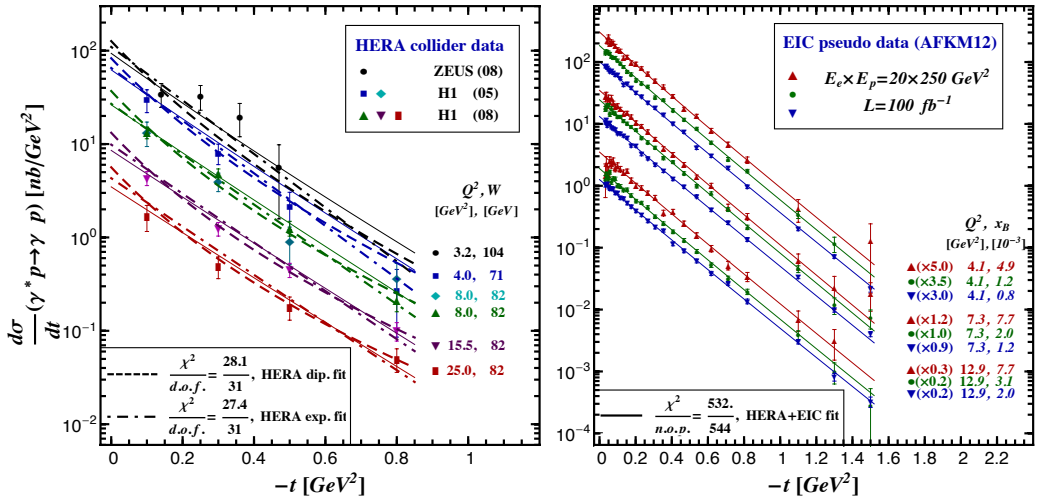

Figure 2-13: A model dependent extraction of GPD $H$ from cross section measurements of the H1/ZEUS collaborations (left) and from a combined fit that includes eRHIC pseudo data (right) for 20 GeV x 250 GeV$^2$ e+p collisions. The label "dip. fit" refers to using a fit function with a dipole functional form.

The experimental uncertainties in Figure 2-13 and Figure 2-14 were estimated based on the statistical uncertainties obtained from a simulation using the MILOU generator [25], which are rescaled for the DVCS cross section, include a 5% systematic uncertainty on cross section level, an additional 3% uncertainty due to the subtraction of the Bethe Heitler (BH) background, and a 5% beam polarization uncertainty.

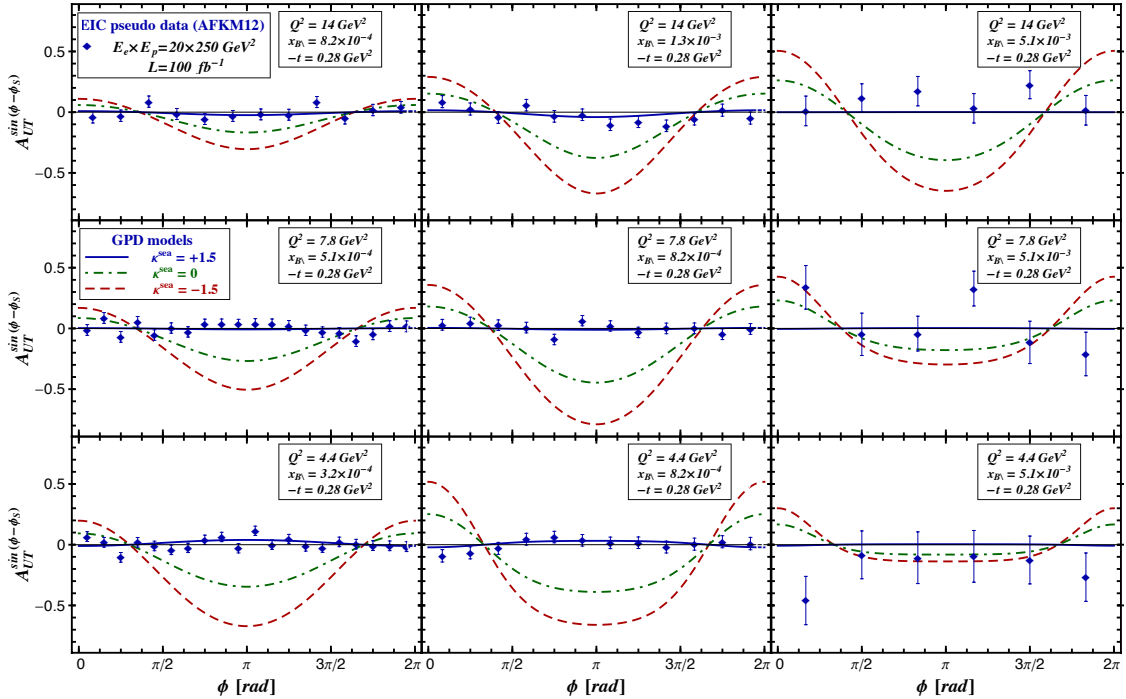

Figure 2-14: eRHIC pseudo data (diamonds) for the transverse target spin asymmetry at beam energies 20 GeV on 250 GeV are shown together with AFKM12 GPD model predictions, where GPD $E^{sea}$ is taken as large positive (solid), vanishing (dot-dashed), and large negative (dashed), respectively.

Lacking any experimental constrains for the GPD $E$, three models with very different predictions for the GPD $E$ (labeled $\kappa^{sea}$= -/+1.5, 0) were used to calculate the transverse proton spin azimuthal asymmetry shown in Figure 2-14 together with simulated pseudo data. Certainly, the predictions of all these three models are experimentally distinguishable at eRHIC.

In Figure 2-15 we compare the resulting GPDs from fits to the HERA data alone and to the combined HERA+eRHIC data at $Q^2$=4 GeV$^2$, $x_B$=10$^{-3}$, and variable -$t$ (covering the HERA region). In the left panel one realizes that the uncertainty of the sea quark GPD $H^{sea}$, which is to certain extent constrained by HERA data, can be strongly reduced in particular, at smaller -$t$ values. The gluon GPD $H^G$, displayed in the middle panel, is extracted by means of the $Q^2$ evolution, and it is rather weakly constrained by HERA DVCS data alone. Here, the inclusion of eRHIC data yields a substantial improvement, even if the available lever-arm in $Q^2$, is still rather limited compared to HERA kinematics. Currently the uncertainties in the forward distributions are not yet included in the uncertainty bands. As emphasized above, information on the GPD $E$ can only be obtained from a new lepton proton scattering experiment with a transversely polarized proton beam. The right panel of Figure 2-15 clearly demonstrates that the sea quark component of this GPD can be extracted with relatively small uncertainties.

In addition, exclusive J/Ψ production provides selective access to the unpolarized gluon GPD through the dominant photon-gluon fusion production mechanism. In this case, the hard scale of the process is Q2+M2J/Ψ rather than Q2, so that both photo- and electro-production can be used to probe GPDs. Electro-production has smaller rates but reduced theoretical uncertainties. Furthermore, the cross sections sL and sT for longitudinal and transverse photon polarization, which can be separated experimentally from the angular distribution in the decay J/Ψ→l+l-, provide two independent observables to validate the theoretical description.

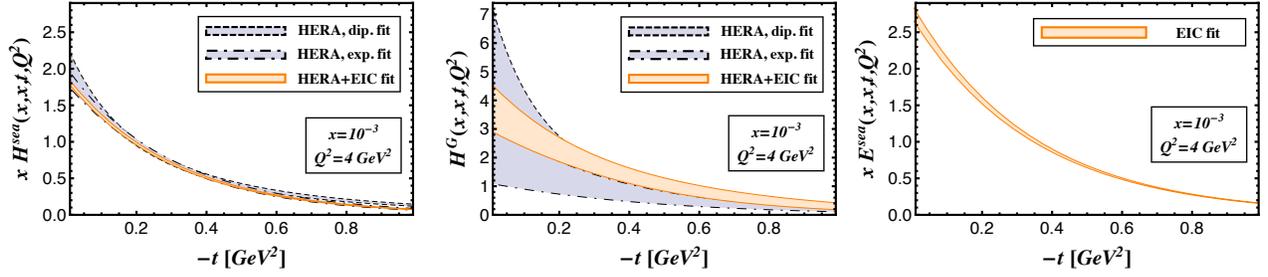

Figure 2-15: Extraction of sea quark GPD $H^{sea}$, (left) and gluon GPD $H^G$ (middle) (gray area) using only the HERA collider data. The results of a combined HERA/eRHIC fit including pseudo data for the unpolarized DVCS cross section, c.f. Figure 2-13, and the transverse target spin asymmetry, c.f. Figure 2-14, are shown as light orange area. In addition for v the first time the sea quark GPD $E^{sea}$ could be extracted (right panel).

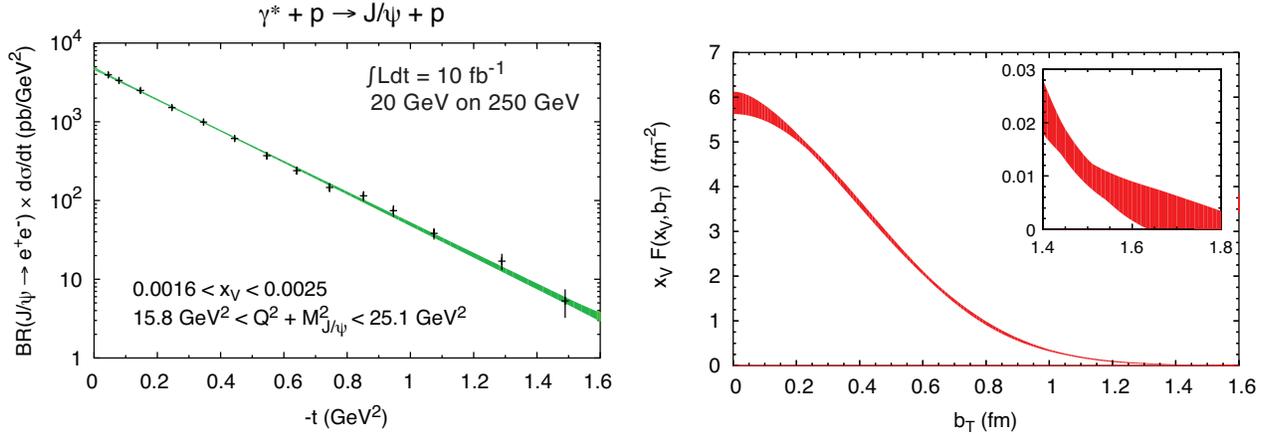

Figure 2-16: Left: Expected experimental accuracy for a cross-section measured for $\gamma^*p \to J/\Psi + p'$ for one bin in $x_V$ and $Q^2$. Right: the distribution of gluons in impact parameter $b_T$ obtained from the J/Ψ production cross section. The bands represent the parametric errors in the fit of $d\sigma_{DVCS}/dt$ and the uncertainty from different extrapolations to the regions of unmeasured (very low and very high) t.

An example for the expected spectrum in t for J/Ψ production is shown in Figure 2-14. Also shown are the impact parameter b space images obtained from the $\gamma^*p \to J/\Psi+p'$ scattering amplitude by a Fourier transform. The distributions thus contain a contribution from the small but finite size of the J/Ψ meson, which needs to be disentangled in a full GPD analysis. We see from the Figure that the data will enable us to accurately probe the spatial distribution of gluons over two orders of magnitude in x, up to the region where the dominant partons are valence quarks. The transverse proton spin asymmetry will, in addition, give constraints on the GPD E for gluons and thus strongly complement what can be achieved with DVCS.

Finally, we shortly discuss the role of the eRHIC measurements in elucidation of the Ji spin sum rule. Ji's decomposition implies that the partonic total angular momenta $J_{q,g}$ are given by the expectation values of the corresponding gauge invariant parts of the energy momentum tensor and might be further decomposed in spin and orbital angular momenta. Most crucially, $J_{q,g}$ can be expressed in terms of moments of GPDs H and E,

$$J^q(Q^2) = \int_0^1 dx\, x[q(x,Q^2) + \bar{q}(x,Q^2)] + \tfrac{1}{2}\lim_{t=0,\xi=0}\int_{-1}^1 dx\, x\, E^q(x,\xi,t),$$

The averaged momentum fractions of unpolarized partons is already phenomenologically well constrained by inclusive DIS measurements and momentum conservation guarantees that they are normalized to one and the angular momentum sum rule implies then that the anomalous gravitomagnet-ic nucleon moment vanishes. To verify this theoretical prediction for the anomalous gravitomagnetic moment, the GPD *E* needs to be determined from experimental data [26].

Certainly, it will be challenging to measure the $\mathcal{I}m\,\mathcal{E}$, but through measuring DVCS on an effective neutron target as He$^3$, the transverse proton spin asymmetry or the beam charge asymmetry at small $\xi$ it will be possible to constrain the currently unconstrained GPD *E*. As we have seen, DVCS measurements at an eRHIC will allow one to access the GPD *E* and, in a model dependent manner, to extract also its normalization in the forward kinematics. In fact, what is often also called the anomalous magnetic moment of sea quarks is simply related to their angular momentum:

$$J^{sea} = \tfrac{1}{2}(1 + \kappa^{sea})A^{sea},$$

where the phenomenological value of the momentum fraction is at $Q^2 = 4$ GeV$^2$ given by $A^{sea}(Q^2=4\ GeV^2)\sim 0.15$.

## 2.2 The Nucleus as a Laboratory for QCD

Theoretical breakthroughs and experimental results in the past decade suggest that both nucleons and nuclei, when viewed at high energies, appear as dense systems of gluons creating fields whose intensity may be the strongest in nature. These high densities will possibly lead to the phenomenon of parton (gluon) saturation, also known as the Color Glass Condensate (CGC) [27,28]. The transition to this non-linear regime is characterized by the saturation momentum, $Q_S$, which can be large for heavy ions. eRHIC will allow us to probe the wave functions of high energy nuclei with an energetic electron: by studying these interactions, one will probe the strong gluon fields of the CGC. While experiments at HERA, RHIC, and the LHC found first evidence for saturation, eRHIC will have the potential to unambiguously identify this new regime and quantify its relevant parameters.

The exploration of the unknown nature of glue in general and particularly the unambiguous discovery and study of parton saturation drives the development of the $e$+A physics program at eRHIC. Investigating gluons in nuclei instead of protons has multiple advantages:
- The nucleus is an efficient **amplifier** of the physics of high gluon densities. Simple considerations suggest that $Q_s^2 \propto (A/x)^{1/3}$. Therefore, DIS with large nuclei probes the same universal physics as seen in DIS with protons at $x$'s at least two order of magnitude lower or equivalently an order of magnitude larger $\sqrt{s}$.
- The nucleus is also a powerful **analyzer** of physics across the full range of $x$, $Q^2$, and A. In e+A collisions at high energies viewed in the rest frame of the nucleus, the virtual photon mediating the interaction splits into a compact $q\bar{q}$ dipole, which scatters off the nuclear medium. The interaction of these fast, compact dipoles with an extended gluon medium provides insight into how partons lose energy, are absorbed, and how hadron formation is modified in the presence of a colored medium.

While for many studies the nuclei serve plainly as "vessels" of gluons, electron-ion collisions at eRHIC will allow us also to gain insight into the **short-range structure** of nuclei. With their capability to measure a wide range of processes, experiments at eRHIC will be able provide the first 3-dimensional images of sea quarks and gluons in the nucleus with sub-femtometer resolution.

Nuclei are made out of nucleons, which in turn, are bound states of the fundamental constituents, quarks and gluons, probed in high-energy scattering. The binding of nucleons into a nucleus must be sensitive to how these quarks and gluons are confined into nucleons, and must influence how they are distributed inside the bound nucleons. EMC at CERN [29,30] and many follow-up experiments revealed a peculiar pattern of nuclear modification of the DIS cross-section as a function of Bjorken $x$, giving us clear evidence that the momentum distributions of quarks in a fast-moving nucleus are strongly affected by the binding and the nuclear environment. With much wider kinematic reach in both $x$ and $Q$, and unprecedented high luminosity, experiments at eRHIC not only can explore the influence of the binding on the momentum distribution of sea quarks and gluons, but also, for the first time, determine the spatial distribution of quarks and gluons in a nucleus by diffractive or exclusive processes. In addition, the wealth of semi-inclusive probes at eRHIC provides direct and clean access to the fluctuations of color or density of quarks and gluons in nuclei.

The kinematic acceptance in $e$+A of eRHIC compared to world's data collected in nuclear DIS and in Drell-Yan (DY) experiments is shown in Figure 2-2. eRHIC, with its high luminosity, wide energy range, and its possibility to accelerate heavy ions up to Uranium, allows for an $e$+A program that is perfectly suited to address the fundamental questions raised in the EIC White Paper [1]:
- Can we experimentally find evidence of a novel universal regime of non-linear QCD dynamics in nuclei?

- What is the role of saturated strong gluon fields, and what are the degrees of freedom in this strongly interacting regime?
- What is the fundamental quark-gluon structure of light and heavy nuclei?
- Can the nucleus, serving as a color filter, provide novel insight into propagation, attenuation and hadronization of colored probes?

In the following sections, we discuss in more detail the physics and a comprehensive set of key measurements of an $e$+A program at eRHIC. In Sec. 2.2.1 we describe those that are relevant at small-$x$, in Sec. 2.2.2 those relevant at medium to large $x$.

Some of these measurements have analogs in $e$+$p$ collisions but have never been performed in nuclei, others have no analog in $e$+$p$ collisions and nuclei provide a completely unique environment to explore these. For the former, the comparison of results in $e$+A to those in $e$+$p$ is crucial. As was the case for the heavy-ion program at RHIC, a successful $e$+A program will require an $e$+$p$ program at matching beam energies as a baseline.

## 2.2.1 Physics of High Gluon Densities and Low-x in Nuclei

The simplest view of a nucleon is that of three quarks interacting via the exchanges of gluons that bind the quarks together. However, as experiments probing the proton structure at the HERA collider at DESY showed, this picture is far too simple. Countless other gluons and a "sea" of quarks and anti-quarks pop in and out of existence within each nucleon. These quantum fluctuations can only be probed in high-energy scattering experiments because the Lorentz time dilation freezes the cascading of partons in the lab frame. The higher the energy of the nucleon, the more the gluon fluctuations slow down so that it is possible to "take snapshots" of them with a probe particle sent to interact with the high-energy proton.

The wave function of the nucleon depends on both $x$ and $Q^2$. An example of the $x$-dependence is shown in Figure 2-1, extracted from the data measured at HERA for DIS on a proton. The PDFs of the "sea"- quarks and gluons, denoted by $xG$ and $xS$ in Figure 2-1, grow dramatically towards low $x$. Remembering that low $x$ means high energy, we see that the part of the proton wave function responsible for the interactions in high-energy scattering consists, for $x <$ 0.01, almost entirely of gluons.

These gluons populate the transverse area of the proton, creating a high density of gluons. This dense small-$x$ wave function of an ultra-relativistic proton (or nucleus) is referred to as the Color Glass Condensate [27,28] (CGC). To understand the onset of the dense regime, one needs to employ QCD evolution equations. While the current state of QCD theory does not allow for a first-principles calculation of the quark and gluon distributions, the evolution equations allow one to determine these distributions at some values of $(x, Q^2)$ if they are initially known at some other $(x_0, Q_0^2)$. The most widely used evolution equation at large $x$ and at large $Q^2$ are the linear evolution equations DGLAP [31,32,33] (along $Q^2$) and BFKL [34,35] (along $x$) as illustrated in Figure 2-17. The rapid growth in gluon densities with decreasing $x$ is understood to follow from a self similar Bremsstrahlung cascade in the BFKL evolution where harder, large $x$, parent gluons successively radiate softer daughter gluons.

However, gluon and quark densities cannot grow without limit at small-$x$. While there is no strict bound on the number density of gluons in QCD, there is a bound on the scattering cross-sections stemming from unitarity [36]. A proton or nucleus with a lot of "sea" gluons is more likely to interact in high-energy scattering, which leads to larger scattering cross-sections. Gluon saturation is a simple mechanism for nature to tame this growth. When the density of gluons becomes large, softer gluons can recombine into harder gluons. The competition between linear QCD Bremsstrahlung and non-linear gluon recombination causes the gluon distributions to saturate at small $x$.

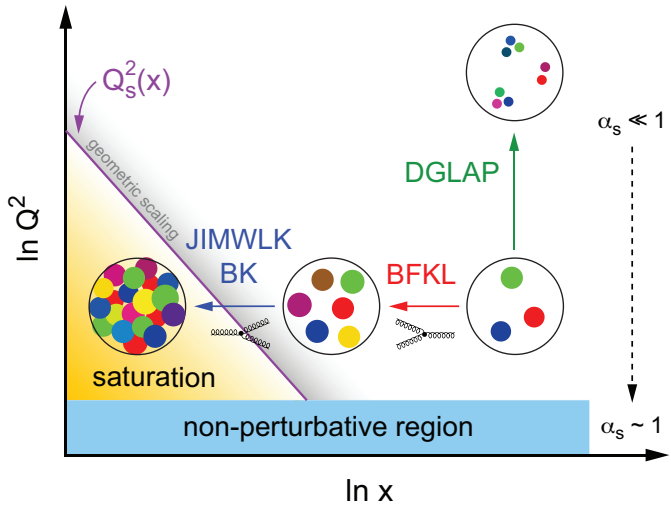
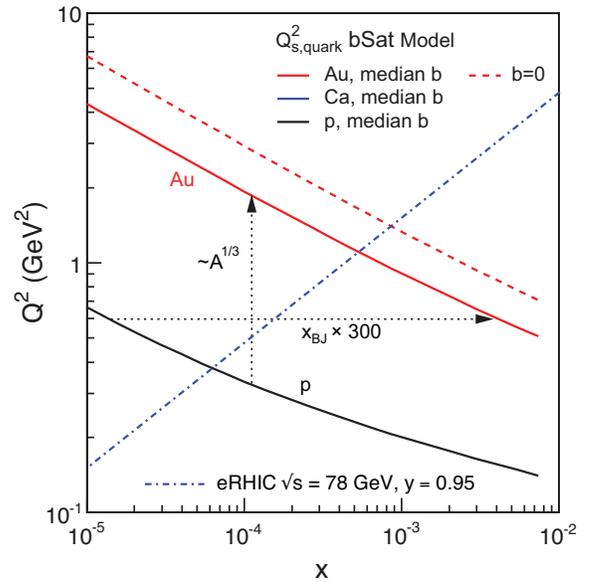

Figure 2-17: A schematic map of high energy QCD in the $x, Q^2$ plane depicting regions of non-perturbative and perturbative QCD, including in the latter, low to high saturated parton density, and the transition region between them. The line indicating the saturation regime reflects a line of constant gluon density. It represents not a sharp transition but rather indicates the approximate onset of saturation phenomena.

Figure 2-18: The theoretical expectation of the saturation scale, $Q_s^2$, as a function of $x$ for protons, Ca, and Au. While the increase of the saturation scale from $p$ to Au is only a factor of ~6, the effect in $x$ is dramatic. This allows one to study saturation effects with $e$+A at eRHIC that would be otherwise inaccessible in $e$+$p$

The non-linear, small-$x$ renormalization group equations, JIMWLK [37,38,39,40] and its mean field realization BK [41,42,43], propagate these non-linear effects to higher energies leading to saturation (see Figure 2-17).

The onset of saturation and the properties of the saturated phase are characterized by a dynamical scale, the saturation scale [44,45,46], $Q_S^2$, which grows with increasing energy (smaller $x$). The nature of gluon saturation at high energies is terra incognita in QCD.

The advantage of using nuclei to explore this regime is the enhancement of the saturation phenomena with increasing A, making it easier to observe and study experimentally. The reason for this dependence is simple: any probe interacting over the distance $L \sim (2m_N x)^{-1}$ cannot distinguish between nucleons in the front or back of the Lorentz contracted nuclei once $L > 2\,RA \sim A^{1/3}$; the probe then interacts *coherently* with all nucle-ons. These considerations suggest that $Q_S^2 \propto (A/x)^{1/3}$. This dependence is supported by detailed studies [47,48,49,50,51] and is often referred to as the nuclear "oomph" factor, since it reflects the enhancement of saturation effects in the nucleus as compared to the proton. For heavy nuclei such as Au and Pb, the nuclear oomph factor is ~6. DIS with large nuclei probes the same universal physics as seen in DIS with protons at $x$'s at least two orders of magnitude lower (or equivalently an order of magnitude larger √s) as illustrated in Figure 2-18. When $Q^2 \gg Q_S^2$, one is in the well understood "linear" regime of QCD, while we have little theoretical control over the non-perturbative regime at $Q^2 \lesssim \Lambda_{QCD}^2$. For large nuclei, there is a significant window at small $x$ where $Q_S^2 \gg Q^2 \gg \Lambda_{QCD}^2$. This is in the domain of strong non-linear gluon fields indicated in Figure 2-19.

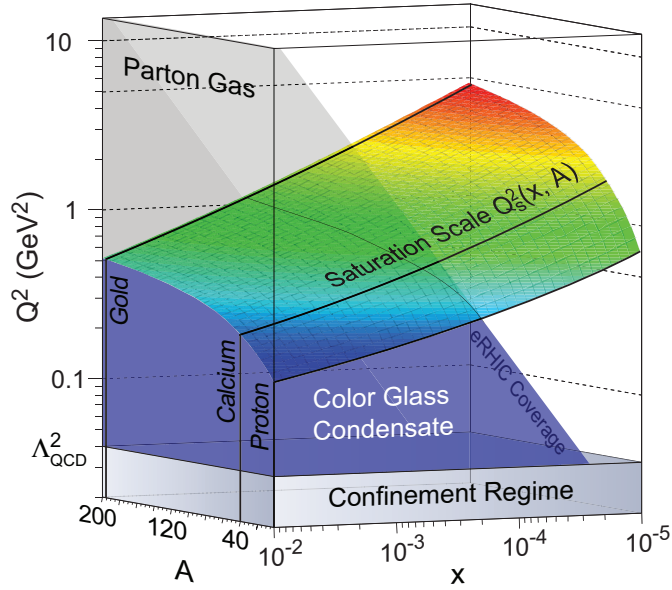

Figure 2-19: The theoretical expectations for the saturation scale, $Q_s^2$, at medium impact parameter as a function of Bjorken-x and the nuclear mass number A.

### *Nuclear Structure Functions*

Figure 2-19 suggests a straightforward way to study saturation: perform the DIS experiment on nuclei, and measure the DIS scattering cross-section at sufficiently low $x$ and $Q^2$ where effects of saturation should become pronounced. The invariant cross-section in $e$+A collisions can be expressed as a function of two structure functions $F_2^A(x, Q^2)$ and $F_L^A(x, Q^2)$. These fully inclusive structure functions offer the most precise determination of quark and gluon distributions in nuclei: $F_2^A$ is sensitive to the sum of quark and anti-quark momentum distributions $xq(x, Q^2)$ while $F_L^A$ measures the gluon momentum distribution $xg(x, Q^2)$. Saturation effects can been seen in both at low $x$, although they should be substantially more pronounced in the latter since $F_L \propto xg(x, Q^2)$ [31,52].

Parton distribution functions such as the one shown in Figure 2-1 are largely derived from our knowledge of the structure function $F_2$ and $F_L$. The quark distributions $xq(x, Q^2)$ are extracted from pQCD fits to $F_2$, the gluon distributions are either derived through scaling violations of $F_2$ ($\partial F_2/\partial \ln(Q^2) \neq 0$) with $Q^2$ or directly from $F_L$ when available. In Figure 2-1, one can see the PDFs of the valence quarks in the proton, $xu_v$ and $xd_v$, which decrease with decreasing $x$. The PDFs of the "sea" quarks and gluons, denoted by $xG$ and $xS$, appear to grow very strongly towards low $x$.

DIS experiments with nuclei have established that PDFs (or structure functions) in nuclei exhibit various nuclear effects, not surprisingly most prominently for gluons: a strong suppression of the gluon distribution function in nuclei compared to that in nucleons for $x < 0.01$ (shadowing), a slight enhancement around $x \sim 0.1$ (anti-shadowing), followed again by a suppression (EMC effect [29,30]) at large $x$. In sharp contrast to the proton, the gluonic structure of nuclei is not known for $x < 0.01$. The nuclear effects in the structure functions can be quantified using their expansion in powers of $1/Q^2$. The standard linear perturbative QCD approaches calculate the leading term in the $1/Q^2$ expansions of structure functions. The order-one contribution is referred to as the 'leading twist' term; hence the name Leading Twist Shadowing (LTS) for models that attempt to describe shadowing via nuclear PDFs in pQCD. However, the effects that cause saturation contribute to all orders in the $1/Q^2$ expansions. Of particular interest is their contribution to the non-leading powers of $1/Q^2$, known as 'higher twists': the main parts of those corrections are enhanced by the nuclear "oomph" factor $A^{1/3}$ and by a power of $(1/x)^\lambda$, where $\lambda = 0.2 - 0.3$.

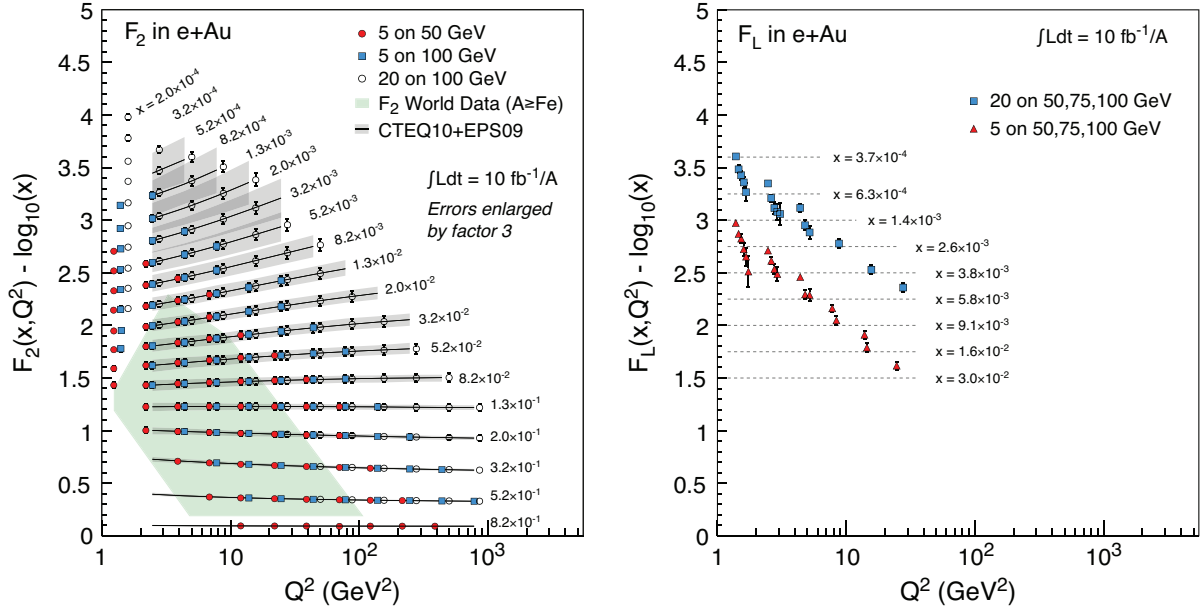

Figure 2-20: The structure functions $F_2$ (left) and $F_L$ (right) as a function of $Q^2$ for various $x$-values in $e$+Au collisions at eRHIC. Note that in the figure, data points from different energies at the same $Q^2$ are slightly offset along the abscissa for better visibility. For details see text.

The telltale signs of saturation physics are the higher twist corrections, which are enhanced in DIS on a nucleus, and at smaller-$x$. Higher twists tend to decrease the structure function with decreasing $x$.

To verify eRHIC's capability to measure the structure functions $F_2$ and $F_L$, we conducted simulations of inclusive events in $e$+Au collisions. Figure 2-20 shows the resulting structure functions $F_2$ (left) and $F_L$ (right) as a function of $Q^2$ for various $x$ values. For clarity, $F_2$ and $F_L$ are offset by $\log_{10}(x)$. The simulations of $F_2$ were conducted for 5 on 50 GeV, 5 on 100 GeV, and 20 on 100 GeV, the highest eRHIC energy. The referring errors bars are based on an assumed 3% systematic normalization uncertainty added in quadrature to the statistical errors. The latter are evaluated for an integrated luminosity of 10 fb$^{-1}$/A for all three energies combined. Since the resulting errors on $F_2$ are barely visible, they are enlarged by a factor of 3. For $F_2$, we also depict the curves and respective uncertainty bands from NLO calculations using the EPS09 parameterization of the nuclear parton distribution functions [53]. The comparison of the current EPS09 uncertainty bands (also enlarged by a factor of 3) with the errors of the respective data points demonstrates that for $x < 0.01$, eRHIC will have a substantial impact on reducing the uncertainty of nuclear PDFs. The green shaded region denotes the kinematic region for which $F_2$ measurements for heavy nuclei exist.

Any measurement of $F_L$ requires runs at various $\sqrt{s}$. For eRHIC we can make use of the flexibility in the ion beam energy. No direct measurements of $F_L$ for nuclei were ever conducted. In our studies (Figure 2-20 right), we varied the ion beam energy from 50 to 100 GeV for electron energies of 5 and 20 GeV. The final values for $F_L$ were extracted using the established Rosenbluth method, which is sensitive to the quality of the absolute normalization achieved at the various energies. Since systematic uncertainties depend on the quality of the final detectors and the accuracy of luminosity measurements their ultimate magnitude is hard to estimate. Here we assumed systemic normalization uncertainties of 3% per energy, values that were achieved at HERA. The presented errors include both systematical and statistical contributions. We derived the statistical errors using an integrated luminosity of 10 fb$^{-1}$/A divided among all energies listed.

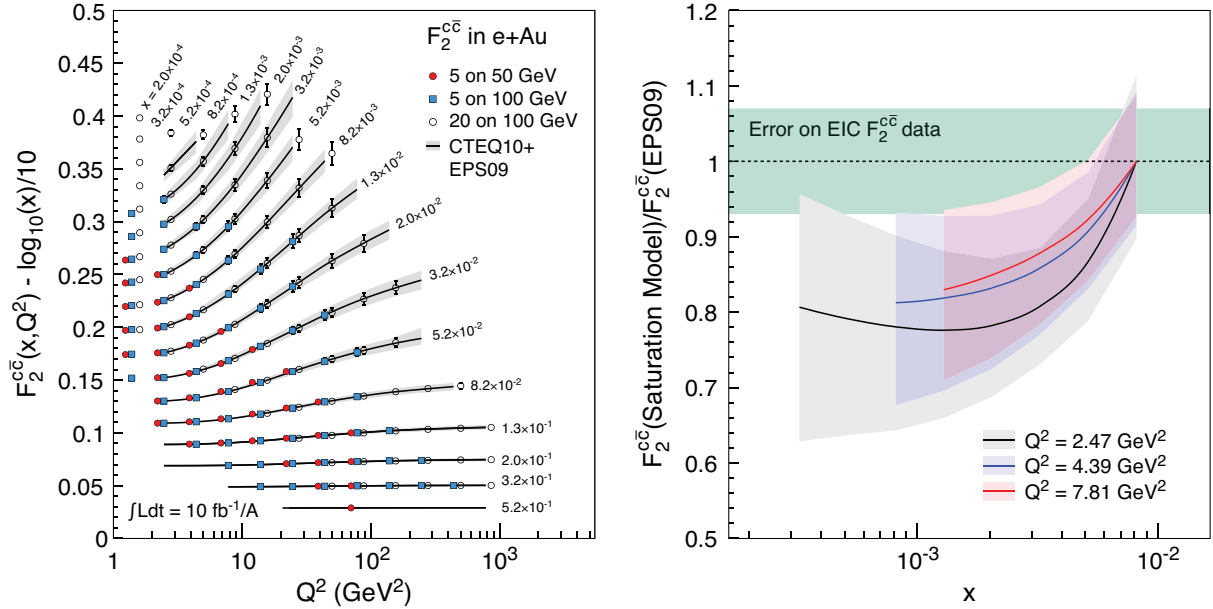

Figure 2-21: Left panel: The charm structure function $F_2^{c\bar{c}}$ as a function of $Q^2$ for various $x$ values in $e$+Au collisions at eRHIC. Data points from different energies at the same $Q^2$ are slightly offset along the abscissa for better visibility. Right panel: Ratio of $F_2^{c\bar{c}}$ predictions from a saturation model and leading twist shadowing pQCD predictions using the EPS09 nuclear PDFs for three different $Q^2$ values. The uncertainty band for each $Q^2$ value reflects the combined uncertainties in both models. For details see text.

A comparison of $F_2$ and $F_L$ clearly shows the intricacy of $F_L$ studies. While of enormous importance for the study of gluons, the kinematic reach of $F_L$ measurements is much narrower than that of $F_2$ and errors are substantially larger. We therefore studied an additional, complementary method for determining the gluon density through the charm structure function $F_2^{c\bar{c}}$. The left plot in Figure 2-21 shows $F_2^{c\bar{c}}$ as a function of $Q^2$ for various $x$ values in $e$+Au collisions at eRHIC. For clarity, values are offset by $\log_{10}(x)/10$. Depicted are measurements and corresponding errors for three different energies to illustrate the respective kinematic reach, 5 on 50 GeV, 5 on 100 GeV, and 20 on 100 GeV. Statistical errors are based on 10 fb$^{-1}$/A for all three energies combined. The depicted errors are derived from the statistical errors and a 7% systematic uncertainty added in quadrature. Also shown are curves and respective uncertainty bands resulting from the EPS09 parameterization of nuclear parton distribution functions [53]. While an EIC will certainly constrain these parameterizations further, one has to keep in mind that with $F_2^{c\bar{c}}$, one probes the PDFs at $x' \approx x(1 + (4m_c^2)/Q^2)$, where the PDFs are typically better constrained by the existing data. The fact that $F_2^{c\bar{c}}$ is relatively well predicted in DGLAP-based approaches can be used to test for differences between the traditional leading-twist shadowing models (such as EPS09) and saturation models. The right plot in Figure 2-21 compares one such model [54] to NLO pQCD calculations (using EPS09 nuclear PDFs) by depicting the ratio of their predictions for $F_2^{c\bar{c}}$ for three different $Q^2$ values as functions of $x$, where we expect these non-linear dynamics to be important. As one can clearly see, saturation models predict a markedly different $x$-dependence than NLO pQCD calculations based on EPS09: importantly, the difference between the models (together with the combined uncertainty of both models) exceeds the expected uncertainty of EIC measurements (the green band). This comparison demonstrates that eRHIC experiment with charm capabilities will be able to distinguish between saturation and leading-twist shadowing predictions for $F_2^{c\bar{c}}$, providing us with yet another measurement capable of identifying saturation dynamics.

For a better discrimination between models, especially involving non-linear dynamics, several observables sensitive to the gluon distribution will be essential: *(i)* scaling violation of $F_2$, *(ii)* the direct measurement of $F_L$, and *(iii)* $F_2^{c\bar{c}}$.

Note that all three observables, $F_2$, $F_L$, and $F_2^{c\bar{c}}$, can be measured already at moderate luminosities with good statistical precision at eRHIC. The final experimental errors for the structure functions to be measured will be dominated by systematic uncertainties. High luminosities are not required for the measurement of structure functions, while precise knowledge of the actual luminosity is paramount.

The nuclear effects on any structure function can be quantified by the ratio $R_{2,L} = xF_{2,L}^A(x, Q^2)/AxF_{2,L}^p(x, Q^2)$ of the structure function in nuclei and protons. However, more intuitive and physically relevant are the ratios of the PDFs in nuclei and protons *derived* from these structure functions: $R_i = x f_i^A(x, Q^2)/Ax f_i^p(x, Q^2)$. Here the subscript $i$ labels valence quarks, sea quarks, or gluons.

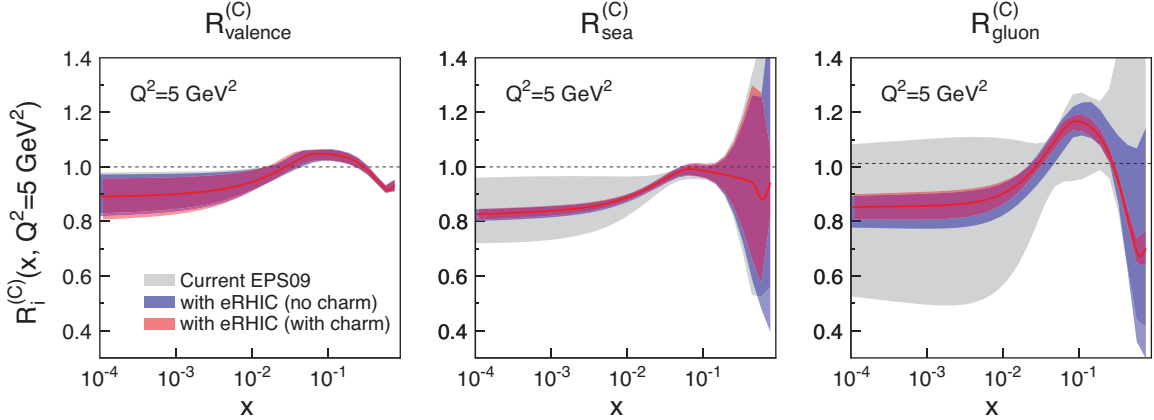

Figure 2-22: Theoretical predictions from EPS09 [53] for the ratio of PDFs in Carbon and protons at $Q^2 = 5$ GeV$^2$ for valence quarks, sea quarks, and gluons, respectively. The blue and red bands reflect the improvement in the predictions when including eRHIC data.

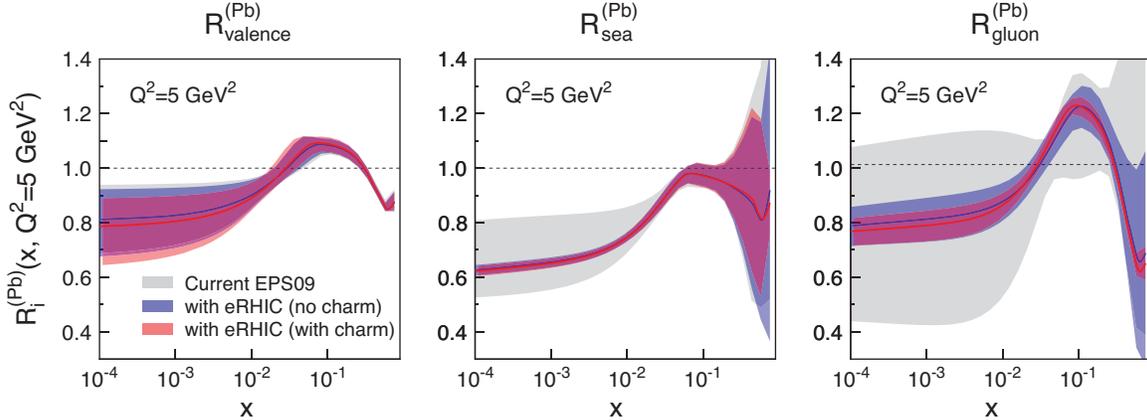

Figure 2-23: Theoretical predictions from EPS09 [53] for the ratio of PDFs in Carbon and protons at $Q^2 = 5$ GeV$^2$ for valence quarks, sea quarks, and gluons, respectively. The blue and red bands reflect the improvement in the predictions when including eRHIC data.

It is instructive to see how an eRHIC will constrain our current knowledge of these ratios given the measurements described above. We again use EPS09 [53] as a common model for nuclear PDFs. In this DGLAP-based description of nuclear PDFs, shadowing is included in the parameterizations of the initial conditions for DGLAP evolution. The parameters are obtained by global fits to currently existing data from $e$+A and $p$+A collisions. The DGLAP equation that is used, describing evolution in $Q^2$, cannot predict the $x$ dependence of distribution functions at low-$x$ without data at comparable values of $x$ and at lower $Q^2$. It is therefore not surprising that the DGLAP-based "predictions" suffer from large uncertainties especially for gluons where few data are available. In order to evaluate the impact of eRHIC on our knowledge of nuclear PDFs, we provided the EPS09 authors with the simulated pseudo-data (such as shown in Figure 2-20 and Figure 2-21) who then repeated their global fit procedure and error evaluation. The result of these studies is depicted in Figure 2-22 for Carbon and Figure 2-23 for lead. The ratios at $Q^2 = 5 \text{ GeV}^2$ for valence quarks, sea quarks, and gluons, respectively, are shown. The grey band illustrates the uncertainties in the ratio of the current EPS09 parameterization, the blue band the one with eRHIC data but without charm, and the red band and central curve depicts the case where all information from eRHIC, including charm, is used to constraint the nuclear PDFs. While for valence quarks the effect is subtle, the improvements for sea quarks and gluons are truly dramatic. For gluons, especially, this effect is observed for all values of $x$.

Clearly, the EIC will reach into unexplored regions with unprecedented precision and will be able to distinguish between traditional and non-linear QCD models. These measurements will have a profound impact on our knowledge of nuclear structure functions and the underlying evolution scheme.

### *Di-hadron Correlations*

Quite generically, multi-parton correlations are more sensitive to the detailed dynamics of the probed objects than single parton distributions. One of the most captivating measurements in $e$+A is that of the azimuthal correlations between two hadrons $h_1$ and $h_2$ in $e + A \rightarrow e' + h_1 + h_2 + X$ processes. These correlations are not only sensitive to the transverse momentum dependence of the gluon distribution, but also to that of gluon correlations for which first principles CGC computations are now becoming available. The precise measurements of these di-hadron correlations at eRHIC would allow one to extract the spatial multi-gluon correlations and study their non-linear evolution.

Experimentally, di-hadron correlations are relatively simple to study at eRHIC. They are usually measured in the plane transverse to the beam axis, and are plotted as a function of the azimuthal angle $\Delta\varphi$ between the momenta of the produced hadrons in that plane. Back-to-back correlations are manifested by a peak at $\Delta\varphi = \pi$ (see Figure 2-24). Saturation effects in this channel correspond to a progressive disappearance of the back-to-back correlations of hadrons with increasing atomic number A. In the conventional linear QCD picture, one expects from momentum conservation that the back-to-back peak will persist as one goes from $e$+$p$ to $e$+A. In the saturation framework, due to multiple re-scatterings and multiple gluon emissions, the large transverse momentum of one hadron is balanced by the momenta of several other hadrons, effectively washing out the correlation [55]. A comparison of the heights and widths of the di-hadron azimuthal distributions in $e$+A and $e$+$p$ collisions respectively would clearly mark out experimentally such an effect. An analogous phenomenon has already been observed at RHIC for di-hadrons produced at forward rapidity in comparing central $d$+Au with $p$+$p$ collisions at RHIC [63,64]. In that case, di-hadron production originates from valence quarks in the deuteron scattering on small-$x$ gluons in the target Au nucleons.

However, the analysis and the interpretation of these studies in $p(d)$+A are by far not as straight forward as in $e$+A. The background in the former is large and the actual $x$ of the gluon probes cannot be derived. Di-hadrons studied in DIS are essentially background free and the measurement of the scattered electron allows us to determine the required kinematic variables $x$ and $Q^2$, which is essential for precision studies of saturation phenomena.

The three curves in Figure 2-24 show predictions in the CGC framework at eRHIC energies for di-hadron $\Delta\varphi$ correlations in deep inelastic $e$+$p$, $e$+Ca, and $e$+Au collisions, respectively [56,57]. The calculations are made for $Q^2 = 1 \text{ GeV}^2$ and include a Sudakov form factor to account for generated radiation through parton showers; only $\pi^0$s were used. The highest transverse momentum hadron in the di-hadron correlation function is called the "trigger" hadron, while the other hadron is referred to as the "associated" hadron. The trigger hadrons have transverse momenta of $p_T^{trig} > 2 \text{ GeV}/c$ and the associated hadrons were selected with $1 \text{ GeV}/c < p_T^{assoc} < p_T^{trig}$. The model predictions show clearly the "melting" of the correlation peak with increasing nuclear mass due to saturation effects.

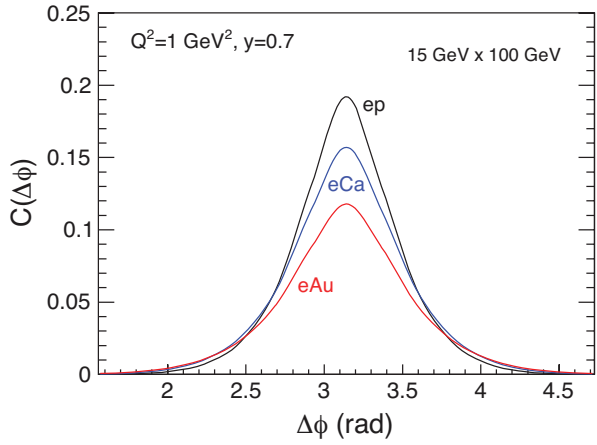 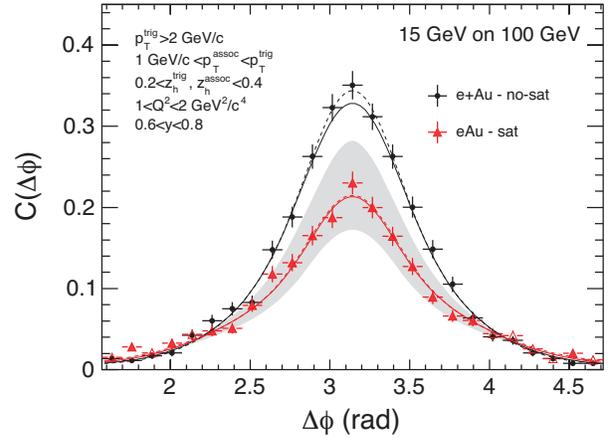

Figure 2-24: A saturation model prediction of the coincidence signal versus azimuthal angle difference $\Delta\Phi$ between two hadrons in e+p, e+Ca, and e+Au collisions at $Q^2=1$ GeV$^2$ for eRHIC energies [56,57].

Figure 2-25: Comparison of di-hadron correlation function for eRHIC energies for saturation model prediction for e+Au collisions with calculations from a conventional non-saturated model. Statistical error bars correspond to 1 fb$^{-1}$/A integrated luminosity.

It is important to verify how precise the suppression of the away-side peak can be studied at eRHIC and how clearly the saturation predictions can be distinguished from a conventional leading twist shadowing (LTS) scenario [58,59]. To derive the latter, we use a hybrid Monte Carlo generator, consisting of PYTHIA-6 [60] for parton generation, showering and fragmentation, DPMJet-III [61] for the nuclear geometry, and a cold matter energy-loss afterburner [62]. The EPS09 [66] nuclear parton distributions were used to include leading twist shadowing. The resulting correlation function is shown in Figure 2-25 as the black curve/points. The error bars reflect the statistical uncertainties for 1 fb$^{-1}$/A integrated luminosity. The solid black curve includes detector smearing effects, the dashed curve shows the result without taking into account any detector response. The red curve in Figure 2-25 represents the CGC predictions. While the underlying model is identical to that shown in Figure 2-24, the simulations include all charged hadrons as well as the quark channel contributions. The solid and dashed red lines represent detector response effects switched on and off, respectively. The shaded region reflects uncertainties in the CGC predictions due to uncertainties in the knowledge of the saturation scale, $Q_S$. This comparison nicely demonstrates the discriminatory power of these measurements. In fact, already with a fraction of the statistics used here one will be able to exclude one of the scenarios conclusively.

Another way of studying the di-hadron correlation function in more detail is illustrated in Figure 2-26. Here, the predicted suppression is expressed through $J_{eAu}$, the relative yield of correlated away-side hadron pairs in e+Au collisions compared to e+p collisions scaled down by $A^{1/3}$ (the number of nucleons at a fixed impact parameter):

$$J_{eA} = \frac{1}{A^{1/3}} \frac{\sigma_{eA}^{pair}/\sigma_{eA}}{\sigma_{ep}^{pair}/\sigma_{ep}}$$

The absence of collective nuclear effects in the pair production cross section would correspond to $J_{eAu} = 1$, while $J_{eAu} < 1$ would signify the suppression of di-hadron correlations.

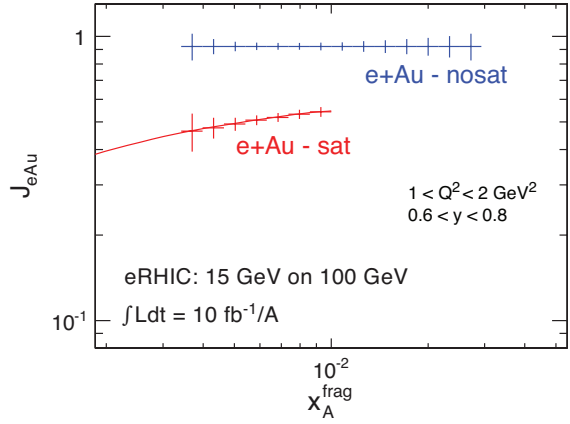 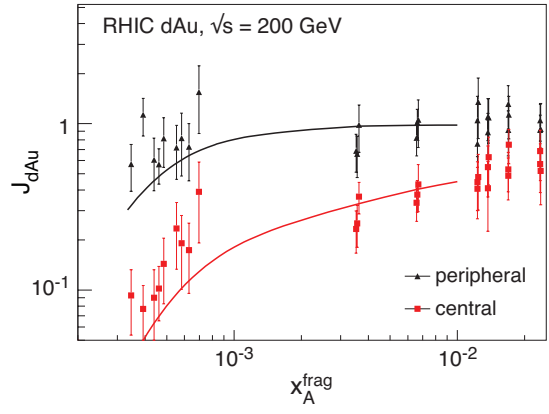

Figure 2-26: The relative yield of away-side di-hadrons in e+Au compared to e+p collisions, $J_{eAu}$, plotted versus $x_A^{frag}$, which is an approximation of the average momentum fraction of the struck parton in the Au nucleus, derived from the kinematics of the measured hadrons assuming they carry the full parton energy. Predictions for linear (nosat) and non-linear (sat) QCD models for eRHIC energies are presented. The statistical error bars correspond to 10 fb$^{-1}$/A integrated luminosity.

Figure 2-27: The corresponding measurement in √s = 200 GeV per nucleon d+Au collisions at RHIC [63]. The curves depict calculations in the CGC framework.

In Figure 2-26, $J_{eAu}$ is plotted as a function of $x_A^{frag}$, which is an approximation of the average momentum fraction of the struck parton in the Au nucleus, $x_g$, derived from the kinematics of the measured hadrons assuming they carry the full parton energy. Compared to the measurement shown in Figure 2-25 this study requires the additional e+p baseline and higher statistics since the data sample has to be divided in bins of $x_g$. The error bars reflect the statistical uncertainties for 10 fb$^{-1}$/A integrated luminosity. It is instructive to compare this plot with the equivalent measurement in d+Au collisions at RHIC shown in Figure 2-27 [63,64]. Note that here $A^{1/3}$ is the definition of $J_{dAu}$ is replaced by the number of the binary nucleon–nucleon collisions at a fixed impact parameter. In both collisions systems, e+A and p+A, the exact momentum fraction of the struck parton, $x_g$, cannot experimentally be measured but has to be ultimately modeled. However, these calculations are much better constrained in DIS where the key kinematic variables $x$ and $Q^2$ are known precisely. The two curves in the right panel of Figure 2-27 represent the same CGC calculations used in our simulations but without the Sudakov factor. This example nicely illustrates the correspondence between the physics in p(d)+A and e+A collisions but also shows superior control of the underlying kinematics in DIS.

## Diffractive Events

Diffractive interactions result when the electron probe in DIS interacts with a proton or nucleus by exchanging a colorless combination of partons referred to as "pomeron". The simplest model of pomeron exchange is that of a colorless combination of two gluons. One of the key signatures of these events is a large rapidity gap between the scattered proton or nuclei traveling at near-to-beam energies and the final-state particles produced at mid-rapidity that can measured in a central detector. A schematic diagram of a diffractive event is depicted in Figure 2-28.

At HERA, an unexpected discovery was that 15% of the $e+p$ cross-section is from diffractive final states; the naive expectation was that such gaps in rapidity are exponentially suppressed. Linear QCD is able to describe several aspects of the behavior of diffractive events such as their $Q^2$ dependence, which is well understood by conventional DGLAP evolution. Other features, however, especially the observation that the ratio of the diffractive to the total cross-section is constant with energy, cannot be easily reconciled in a linear QCD picture but can be naturally explained assuming parton saturation.

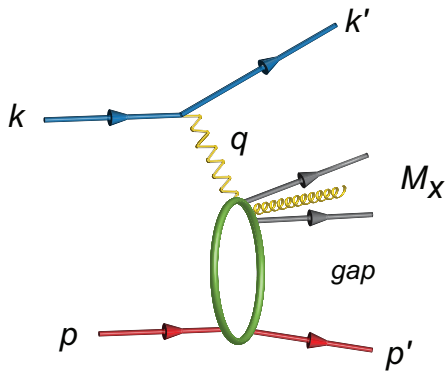

Figure 2-28: Kinematic quantities for the description of a diffractive event. $t = (p - p')^2$ is the square of the momentum transfer at the hadronic vertex, $M_X$ is the invariant mass of the final state.

What makes the diffractive processes so interesting is that that they are most sensitive to the underlying gluon distribution, in some cases up to $\sigma \propto xg(x, Q^2)^2$ [65]. Furthermore, exclusive diffractive events are the only known class of events that allows one to study the spatial distribution of gluons in nuclei. It is therefore anticipated that the strongest hints for manifestations of new, non-linear effects in $e+A$ collisions are likely to come from inclusive as well as exclusive diffractive measurements. However, while the physics goals are golden, the measurement of these events is technically challenging, but not insurmountable, and requires careful planning of the detector and interaction region.

For nuclei one distinguishes two kinds of diffractive events: *coherent* (nucleus stays intact) and *incoherent* (nucleus breaks up, but nucleon stays intact). Both are interesting in their own right. Coherent diffraction is sensitive to the space-time distribution of the partons in the nucleus, while incoherent diffraction (dominating at larger $t$ and thus small variation in impact parameter $b_T$) is most sensitive to high parton densities where saturation effects are stronger. While in coherent $e+p$ collisions, the scattered protons can be detected in a forward spectrometer placed many meters down the beam line, scattered heavy nuclei stay too close to the ion beam. However, studies showed that the nuclear breakup in incoherent diffraction can be detected at eRHIC with close to 100% efficiency by measuring the emitted neutrons in a zero degree calorimeter placed after the first dipole magnet that bends the hadron beam. This tagging scheme could be further improved by using a forward spectrometer to detect charged nuclear fragments. A rapidity gap and the absence of any break-up fragments were found to be sufficient to identify coherent events with very high efficiency.

## Inclusive Diffractive Events

One of the first studies that potentially could yield clear evidence for saturation at eRHIC is the measurement of the ratio of diffractive to total cross-sections. While in $e+p$ collisions at HERA this ratio was about ~15%, CGC calculations predict this ratio to be significant larger in $e+A$ collisions at eRHIC. The upper panels in Figure 2-29 and Figure 2-30 show the rate of diffractive over total cross-section as a function of the produced invariant mass of the diffractive system, $M_X^2$, for $x = 10^{-3}$ and $Q^2 = 1$ and $5$ GeV$^2$, respectively. For fixed $Q^2$ and $x$, $M_x^2$ can also be expressed as the fraction of the momentum of the pomeron that is carried by the struck parton within the proton or nucleus, $\beta$, shown along the alternative abscissa on the top of each plot where $\beta \sim xQ^2/(Q^2 - M_x^2)$. The red curves represent the predictions of the IPSat saturation model [51,67], which clearly shows that the relative predicted rate for diffractive events in $e+Au$ collisions is significantly larger than that in $e+p$. In Figure 2-30, for $Q^2 = 5$ GeV$^2$, we also included predictions of a leading-twist shadowing (LTS) model (blue curve/band) [58,59]. In $e+Au$ collisions, the LTS predictions depend on the amount of shadowing which is currently little constraint by data [66]. The blue band reflects this uncertainty. For $Q^2 = 1$ GeV$^2$ LTS calculations are not applicable.

To better illustrate the difference in the predictions from the saturation and LTS models we plot the ratio between the relative diffractive cross-section in $e+Au$ over that in $e+p$ in the lower panel of Figure 2-30 for $Q^2 = 5$ GeV$^2$. Figure 2-29 depicts this double ratio at $Q^2 = 1$ GeV$^2$ only for the saturation model.

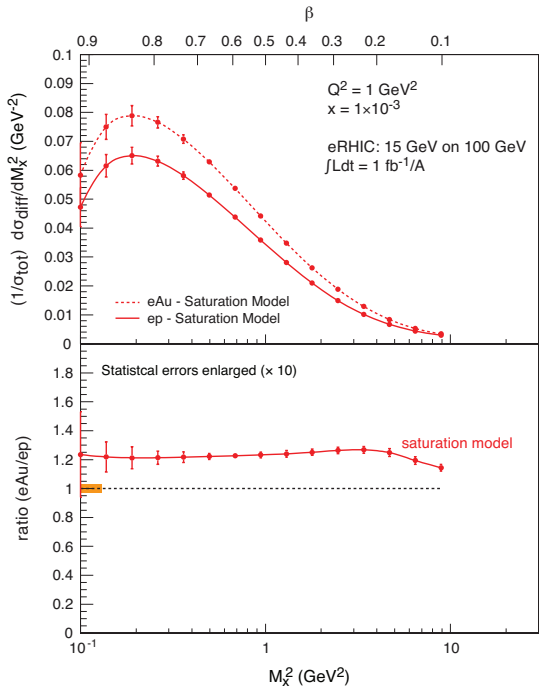

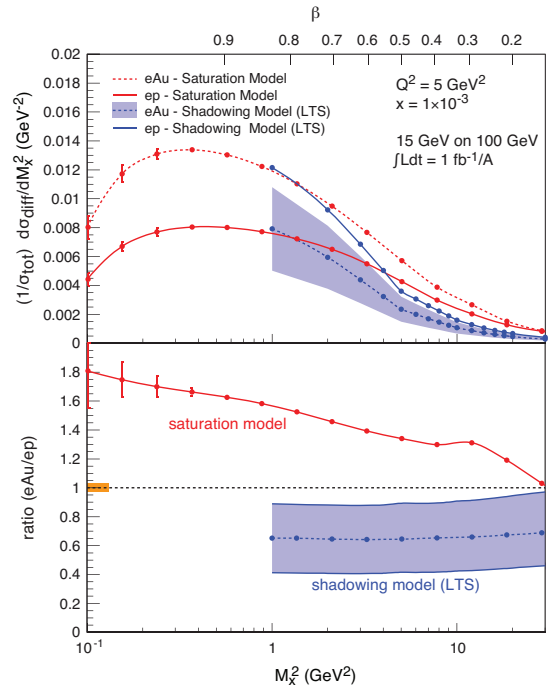

Figure 2-29: The top panel depicts the ratio of diffractive over total cross-sections at $Q^2 = 1$ GeV$^2$ and $x = 10^{-3}$, plotted as a function of the invariant mass of the produced particles, $M_X^2$, for nominal eRHIC energies. The bottom panel shows the corresponding double ratio $[(d\sigma_{\text{diff}}/dM_X^2)/\sigma_{\text{tot}}]_{eA} / [(d\sigma_{\text{diff}}/dM_X^2)/\sigma_{\text{tot}}]_{ep}$. The statistical error bars for an integrated luminosity of 1 fb$^{-1}$/A are too small to depict and are enlarged by a factor 10. Predictions are from a saturation model [51,67]. The orange bar indicates systematic uncertainties (to scale).

Figure 2-30: Same as the left figure but for $Q^2 = 5$ GeV$^2$. Also shown are predictions from the LTS model [58,59]. The error band on the $e+Au$ curve reflects the uncertainty in the amount of shadowing in this kinematic range.

Detailed studies in the saturation framework show that the enhancement of the double ratio is most pronounced at large $\beta$ [67]. At small $\beta$ values (large $M_X^2$) the relative diffractive cross-section in $e+Au$ is similar or even less than that in $e+p$. This behavior is clearly seen for $Q^2 = 5$ GeV$^2$: the double-ratio is largest for small $M_X^2$ and falls monotonically with increasing $M_X^2$. For $Q^2 = 1$ GeV$^2$ (Figure 2-29) the double ratio is closer to unity for all $M_X^2$ since the saturation effects are suppressing the diffractive more than the total cross-section in $e+Au$. As can be seen in Figure 2-30, the LTS model shows a different behavior for the double ratio. While the saturation model predicts an *increased* rate of diffraction in $e+Au$, LTS is predicting a *decreased* rate and little to no dependence on $M_X^2$ (or $\beta$).

The statistical error bars on the double ratios, shown in Figure 2-29 and Figure 2-30 correspond to an integrated luminosity of 1 fb$^{-1}$/A. We conclude that the errors of the actual measurement would be dominated by the systematic uncertainties dependent on the quality of the detector and of the luminosity measurements. The orange bar reflects this uncertainty assuming a 3% uncertainty per collision system. Our studies confirm that the two scenarios can be clearly distinguished over a wide range in $x$ and $Q^2$, allowing for a clear day-1 measurement aimed at finding evidence of parton saturation.

### *Exclusive Diffractive Vector Meson Production*

Perhaps the best analogy to diffraction in high-energy QCD comes from optics: the diffractive pattern of the light intensity on a screen behind a circular obstacle features the well-known diffractive maxima and minima. The positions of the diffractive minima are related to the size of the obstacle by $\theta_i \propto 1/R$ for small-angle diffraction. Elastic scattering in QCD has a similar structure. The elastic process is described by the differential scattering cross-section $d\sigma_{\text{elastic}}/dt$ with the variable $t$ describing the momentum transfer between the target and the projectile. The essential difference to QCD is: (*i*) The proton/nuclear target is not always an opaque "black disk" obstacle of geometric optics. A smaller projectile, which interacts more weakly due to color-screening and asymptotic freedom, is likely to produce a different diffractive pattern from the larger, more strongly interacting, projectile. At small-$x$ the spectrum of the cross-section with respect to $t$ is related to the transverse spatial distribution of the gluons in the ion through a Fourier transform [73]. (*ii*) The scattering in QCD does not have to be completely elastic: the projectile or target may break up. The event is still called diffractive, as long as there is a rapidity gap. In these so-called incoherent diffractive events, the typical diffractive pattern of minima and maxima in $d\sigma_{\text{elastic}}/dt$ seen in coherent diffractive events is absent (see Figure 2-31). Nevertheless, these events are of great interest since the incoherent cross-section is a direct measure of the lumpiness of the gluons in the nucleus [67]. The property (*i*) is very important for diffraction in DIS in relation to saturation physics. At larger $Q^2$, the virtual photon probes shorter transverse distances, and is less sensitive to saturation effects. Conversely, a virtual photon with lower $Q^2$ is likely to be more sensitive to saturation physics.

Diffractive vector meson production, $e + A \rightarrow e' + A' + V$ where $V = J/\psi, \phi, \rho$, or $\gamma$, is a unique process, since it allows the measurement of the momentum transfer, $t$, at the hadronic vertex even in $e+A$ collisions where the 4-momentum of the outgoing heavy nuclei cannot be measured. Since only one new final state particle is generated, the process is experimentally clean and can be unambiguously identified by the presence of a rapidity gap. The study of various vector mesons in the final state allows a systematic exploration of the saturation regime [68]. The $J/\psi$ is the vector meson least sensitive to saturation effects due to the small size of its wave function. Larger mesons such as $\phi$ or $\rho$ are considerably more sensitive to saturation effects [71].

The key to the spatial gluon distribution is the measurement of the d$\sigma$/dt distribution. As follows from the optical analogy, the Fourier-transform of the square root of this distribution is the source distribution of the object probed. In Figure 2-31, we show the differential cross-section dσ/dt for both $J/\psi$- and $\phi$-meson production for saturation and non-saturation models. Both curves were generated with the Sartre event generator [69,70], an $e+A$ event generator specialized for diffractive exclusive vector meson production based on the bSat dipole model [71] and its linearization, the bNonSat model [72]. The parameters of both models were tuned to describe the $e+p$ HERA data. The generated energies correspond to nominal eRHIC energies $E_e$ =15 GeV and $E_A$ =100 GeV. We limit the calculation to $1 < Q^2 < 10$ GeV² and $x < 0.01$ to stay within the validity range of saturation *and* non-saturation models.

As the J/ψ is smaller than the φ, one sees little difference between the saturation and no-

The produced events were passed through an experimental filter and scaled to reflect an integrated luminosity of 10 fb⁻¹/A; experimental cuts are listed in the figures. We assume a conservative $t$-resolution of 5%, which should be easily achievable with eRHIC detectors. Experimentally, the sum of the coherent and incoherent parts of the cross-section is measured. Through the detection of emitted neutrons (e.g. by zero-degree calorimeters) from the nuclear breakup and, optionally, the breakup products in detectors along the beam-line (Roman-Pots) in the incoherent case, we verified that is experimentally feasible to disentangle the two contributions unambiguously. saturation scenarios for exclusive $J/\psi$ production but a pronounced effect for the $\phi$, as expected.

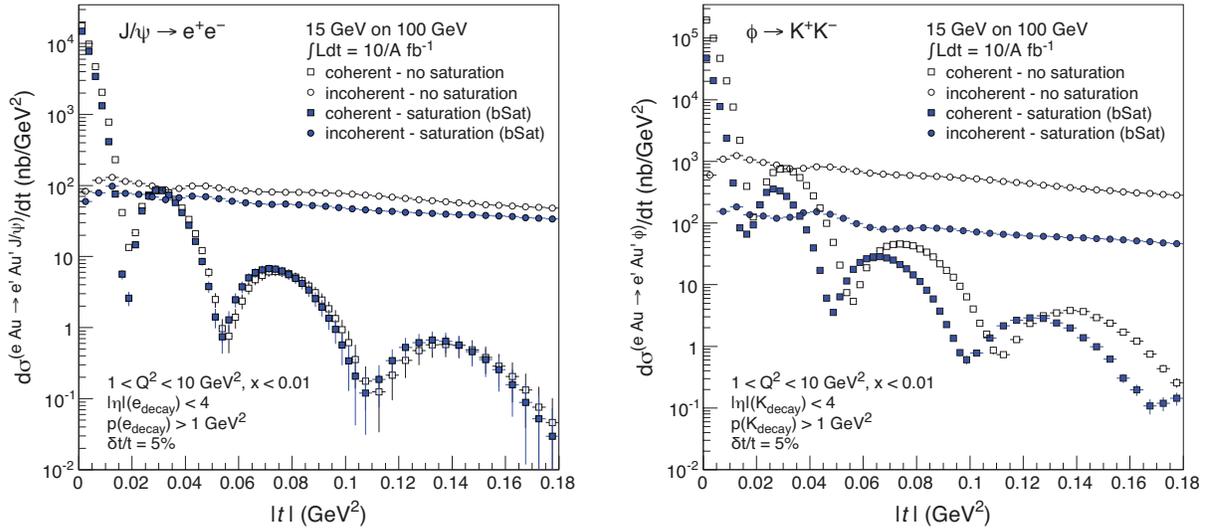

Figure 2-31: $d\sigma/dt$ distributions for exclusive $J/\psi$ (left) and $\phi$ (right) production in coherent and incoherent events in diffractive $e$+Au collisions at eRHIC. Predictions from saturation and non-saturation models are shown.

The coherent distributions in Figure 2-31 can be used to obtain information about the gluon distribution in impact parameter space through a Fourier transform [73]. In Figure 2-32 we show the resulting Fourier transforms of the coherent curves in Figure 2-31, using the range $-t < 0.36$ GeV². As a reference, we show (dotted line) the original input source distribution used in the generator, which is the Woods-Saxon function integrated over the longitudinal direction. The obtained distributions have been normalized to unity. For testing the robustness of the method, we used the statistical errors in $d\sigma/dt$ to generate two enveloping curves, $d\sigma/dt(t_i) \pm \delta(t_i)$ where $\delta$ is the one sigma statistical error in each bin $t_i$. The curves are then transformed individually, and the resulting difference defines the uncertainty band on $F(b)$. Surprisingly, the uncertainties due to the statistical error are negligible, and are barely visible in Figure 2-32.

The non-saturation curves for $\phi$ and $J/\psi$-meson production reproduce the shape of the input distribution perfectly. For the saturation model, the shape of the $J/\psi$ curve also reproduces the input distribution, while the $\phi$ curve does not. As explained above, this is expected, as the size of the $J/\psi$ meson is much smaller than that for $\phi$, making the latter more susceptible to non-linear effects as already observed in Figure 2-31. We conclude that the $J/\psi$ is better suited for probing the transverse structure of the nucleus. However, by measuring $F(b)$ with both $J/\psi$ and $\phi$ mesons, one can obtain valuable information on how sensitive the measurement is to non-linear effects.

Thus, both measurements are important and complementary to each other. The results in Figure 2-32 provide a strong indication that eRHIC will be able to obtain the nuclear spatial gluon distribution from the measured coherent $t$-spectrum from exclusive $J/\psi$ production in $e+$A, in a model independent fashion.

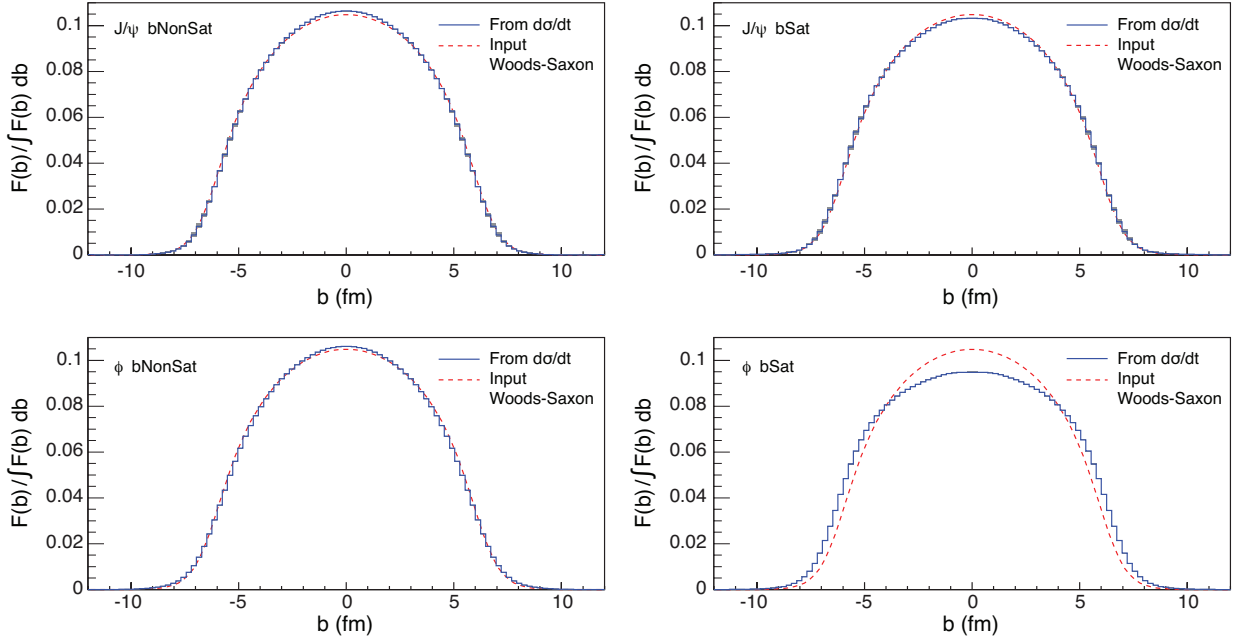

Figure 2-32: The Fourier transforms obtained from the distributions in Figure 2-31 for $J/\psi$-mesons in the upper row and $\phi$-mesons in in the lower row. The results from both saturation and non-saturation are shown. The used input Woods-Saxon distribution is shown as a reference in all four plots.

Strictly, the integral over $t$ in the Fourier transformation should be performed up to $|t| = \infty$. We studies the effects by varying the upper integration limit and found fast convergence towards the input Woods-Saxon distribution already for $|t|{\sim}0.1$ GeV$^2$. Another interesting aspect of diffractive vector meson production is its $Q^2$-dependence. The two panels in Figure 2-33 show the ratios $(d\sigma_{eAu}/dQ^2)/(d\sigma_{ep}/dQ^2)$ of the cross-sections in $e+$Au over that in $e+p$ for exclusive $J/\psi$ (left panel) and $\phi$ (right panel) production in coherent diffractive events. The ratios, plotted as functions of $Q^2$ for saturation and non-saturation models, are scaled by a factor $A^{4/3}$. In the dilute limit (large $Q^2$) this scaling is expected to hold for the integral of the coherent peak, which dominates the cross-section, while deviations from it at lower $Q^2$ are due to the denser gluon regime. For large $Q^2$, the ratios asymptotically approach unity.

All curves were generated with the Sartre event generator as discussed earlier. We again limit the calculation to $1 < Q^2 < 10$ GeV$^2$ and $x < 0.01$ to stay within the validity range of both models. The basic experimental cuts are listed in the legends of the panels in Figure 2-33. As expected, the difference between the saturation and non-saturation curves is small for the smaller-sized $J/\psi$ ($< 20\%$), which is less sensitive to saturation effects, but is substantial for the larger $\phi$, which is more sensitive to saturation. For both mesons, the difference is larger than the statistical errors. In fact, the small errors for diffractive $\phi$ production indicate that this measurement can already provide substantial insight into the saturation mechanism after only 1 fb$^{-1}$/A or less.

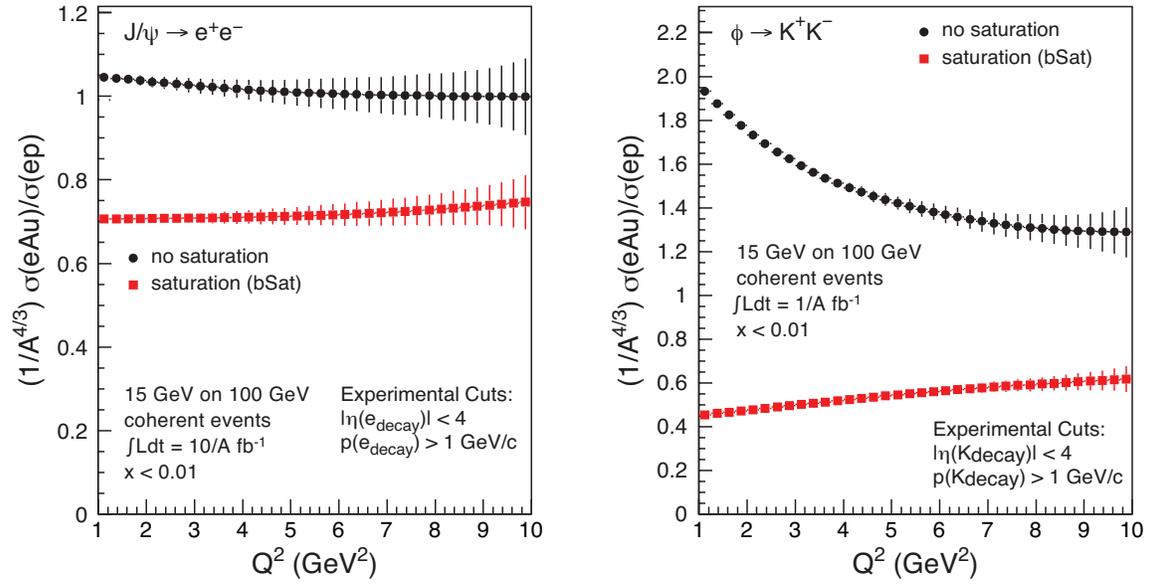

Figure 2-33: Ratios of the cross-sections for exclusive $J/\psi$ (left panel) and $\phi$ (right panel) meson production in coherent diffractive e+A and e+p collisions as a function of $Q^2$. Predictions for saturation and non-saturation models for eRHIC energies are presented. The ratios are scaled by $1/A^{4/3}$.

## 2.2.2 Hadronization and Energy Loss

In DIS on nuclear targets, one observes a suppression of hadron production analogous to, but weaker than, the quenching in the inclusive hadron spectrum observed in heavy-ion collisions at RHIC and the LHC [74]. The cleanest environment to address nuclear modifications of hadron production is clearly nuclear DIS. Semi-inclusive DIS in $e$+A collisions provides a known and stable nuclear medium, well-controlled kinematics of hard scattering, and a final state particle with well-known properties. It allows one to experimentally control many kinematic variables; the nucleons act as femtometer-scale detectors allowing one to experimentally study the propagation of a parton in this "cold nuclear matter" and its space-time evolution into the observed hadron.

The time for the produced parton to shed off its color depends on its momentum and virtuality when it was produced. The process could take place entirely inside the nuclear medium, outside the medium, or somewhere in-between, as illustrated in the cartoon in Figure 2-34.

By facilitating studies on how struck partons propagate through cold nuclear matter and evolve into hadrons, eRHIC would provide independent and complementary information essential for understanding the response of the nuclear medium to a colored fast moving (heavy or light) quark. With its collider energies and its large range of $\nu$, the energy of the exchanged virtual photon, eRHIC is unique for providing clean measurements of medium induced energy.

Experimental data on hadron production in nDIS are typically presented in terms of the ratio of the single hadron multiplicity per DIS event on a target of mass number A, normalized to the multiplicity on a proton or deuterium target [75,76,77,78]. This ratio can be studied as a function of the virtual photon energy $\nu$, the virtuality $Q^2$, the hadron transverse momentum $p_T$, and $z_h$, the fractional energy carried by the hadron with respect to the virtual photon energy in the target rest frame, i.e., $z_h = E_h/\nu$. The basic question to be answered is on what time scale the color of the struck quark is neutralized, acquiring a large inelastic cross-section for interaction with the medium. Energy loss models [79,80,81] assume long color neutralization times, with "pre-hadron" formation outside the medium and parton energy loss as the primary mechanism for hadron suppression. Absorption models [82,83,84,85] assume short color neutralization times with in medium pre-hadron formation and absorption as the primary mechanism.

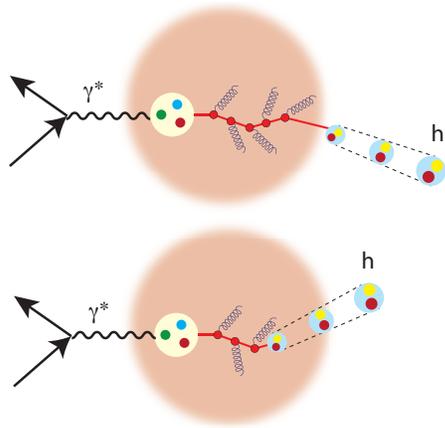

Figure 2-34: Illustration of the interactions of a parton moving through cold nuclear matter when the produced hadron is formed outside (upper) and inside (lower) the nucleus.

At nominal eRHIC energies, the range of photon energies would be 30 GeV $< \nu <$ 2800 GeV, much wider than those at HERMES (2–25 GeV), which provided the most detailed existing studies so far. It therefore offers more channels to study hadronization inside and outside of the nucleus and reaches into a region relevant for the $p$+A and A+A program at the LHC. eRHIC's high luminosity will allow us to conduct multi-differential measurements in all kinematic variables. A novel feature of these studies at eRHIC would be measurements providing insight into the energy loss and hadronization of heavy quarks to form *charm* and possibly *bottom* mesons.

In Figure 2-35 we show simulations results for the multiplicity ratio of semi-inclusive DIS cross-sections for producing a single pion in $e$+Pb collisions over that in $e$+d as a function of $z$ at two different photon energies: $\nu$ = 35 GeV at $Q^2$ = 10 GeV$^2$ (solid line and square symbols) and $\nu$ = 145 GeV at $Q^2$ = 35 GeV$^2$ (dashed line and open symbols) [1]. The $p_T$ of the observed hadrons is integrated.

The ratio for pions (red square symbols) was taken from the calculation in [86,87] but extended to lower $z$. In this model approach, pions are suppressed in $e$+A collisions due to a combination of the attenuation of pre-hadrons as well as medium-induced energy loss. In Figure 2-35, the solid lines are predictions of pure energy loss calculations using the energy loss parameters of [88]. The large differences in the suppression between the square symbols and solid lines are immediate consequences of the characteristic time scale for the color neutralization and the details of the attenuation of pre-hadrons, as well as the model for energy loss. The error bars reflect the statistical uncertainties for 10 fb$^{-1}$/A integrated luminosity. With the size of the systematic errors shown by the yellow bar on the left of the unity ratio, the multiplicity ratio of pion production at eRHIC will provide an excellent and unique opportunity to study hadronization by using the nucleus as a femtometer scale detector.

The multiplicity ratios of $D^0$ meson production is shown in Figure 2-36 [1]. The significant difference of the ratio to that in pion production is an immediate consequence of the harder fragmentation function for heavy flavor mesons [89] and the amount of energy loss, or equivalently, the transport coefficient $\hat{q}$ in cold nuclear matter. The energy loss used in the simulation is a factor of 0.35 less than that of light quarks by taking into account the limited cone for gluon radiation caused by the larger charm quark mass. The strong sensitivity of the shape to the value of $\nu$ will be a unique and powerful tool in the understanding of energy loss of heavy quarks in cold nuclear systems. The discovery of such a dramatic difference in multiplicity ratios between light and heavy meson production in Figure 2-35 and Figure 2-36 at eRHIC would shed light on the hadronization process and on what governs the transition from quarks and gluons to hadrons.

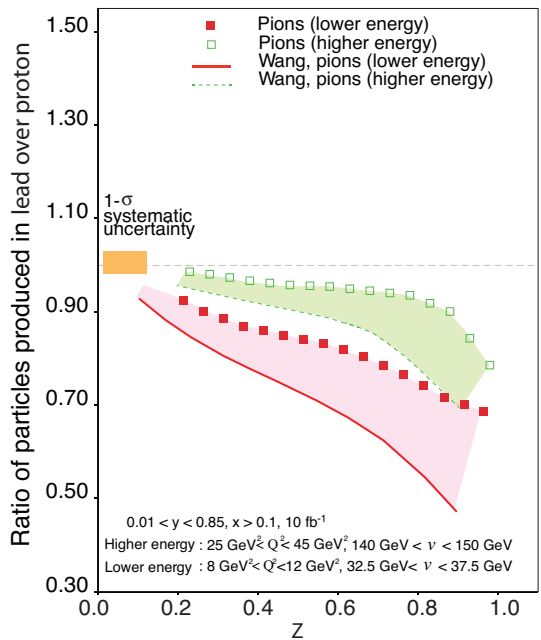 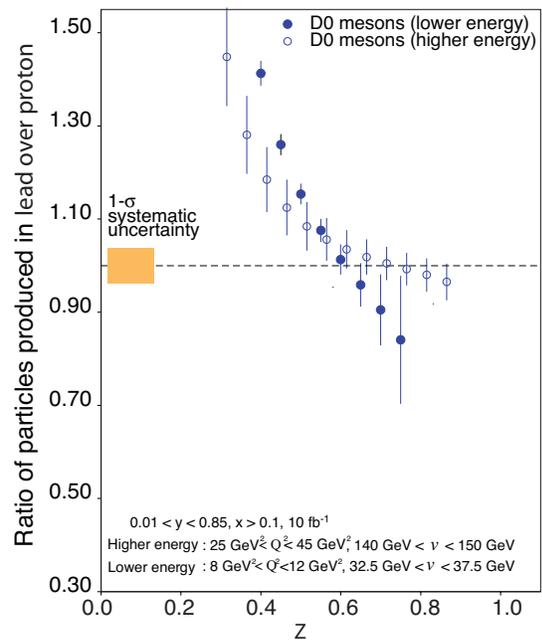

Figure 2-35: The ratio of semi-inclusive cross-sections for producing a single pion in e+Pb collisions over that in e+d collisions as a function of z. Solid symbols depict the ratio for photon energies of n = 35 GeV at Q2 = 10 GeV2 and open symbols $\nu$ = 145 GeV at v = 35 GeV2. Lines are predictions from pure energy loss calculations. Figure taken from [1].

Figure 2-36: Same as Figure 2-35 but for D0 - mesons. The statistical error bars are for 10 fb−1/A integrated luminosity. The orange box depicts the estimated systematic uncertainties for this measurement. Figure taken from [1].

# 3 eRHIC MACHINE DESIGN

## 3.1 The Design Concept

### 3.1.1 Accelerator Concept, Layout and Major Components

The accelerator design of the electron-hadron collider has been developed to fulfill the eRHIC physics goals. It entails the following major features:

- Hadron species: polarized protons (up to 250 GeV), polarized $^3\text{He}^{+2}$ ions (up to 167 GeV/u), heavy ions (typically $^{197}\text{Au}^{+79}$ or $^{238}\text{U}^{+92}$ ions, up to 100 GeV/u)
- Polarized electrons: in the range from 2 GeV up to 21 GeV
- The luminosity: $10^{33}$ - $10^{34}$ cm$^{-2}$s$^{-1}$ in terms of e-nucleon collisions

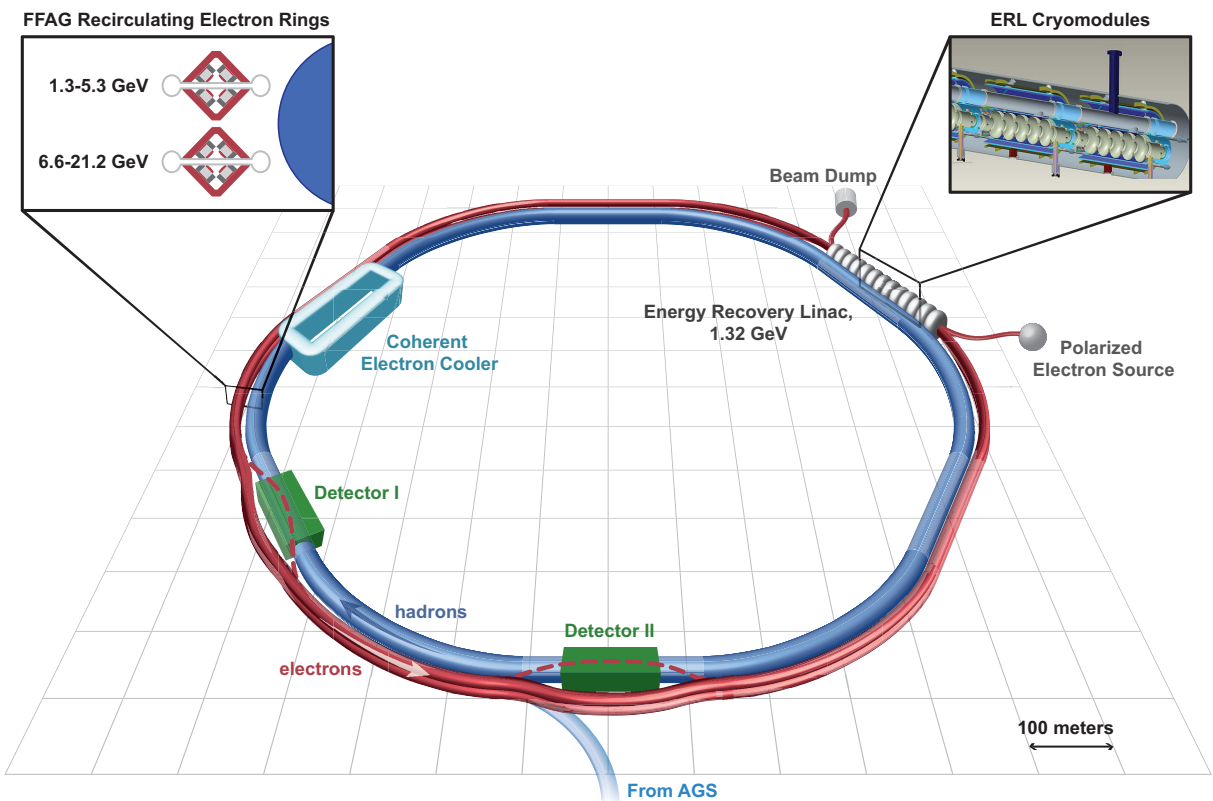

Figure 3-1: The layout of the eRHIC collider.

The key goal of the eRHIC accelerator design has been to achieve the required high-energy, high-luminosity performance at a realizable machine construction cost. For the hadron part of the machine, eRHIC takes advantage of the existing RHIC accelerator complex, including the full suite of injector systems for polarized protons and fully stripped heavy ions. The new electron accelerator is achieved through a cost-effective design, taking advantage of significant recent advances in accelerator technology.

As shown in Figure 3-1, the eRHIC facility uses one of the RHIC hadron beams (the clockwise-moving "blue" beam), with a high energy electron beam counter-rotating in the same tunnel, and collisions occurring in two intersection regions occupying the present experimental areas of the STAR (IR6) and PHENIX (IR8) detectors. The full range of RHIC hadron beams is thus available for eRHIC, up to 250 GeV for polarized protons and 100 GeV/u for Au ions.

The accelerated electrons originate in a new, high-current polarized source and are accelerated to 12 MeV for injection into a 1.32 GeV Energy Recovery Linac (ERL). Using recirculating rings inside the RHIC tunnel the electrons make multiple passes through the ERL, gaining 1.32 GeV of energy with each pass. The electrons can be extracted after 12 passes (15.9 GeV) or 16 passes (21.2 GeV) and brought into collision with the hadron beam at either IR6 or IR8. The spent electron bunch is then recirculated back through the ERL, returning its energy to the superconducting RF structure of the linac, after which the decelerated electrons are dumped. Thus, each electron bunch participates in only one collision crossing with the hadron beam, and the process repeats itself for each succeeding bunch. The electron bunches are accelerated and brought into collision with the hadron beam at a frequency of 9.4 MHz. As described below, the luminosity goals are achieved with an electron beam current of 50 mA and tightly focused (small emittance) beams for both the hadrons and electrons. The major eRHIC accelerator components are:

- The 12 MeV injection complex, located at the IR2 area of the RHIC tunnel. It includes a high-current polarized beam injector and 12 MeV linear accelerator. A beam dump for disposing of the 12 MeV decelerated beam is also located in this area.

- The 1.322 GeV Energy Recovery Linac (ERL) is located along the IR2 straight section. The ERL is 120 m long and consists of a string of superconducting 422 MHz cavities. The use of energy recovery technology in the main accelerator linac is essential to reach a high value (50mA) of the electron average current. Additional RF cavities (844 MHz) are used to replenish the beam energy loss caused predominantly by synchrotron radiation. Also, 2.1 GHz cavities are utilized for reducing the beam energy spread.

- Two vertically stacked recirculation beamlines run around the RHIC tunnel circumference, outside of the hadron ring. The optics of each of the beamlines is based on a Fixed Field Alternating Gradient (FFAG) lattice, which is capable of transporting beams of different energies within a common vacuum chamber. The first FFAG beamline transports electrons with energies from 1.3 GeV to 6.6 GeV. The second FFAG beamline is used to pass beams in the 7.9-21.2 GeV range. The magnetic structure of both beamlines is based on permanent magnets. The main idea behind using the FFAG lattice approach and the permanent magnet technology is to lower machine construction and operation costs.

- A spreader and a combiner are placed either side of the ERL for proper distribution and matching of the electron beams of different energies between the ERL and FFAG beamlines. Both the spreader and the combiner have 16 arms used to transport beams of particular energies. The arms also are used for optics matching and path length correction (to make one turn transport completely isochronous and achromatic) as well as for betatron phase adjustments. 16 arms are required for acceleration to 21.2 GeV. For acceleration up to 15.9 GeV only 12 arms are used.

- A cooling device in the IR10 region of the RHIC tunnel achieves cooling of the proton and ion beams. The device will employ the Coherent Electron Cooling technique for efficient cooling in longitudinal and transverse planes.

- The electron-hadron collisions occur in two interaction regions (IR6 and IR8 RHIC areas). Near these interaction regions 15.9 GeV or 21.2 GeV electrons are extracted from the FFAG beamline using a septum magnet and directed into a dedicated beamline towards the experimental detectors. The interaction regions include superconducting magnets and provide strong focusing to achieve the $\beta^*$=5 cm for both beams. The electron and hadron beams are brought into the collision with a 10 mrad crossing angle. Crab cavities are employed to prevent loss of luminosity due to the crossing angle.

The present RHIC accelerator uses superconducting magnets to circulate hadron beams in two rings of 3834 m circumference. The wide energy reach of RHIC provides a natural opportunity to operate eRHIC over a wide range of center-of-mass collision energies. Existing proven accelerator technologies, exploited in RHIC and its injectors to produce and preserve proton beam polarization, will provide the highly polarized proton beam required for the eRHIC experiments. Modifications of the present RHIC machine for the eRHIC era include new quadrupole and dipole magnets in two interaction regions with experimental detectors, copper coated beam pipe and additional Siberian Snakes for acceleration of polarized $^3$He$^{+2}$. A cooling device will be added with the purpose of producing small transverse and longitudinal beam emittances. Also, space charge compensation is planned in order to provide sufficiently high luminosity at lower hadron beam energies.

### 3.1.2 Design Beam Parameters and Luminosities

Based on the fact that electrons, accelerated by the linear accelerator, collide with the protons (or ions) accelerated and stored in the circular machine, the eRHIC collision scheme is called the "linac-ring" scheme. This scheme has been chosen for eRHIC because of several clear advantages it brings in luminosity and electron polarization. On the luminosity side the "linac-ring" scheme overcomes one of the fundamental luminosity limitations of the "ring-ring" scheme from circulating electron beam quality deterioration caused by many repeating beam-beam interactions. Unlike the electron beam circulating in a storage ring, the electron beam from a linac passes through the collision point(s) only once. Hence, a beam-beam interaction of much higher strength can be allowed, paving the way to higher luminosity. The luminosity of the "linac-ring" scheme can be written as a function of the hadron beam parameters:

$$L = f_c \xi_h \frac{\gamma_h}{\beta_h^*} \frac{ZN_h}{r_h} H_{hg} H_p,$$

where $r_h = Z^2 e^2 / Mc^2$ is the hadron classical radius, $\xi_h$ is the hadron beam-beam parameter, $\beta_h^*$ is the hadron beta-function at the interaction point, $N_h$ is the hadron bunch intensity, $\gamma_h$ is the hadron relativistic factor and $Z$ is the hadron charge. $f_c$ is the collision frequency, which is the same as the bunch repetition rate.

The geometric loss factor $H_{hg}$ arises from luminosity loss due to the hour-glass effect and the crossing angle. With a 10 mrad crossing angle at the eRHIC collision points, the crab-crossing technique has to be employed to prevent luminosity loss.

The $H_p$ parameter represents the luminosity enhancement resulting from the pinching of the electron beam size at the collision point caused by the hadron beam focusing force.

The design luminosity and choice of beam parameters are influenced by both physical limits and practical considerations. Some of these limitations, such as the maximum limits for the hadron beam-beam and space-charge parameters for hadrons come from operational and experimental observations at RHIC or other hadron colliders. Others, like the choice of $\beta^*$ or the polarized electron beam current, are defined by the limits of accelerator technology. Considerations of the operational cost of the machine limit the electron beam power loss caused by synchrotron radiation.

The major limits assumed for the beam and accelerator parameters are:

- Polarized electron average current: $I_e \leq 50$ mA
- Minimum $\beta^* = 5$ cm (for both electrons and hadrons)
- Hadron space-charge tune shift: $\Delta Q_{sp} \leq 0.08$
- Hadron beam-beam parameter: $\xi_h \leq 0.015$
- Electron synchrotron radiation power: $P_{SR} < 3$ MW

Table 3-1 lists the beam parameters and design luminosities. The listed values of peak luminosity assume the following H-factors: $H_{hg}=0.84$ and $H_p=1.34$.

|  | e | p | $^2He^3$ | $^{79}Au^{197}$ |
|---|---|---|---|---|
| **Energy, GeV** | 15.9 | 250 | 167 | 100 |
| **CM energy, GeV** |  | 126 | 103 | 80 |
| **Bunch frequency, MHz** | 9.4 | 9.4 | 9.4 | 9.4 |
| **Bunch intensity (nucleons), $10^{11}$** | 0.07 | 3.0 | 3.0 | 3.0 |
| **Bunch charge, nC** | 1.1 | 48 | 32 | 19.6 |
| **Beam current, mA** | 10 | 415 | 275 | 165 |
| **Hadron rms normalized emittance, $10^{-6}$ m** |  | 0.2 | 0.2 | 0.2 |
| **Electron rms normalized emittance, $10^{-6}$ m** |  | 23 | 35 | 58 |
| **$\beta^*$, cm (both planes)** | 5 | 5 | 5 | 5 |
| **Hadron beam-beam parameter** |  | 0.004 | 0.003 | 0.008 |
| **Electron beam disruption** |  | 36 | 16 | 6 |
| **Space charge parameter** |  | 0.08 | 0.08 | 0.08 |
| **rms bunch length, cm** | 0.4 | 5 | 5 | 5 |
| **Polarization, %** | 80 | 70 | 70 | none |
| **Peak luminosity, $10^{33}$ cm$^{-2}$s$^{-1}$** |  | 4.1 | 2.8 | 1.7 |

Table 3-1: Beam parameters and luminosities.

The eRHIC bunch frequency is 9.4 MHz is equal to the bunch frequency of the present RHIC hadron beam. This choice of bunch frequency not only allows us to avoid modifications of the RHIC injection system but also suits eRHIC detector requirements.

The eRHIC accelerator design employs transverse and longitudinal cooling of hadron beams. The transverse cooling helps to reach the high peak luminosity by reducing the transverse beam size and is essential for achieving the small angular spread at the interaction points, which is required for efficient detection of collision products propagating at small angles to the hadron beam. The longitudinal cooling shrinks the bunch length to the scale of $\beta^*$ in order to maximize the $H_{hg}$ factor. Also, the crab-crossing system benefits from the shorter hadron bunch length in terms of the required voltage and, hence, the cost of the system. Both transverse and longitudinal cooling will be used to counteract beam emittance growth and related particle losses produced by intra-beam scattering, extending the luminosity lifetime and maximizing the average luminosity. Normally the cooling process will be activated during the store, after the hadron beam has been accelerated.

At a given bunch intensity the eRHIC hadron bunch is much denser than the RHIC bunch due to cooling. To operate with the high charge density beam the moderate cost hadron ring upgrades will be realized, such as copper coating of the beam pipe, an additional high-frequency RF system and BPM system modification.

A luminosity upgrade is possible in the future. This 'ultimate capability' upgrade involves accelerator and detector modifications to allow for increased bunch repetition rate and for further increase of beam intensities, which should bring the luminosity level to $10^{35}$ cm$^{-2}$s$^{-1}$.

### 3.1.3 Luminosity vs. Beam Energy

eRHIC will be able to produce collisions spanning a wide range of the center-of-mass-energies. The hadron machine can deliver 20-255 GeV proton beams and 5-100 GeV/u Au ions. The electrons can also be provided at different energies. The FFAG beamline contains and transports the electron beam up to a maximum energy 21.2 GeV. The 16-arm spreader and combiner have been designed for this maximum energy. Using a dedicated extraction scheme it will be possible to extract the electrons at the top energies of either 15.9 GeV or 21.2 GeV into the IR beamlines that go through the experimental detectors.

At the space charge limit the luminosity would drop sharply at lower hadron energy, as $E_h^3$, since the hadron bunch intensity has to be reduced. In order to prevent the luminosity drop at proton energies below 250 GeV the space charge compensation is applied.

An additional aspect of operating at different hadron energies is bunch frequency matching between the electron and hadron beams. The frequency matching scheme (Sec. 3.3.7) uses a hadron circumference lengthening and, for operation at low hadron energies, the harmonic switching method. The present frequency-matching scheme allows operation in several areas of hadron energies below 50 GeV/u, but the energies between 46 and 98 GeV/u cannot be used.

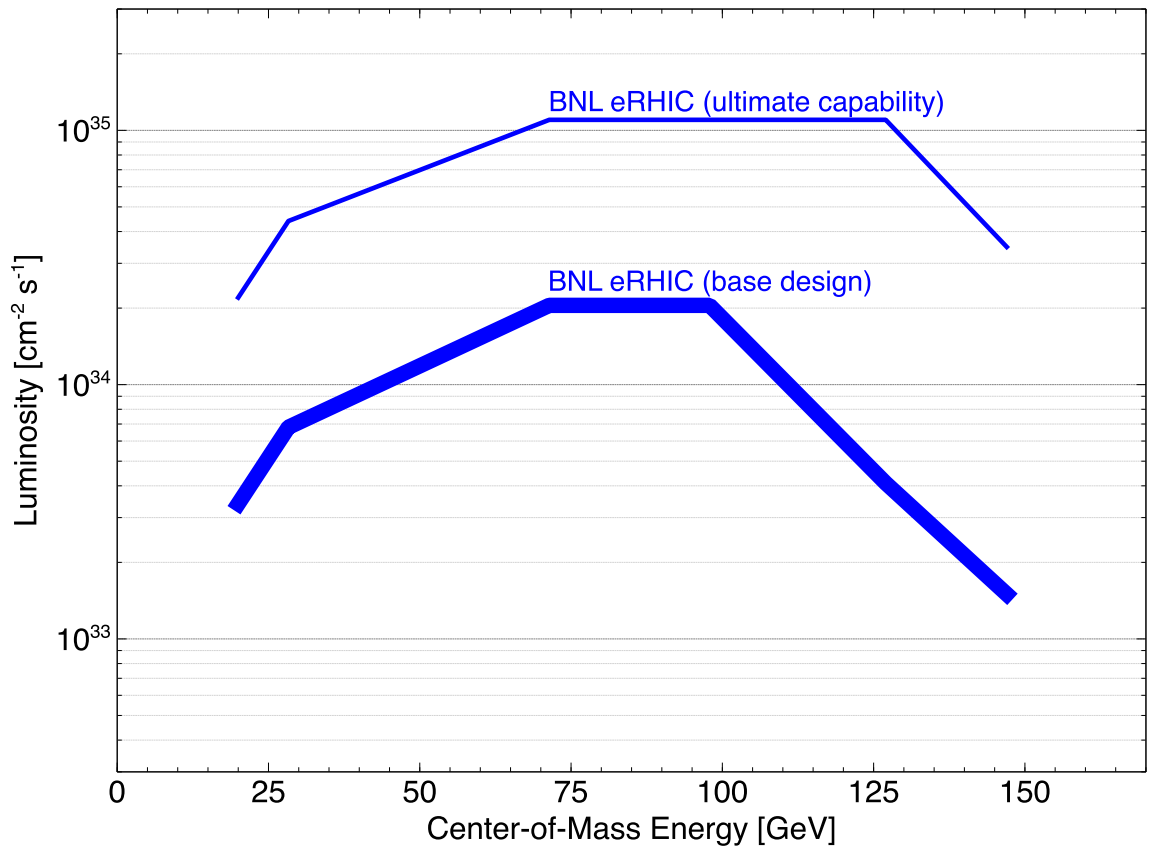

Figure 3-2: The optimal dependence of eRHIC peak luminosity of e-p collisions on the center-of-mass energy. Beside the base design, the luminosity curve corresponding to possible future upgrade, described at the end of section 3.1.2 is shown.

The limit on acceptable synchrotron radiation (SR) power (3 MW) results in a reduction in luminosity for operation with 21.2 GeV electrons as compared to 15.9 GeV electrons. To satisfy the SR power limit, the electron beam current accelerated to 21.2 GeV has to be decreased and the luminosity is reduced by a factor of 2.7 compared with collisions involving 15.9 GeV electrons. On the other side, the electron beam current at the electron energies lower than 15.9 GeV can be increased up to 50 mA, with correspondingly higher luminosity. On the other side, the electron beam current at the electron energies lower than 15.9 GeV can be increased up to 50 mA, with correspondingly higher luminosity.

Taking into account all luminosity-defining limits mentioned in preceding paragraphs, Figure 3-2 presents the optimal luminosity dependence on the center-of-mass energy. Highest luminosity in the base design is achieved when colliding 250 GeV protons with the electrons in 5-10 GeV energy range (at 50 mA electron current). At center-of-mass energies below 75 GeV the luminosity decrease is defined by the hadron beam-beam limit, capability of the space compensation system and by a detector limit on acceptable hadron-electron energy asymmetry.

### 3.1.4 Luminosity Sharing

Two experimental detectors will be located at the IR6 and IR8 areas. So-called luminosity sharing will be used to distribute the luminosity between these detectors. Since the trajectory of the top energy electron beam between the IR6 and IR8 collision points is longer than that of the hadron beam, collisions cannot be carried out simultaneously at IR6 and IR8. Instead, the machine switches back and forth between collisions at the individual IRs. This collision switching is realized by adjustment of the electron longitudinal phase with respect to the hadron beam. Proper longitudinal phase variation of the electron beam can provide any desired pattern of luminosity sharing. That is, one detector can be given more average luminosity than the other if required.

### 3.1.5 Cost Efficient Design Choices

Several special design choices have been made to minimize the construction and operation cost of the accelerator.

All major electron accelerator components are placed in the existing RHIC tunnel, greatly reducing the civil construction component of the machine construction cost.

The FFAG lattice allows 16 beam re-circulations using only two magnet beamlines, thereby reducing the number of magnets, vacuum chambers, peripheral support equipment, and beam instrumentation devices as compared to the more standard case of separate re-circulation passes for individual beam energies.

Larger number of recirculations provided by the use of the FFAG lattice, considerably reduces required length of main SRF linac.

Permanent magnet technology is used in the FFAG beamline magnets. This eliminates the need for a large number of magnet power supplies and power cables, leading to operational cost savings. A reduction in construction cost is also expected.

# 3.2 Technology Developments that Enable eRHIC

In this section the accelerator technologies underpinning the eRHIC machine design are presented. Some of them, like the high average current polarized electron source and the energy recovery linac based on a superconducting RF system, are necessary to achieve the high luminosity of eRHIC. Others, like the FFAG recirculation pass transport and the use of permanent magnets, are used to minimize the machine construction and operating costs.

## 3.2.1 The High Average Current Polarized Electron Source

eRHIC will require a highly polarized electron source with high average current, short bunch length and low emittance (Table 3-1). The current state-of-the-art polarized electron sources deliver either a high peak current, low average current beam such as the case at SLAC (>5A) or a high average current, low peak current beam as produced at JLab (4 mA). eRHIC will require a very high average current (up to 50 mA) with a bunch charge as high as 5.3 nC, with low emittance and a long cathode lifetime.

GaAs was selected as a photocathode because it is well established and widely used as a source of polarized electrons. The current state of the art single GaAs based electron sources cannot deliver the required 50 mA current due to ion back-bombardment and surface charge limits. Therefore, a novel approach to the design of the eRHIC electron source is required. To achieve the high beam current, BNL has adopted the Gatling gun principle: up to 20 photocathodes will generate electron bunches that are funneled onto a single common beam axis. The multiple cathodes increase the (current × lifetime) product of the gun. Each cathode produces the average current of 2.5 mA. Funneling bunches from the 20 cathodes together will produce 50 mA per twenty-cathode gun.

For a dual source scenario, cathode exchange can allow an operational lifetime of about 85 hours per week per source. While the cathodes of one source deliver beam to eRHIC, the cathodes of the other source can be exchanged with freshly activated ones. During this period, with the Gatling gun producing an average beam current of 50 mA, it would deliver a charge of about 15,300 C. Individual GaAs photocathodes can reliably deliver 1000 C at 2.5 mA. Twenty cathodes, each producing 2.5 mA, will need to deliver only about 765 C each to meet eRHIC requirements. Superlattice GaAs will be used to meet the minimum polarization requirement of > 80%.

A conceptual layout of the Gatling gun is shown in Figure 3-3. Twenty lasers deliver sequenced beam pulses to a circular array of photo cathodes. The cathodes are located on the surface of a cathode shroud charged to 220 kV. The repetition frequency of a single cathode is 450 kHz, due to the multiplexing of 20 cathodes producing electron pulses with bunch lengths of 1.5 ns. Solenoids placed within the anode provide focusing. A series of fixed dipole magnets first bend the off-axis electron bunches toward the gun's center axis. Then the bunches are kicked into alignment with the gun's center axis by the rotating magnetic field of the combiner magnet that bends the electron bunches of all the cathodes onto a common axis. The repetition frequency of the funneled bunches is 9.4 MHz for the total average current reaches~ 50 mA.

The Gatling-Gun has significant technical challenges. Field emission and the resulting ion back-bombardment can degrade the quantum efficiency of cathode surfaces. Low vacuum pressure at the cathode surface is critical for cathode life expectancy, GaAs cathode surfaces are highly sensitive and cathode life rapidly degrades with rising pressure. To have a practical operating lifetime between activations the gun cathodes require an operating vacuum pressure in the range of $10^{-12}$ Torr. Even at these extremely high vacuum levels the degradation of quantum efficiency will limit practical operating lifetime, requiring cathode processing and "activation" between periods of photoemission. This will require an extreme-vacuum-compatible mechanism to exchange cath-odes and a means of reprocessing and activation that is part of the gun system. The funneling combiner dipole magnet is a nontrivial development in itself requiring a magnetic field to rotate at 450 kHz. Developing in house expertise in the preparation of high quantum efficiency photocathodes is also challenging. These and other issues are being addressed in Gatling-Gun development at BNL (Sec. 3.4).

This program is complementary to other high-current polarized gun R&D programs at Jefferson Laboratory and at MIT, where single cathode gun systems are being developed. Advances made at these laboratories can be incorporated into the Gatling gun, which in effect serves to amplify the overall performance of these other programs.

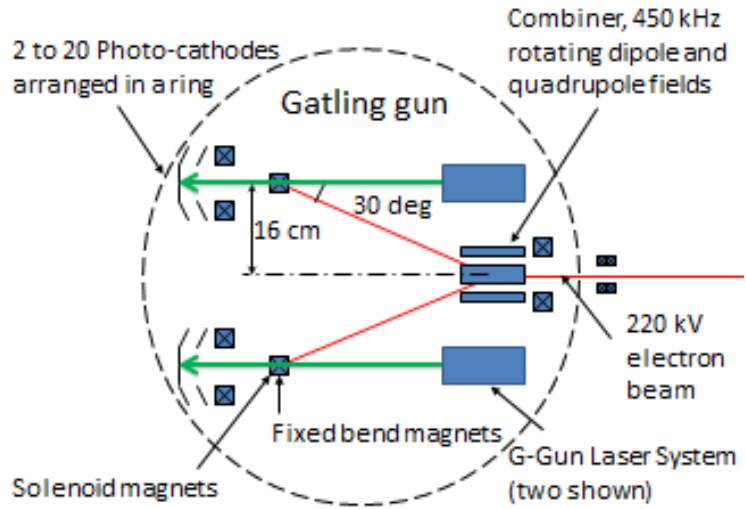

Figure 3-3: The Gatling gun concept for the eRHIC injector.

## 3.2.2 Energy Recovery Linacs and High Current SRF Cavities

High eRHIC luminosity demands high electron beam current accelerated in a linear accelerator operating in the continuous wave (CW) mode. Acceleration of high average current electron beams in a linac can only be practical using energy recovery. Without energy recovery the acceleration of 10 mA electron current to 15 GeV would require at least 150 MW of RF power to be continuously provided to the beam in the linac accelerating cavities. Following the energy recovery method, the top energy electrons in eRHIC are not discarded after passing the collision point. Instead, they are directed again through the linac, but this time in a decelerating phase of the electro-magnetic field. The energy extracted into the linac cavities from decelerating electrons is used again to accelerate other electron bunches.

Operation of the linac in CW mode calls for the use of superconducting RF technology (SRF). Otherwise the power dissipated in the cavity walls would become unacceptably high. A 704 MHz superconducting cavity has been developed at BNL for high-current applications [90,91,92]. The cavity, named BNL3, is shown in Figure 3-4. It has an optimized geometry that supports strong damping of higher order modes (HOMs) while maintaining good properties of the fundamental mode. The damping is accomplished via six antenna-type couplers attached to the large diameter beam pipes [93]. The simulations show that this HOM damping scheme provides sufficient suppression of the parasitic impedance to satisfy eRHIC requirements. The cavity parameters are listed in Table 3-2.

Two BNL3 niobium cavities have been fabricated to date: one by AES, Inc. and one by Niowave, Inc. Both cavities will be tested at the SRF Vertical Test Facility (VTF) at BNL and one of them will be incorpo-rated into a cryomodule under fabrication for the Coherent electron Cooling Proof-of-Principle (CeC PoP) experiment [94]. The second cavity will serve as a vehicle for further eRHIC related R&D.

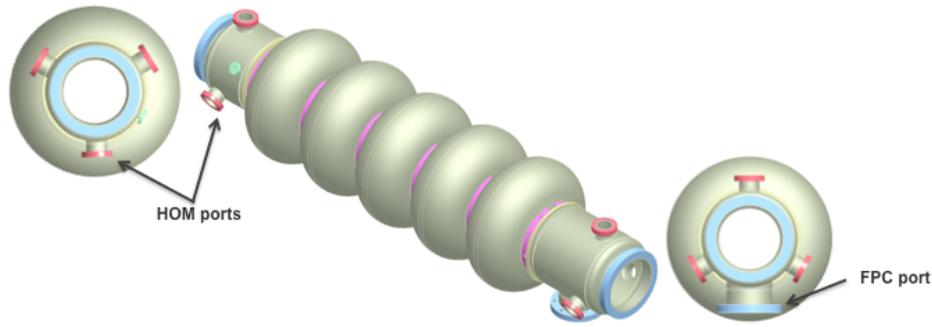

Figure 3-4: Five-cell 704-MHz SRF cavity for eRHIC.

| R/Q | 506.3 Ohm |
|---|---|
| **Geometry factor** | 283 Ohm |
| **Cell-to-cell coupling factor** | 3.02% |
| **Cavity loss factor (at $\sigma_z = 2$ mm)** | 3.96 V/pC |
| **Cavity $Q_0$** | $4 \cdot 10^{10}$ |
| **$E_{peak}/E_{acc}$** | 2.46 |
| **$B_{peak}/E_{acc}$** | 4.27 mT/(MV/m) |
| **Lorentz force detuning coefficient** | 0.45 Hz/(MV/m)$^2$ |

Table 3-2: Parameters of the BNL3 cavity.

### 3.2.3 Coherent Electron Cooling

Small transverse and longitudinal beam emittances of the hadron beam in eRHIC are of critical importance, both for the attainment of high luminosity as well as for separating the products scattered at small angles from the core of the hadron beam required for a number of golden experiments. Specifically, eRHIC requires a 10-fold reduction in transverse and longitudinal emittance of the hadron beams, i.e. about a 1,000-fold increase in brightness, compared with beams currently operating in RHIC. Without such emittance reduction, the eRHIC luminosity would be reduced about 50-fold. There is no established cooling technique capable of this task. The stochastic cooling currently used at RHIC [95] falls a factor of about 100-1,000 short for cooling ion beams to the required density and by a factor of ~$10^4$ short for proton beam cooling. A detailed study of traditional electron cooling of RHIC beams [96] showed that its cooling time will also be insufficient for eRHIC hadron beams by similar factors as above.

There are three advanced, but untested cooling methods: an optical stochastic cooling (OSC) [97], coherent electron cooling (CeC) [98] and recently suggested micro-bunching electron cooling (MBEC) [99], which in principle can satisfy the eRHIC's cooling requirements. Unfortunately OSC is incompatible with eRHIC's need to change the hadron beam energy 5-fold – it would require a 25-fold change of the undulator period in OSC. The two remaining techniques are versions of coherent electron cooling, with CeC theory developed in-depth and MBEC being a new and developing concept. Hence, we present here a CeC cooler as the main approach capable of cooling hadron beams in eRHIC to the designed emittances.

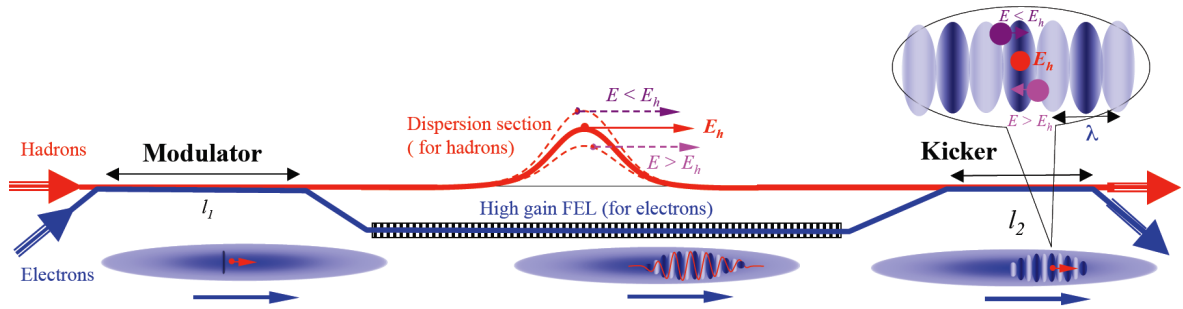

Figure 3-5: A general schematic of the classical Coherent Electron Cooler comprising three sections: A modulator, an FEL plus a dispersion section, and a kicker. For clarity, the size of the FEL wavelength, λ, is exaggerated grossly.

The CeC scheme, shown in Figure 3-5, is based on electrostatic interactions between electrons and hadrons that are amplified ether in a high-gain FEL or by other means. The CeC mechanism bears some similarities to stochastic cooling, but with an enormous bandwidth of the amplifier. In CeC, the electron and hadron beams have the same velocity and co-propagate, in a vacuum, along a straight line in the modulator and the kicker; this is achieved by selecting the energy of electrons such that the relativistic factors of the two beams are identical. CeC works as follows: in the modulator, each hadron induces density modulation in the electron beam, which is amplified in the high-gain FEL; in the kicker, the hadrons interact with the beam's self-induced electric field and experience energy kicks toward their central energy. The process reduces the hadrons' energy spread, i.e. it cools the hadron beam. By coupling the longitudinal and transverse degrees of freedom, the cooling can be shared and the hadron beam cooled in three dimensions: longitudinally, horizontally and vertically.

| **Hadron beam** | | | |
|---|---|---|---|
| **Species** | p | Beam energy, GeV | 250 |
| **Particles per bunch** | $3 \times 10^{10} - 2 \times 10^{11}$ | $\varepsilon_n$, mm mrad | 0.2 |
| **Energy spread** | $10^{-4}$ | RMS bunch length, nsec | 0.27 |
| **Electron beam** | | | |
| **Beam energy, MeV** | 136.2 | Peak current, A | 50 |
| $\varepsilon_n$, **mm mrad** | 1 | RMS bunch length, nsec | 0.27 |
| **CeC** | | | |
| **Modulator length, m** | 10 | Kicker length, m | 10 |
| **FEL wiggler length, m** | 9 | $\lambda_w$, cm | 3 |
| $\lambda_o$, **nm** | 422 | $a_w$ | 1 |
| **g, FEL gain used/max** | 3/44 | CeC bandwidth, Hz | $1.1 \times 10^{13}$ |
| **Cooling time, hours** | 0.12 | | |

Table 3-3: CeC parameters for cooling a 250 GeV proton beam in eRHIC.

With the eRHIC hadron beam parameters the emittance growth time caused by intra-beam-scattering (IBS growth time) is measured not in hours (as in current RHIC) but in seconds. Hence, the cooling should operate at collision energy (e.g. from 40 GeV/u to 250 GeV). Our analytical estimates show that hadron beams (both proton and ion) could be cooled to the required emittances and kept there using the CeC with the parameters listed in Table 3-3.

CeC theory has matured in the last 5-6 years and included all major effects in the modulator, kicker and FEL (including saturation). CeC simulations have also advanced to the stage where we can compute hadron screening and cooling by an inhomogeneous electron beam, including propagating through a modulator or a kicker with quadrupole focusing. A very detailed discussion of this progress as well as numerous references to publications about CeC can be found in [100].

Still, CeC is a new, untested cooling technology and we are proceeding towards its demonstration in a Proof-of-Principle experiment, which is described in detail in Sec. 3.4.

### 3.2.4 Choice of FFAG Using Permanent Magnets for eRHIC

The revival of FFAG accelerators, developed in the 1950s, is very evident today. The concept allows a very large energy acceptance using fixed magnetic fields. The *scaling FFAG* lattice, independently found by four different groups at the time, has exceptional properties: fixed tunes and zero chromaticity at all energies and an infinite energy range. Unfortunately, a weak point of the scaling FFAG is inefficiency in bending, as at least 30% of the bending has to be in the opposite direction, consequently leading to a large overall ring size.

The scaling FFAG has large orbit offsets and requires large aperture magnets to accommodate this. The orbits of different energies have the same shape as each other and differ only in scale, with the highest energy normally having the largest orbit. We have also seriously considered the use of a scaling *Vertical FFAG* (VFFAG) [101] for eRHIC due to a few significant advantages:

- VFFAG orbits are all the same shape and size but stack vertically with energy.

- The whole energy range from injection to the maximum energy can be covered with a single VFFAG.

- The VFFAG is by definition isochronous as the electron beam is very relativistic.

The VFFAG lattice for eRHIC was studied and a preliminary solution for the magnet design with the required exponential magnetic field in the vertical plane was found [102]. There were two main reasons for abandoning further efforts on the VFFAG: one was the large synchrotron radiation due to the required 30% of opposite bending, and the other was a very stringent alignment requirement due to non-linear magnetic fields.

The non-scaling FFAG (NS-FFAG) is a relatively new concept developed during the Muon Collider or Neutrino Factory studies in 1999. The major advantage of the NS-FFAG with respect to the scaling FFAG is a much smaller required aperture due to the very small size of the dispersion function. The "laws" of the scaling FFAG are abandoned, so the tunes change with energy as well as the chromaticity reaching very large values. The magnetic field is linear. This makes it very attractive, as the dynamical aperture is very large and it can be built with standard components. Unfortunately, due to the tune variation with energy and requirement for the tunes to be between integer and half integer resonances (0.5 – 0.0), the energy range is smaller than in the scaling FFAG. The orbits in the NS-FFAG are not parallel to each other. There are just a few options for NS-FFAG lattices with linear magnets: triplet, doublet, or FODO (Focusing-drift-Defocusing-drift) but the magnets are combined function such also contain dipoles that bend the beams. The orbits are roughly circular for the reference energy. For energies larger than this the orbits show positive curvature within the focusing combined function magnet, with an opposite bend in the defocusing magnet. For energies below the reference energy the defocusing magnet has positive curvature while the opposite bending is within the focusing magnets. The time of flight is a parabolic function of energy, as previously analytically shown by Craddock [103].

Although NS-FFAGs were originally developed for muon acceleration, there are many additional ways to use them in other areas like medical applications and non-relativistic ion acceleration. It is important to note that NS-FFAG use for eRHIC is one of the most beneficial cases because:

- The betatron tune variation is not of concern as electrons with a single-energy pass through the NS-FFAG arc only once before their energy is changed by the linac.

- The difference in time of flight is easy to correct with spreaders/combiners that separate the different energies, so RF phase is not affected by the NS-FFAG properties.

- The alignment tolerances are very reasonable to achieve.

- The synchrotron radiation power with the 10 mA electron beam can be kept less than 3 MW in the arcs or energies up 15.9 GeV.

- Orbit and gradient corrections are obtained by small copper correction coils between the iron and the vacuum pipe.

In the eRHIC design, using two NS-FFAG rings allows acceleration of the electrons up to 21.2 GeV. The first, low energy FFAG ring, transports electrons in the energy range 1.3-6.6 GeV. The second, high energy FFAG ring, transports electron beams with energies from 7.9 GeV to 21.2 GeV. The lattice of the FFAG rings is presented in Sec.3.3.3.

## *Permanent Magnets*

It is possible to use permanent magnets in the FFAG recirculation passes, since the field does not vary with time. This simplifies operation and reduces expense on large copper coils, power supplies and water-cooling.

This permanent magnet design considers Samarium-Cobalt (SmCo), although other materials like Neodymium-Iron-Boron ($Nd_2Fe_{14}B$) are also possible. To obtain the best solution, a few important characteristics should be considered:

- Magnetic performance – this is especially important for the high-energy ring where large gradient magnets are required.

- Corrosion resistance – it is well known that the RHIC tunnel is not very well insulated from the outside environment and especially during the summer, humid conditions are possible.

- Thermal Stability – this was one of the major concerns for the Fermilab permanent magnet recycler ring. They developed a very sophisticated temperature compensation system by using materials of opposite temperature dependence in their magnets.

- Radiation Resistance – the RHIC tunnel is a high radiation environment. Experience at RHIC has shown that electronic instruments in the tunnel do not have long survival times. Studies by NASA of different magnetic materials have shown that SmCo has superior radiation resistance to every other available material, as shown in Figure 3-6.

- Magnetization direction

- Manufacturability

- Cost

Long-term stability of different SmCo products shows that it is possible to obtain very stable magnetic material during many years of operation. The permanent magnet design for eRHIC FFAG magnets is presented in Sec. 3.3.4

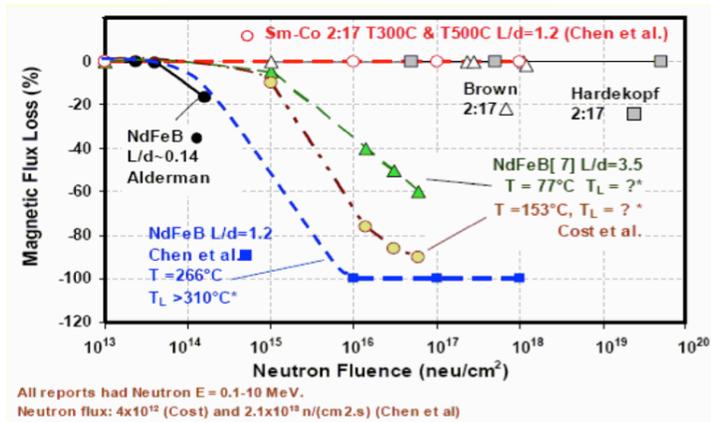

Figure 3-6: Magnetic flux loss due to neutron radiation – results obtained from NASA studies using the Ohio University Research reactor.

### 3.2.5 Space Charge Compensation

The electro-magnetic interactions among charged particles, the so-called space charge forces, play a significant role in modern accelerators. Although the space-charge force does not affect the frequency of the coherent dipolar motion of charged particle bunches as they circulate around the accelerator, it shifts higher order coherent motion frequencies, and may adversely affect the beam's stability. More importantly, this force usually is nonlinear, so introducing an additional tune spread to the circulating particles, and thereby increasing the beam losses due to the machine's non-linear resonances. The space-charge force falls quadratically with the beam's energy, and thus other nonlinear effects, such as beam-beam interactions, usually dominate high-energy colliders. However, future electron-ion colliders, such as eRHIC, are designed to operate with a range of energies. To avoid a significant reduction of the beam's lifetime at lower hadron energies, the bunch intensity must be reduced for low-energy operations.

It would be rewarding to reduce the effects of space charge without sacrificing the bunch's intensity; thus accelerator scientists are motivated to develop novel techniques for compensating for space charge. Techniques based on nonlinear compensating magnets, or the applications of neutralizing charge in an electron column (or electron lenses) have been investigated. However, these approaches face the common difficulty of over-compensation when applied to a bunched beam. In a charged-particle bunch, the space-charge force varies along the bunch, and consequently, without matching the compensation strength with the bunch's longitudinal profile, proper compensation to the bunch's center causes overcompensation at its tail. Recently, a scheme based on a bunched electron beam was proposed to compensate for space-charge effects for positively charged ion-bunches [104].

In this scheme, the electron bunches are launched in the same direction as the ion beam, while mismatching the energy of the compensating electron bunches from that of the circulating ion bunches. This approach significantly lowers the electron beam's current required for space-charge compensation compared to that of the ion beam. In addition, for a given energy of the electron beam, the longitudinal profile of the electron bunches is tailored specifically so that space-charge compensation is optimized for the entire ion bunch.

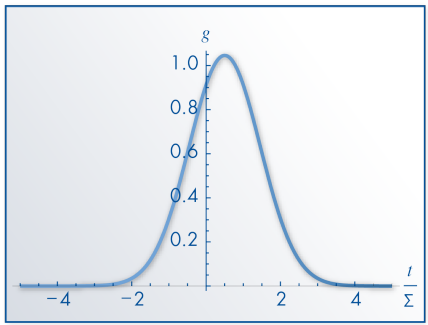

Figure 3-7: Longitudinal profile of the electron bunch required for compensating a positively charged ion bunch with a Gaussian distribution. The abscissa is the longitudinal location along the bunch in unit of R.M.S. bunch length, and the ordinate is the normalized electron instantaneous current.

## 3.3 Machine and IR designs

### 3.3.1 Electron Injector and Dump

*12 MeV Injector*

The electron injector has to produce up to 50 mA polarized electron beam with longitudinal and transverse beam parameters defined in Table 3-1.

Figure 3-8 presents a layout of the 12 MeV electron injector. It consists of an electron gun, energy spread modification cavities, including a bunching cavity and a 3rd harmonic cavity, a drift space for ballistic bunch compression and a booster linac. Long bunches (1.5 ns rms bunch duration) are extracted from the gun to reduce the beam quality degradation caused by space charge effects. A ballistic compression technique is applied to shorten the bunch duration to 13 ps.

A major component of the injector is the polarized electron source (see Sec 3.2.1). The eRHIC Gatling gun is a 20-cathode gun where the bunches originating from different cathodes are merged into one sequence using a magnetic combiner. With a 2.5 mA average current extracted from each cathode, the total average current at the gun exit reaches 50 mA.

The initial long bunch necessitates the use of a low frequency cavity for energy-spread modification. A 84 MHz superconducting RF (SRF) cavity operating at an accelerating voltage of 1.3 MV introduces an energy spread along the bunch, which results in ballistic compression as non-relativistic electrons travel through the drift space. A third harmonic (253 MHz) SRF cavity is used to fine-tune the longitudinal phase space modification. This cavity operates at an accelerating voltage of 0.6 MV. Both energy spread modification cavities are of the quarter wave resonator type.

The booster linac employs a 422 MHz SRF cavity to accelerate the beam to 12 MeV. This cavity is similar to the main linac cavities but has to deliver an RF power of 600 kW. The beam from the booster linac is then transported and injected into the main ERL.

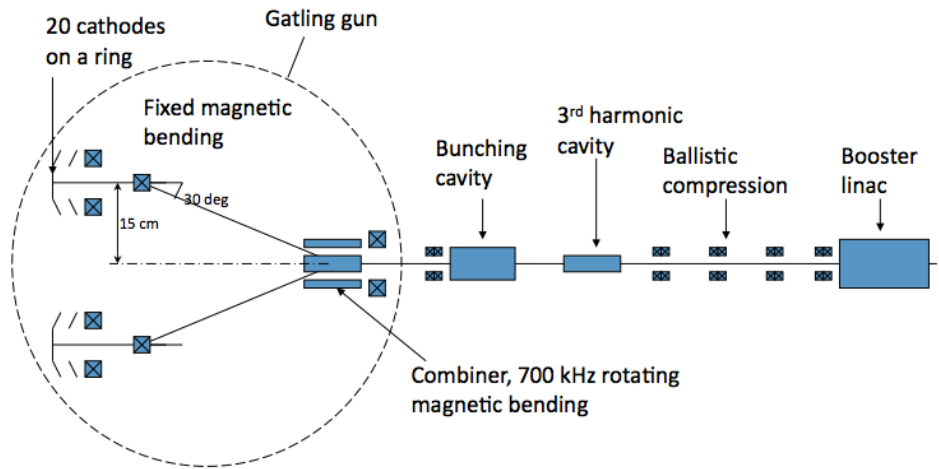

Figure 3-8: 12 MeV injector components.

The IP2 area of the RHIC tunnel is large enough to place, side-by-side, two 12 MeV injectors, if cost considerations allow it. Such an arrangement would minimize the loss of average luminosity caused by limited cathode lifetime and the necessity to replace the cathodes. An expected cathode lifetime at the design current is about 85 h.

*Beam Dump*

A dump beamline transports the decelerated 12 MeV beam from the main ERL to the beam dump. The beamline consists of a dipole magnet, which is a part of the spreader, and two rastering quadrupoles, which disperse the beam over the beam dump surface. The aperture of the dump beamline is large enough to transport the decelerated beam with an energy spread of 2 to 3 MeV.

The beam dump has to be able to absorb a 600 kW heat load from the 12 MeV electron beam. The beam dump of the Cornell ERL Injector has been taken as the basis for the eRHIC dump because of the similarity of the beam parameters [105]. It is made of aluminum instead of copper to reduce neutron production. The dump consists of two sections: the body and an outer shell, containing the cooling water. The interior shape is designed to distribute the scattered electrons as uniformly as possible around the cooled surface.

### 3.3.2 SRF Energy Recovery Linac

An FFAG-based Energy Recovery Linac (ERL) will accelerate an electron beam to 15.9 GeV after 12 passes through a Superconducting RF (SRF) linac or to 21.2 GeV after 16 passes. In both cases the linac energy is 1.322 GeV.

*SRF Systems*

The superconducting RF ERL concept allows recovery of the beam power spent for acceleration of particles by recirculating them after collisions back through the linac at an RF phase offset by 180 degrees with respect to the accelerating phase. Thus the ERL's RF systems will have to provide the power necessary to maintain stable amplitude and phase of the electromagnetic field inside the SRF cavities and to compensate for any parasitic energy losses incurred by the beam (due to synchrotron radiation, resistive wall and higher order modes). The maximum amount of parasitic beam power loss is set to 3 MW, which in turn limits the beam current at 21.2 GeV to 3.6 mA. The linac will be installed in the 200-meter long IP2 straight section of the RHIC tunnel. Parameters of the main SRF linac are listed in Table 3-4.

| **Energy gain** | 1.32 GeV |
|---|---|
| **Bunch length** | 4 mm rms |
| **Bunch repetition frequency** | 9.38 MHz |
| **Number of RF buckets per RHIC revolution** | 120 |
| **Number of RF buckets filled** | 111 |
| **RF frequency** | 422.3 MHz |
| **Number of SRF cavities** | 42 |
| **Linac fill factor** | 0.60 |
| **Cavity type** | elliptical, 5-cell |
| **Accelerating gradient** | 18.4 MV/m |
| **Operating temperature** | 1.9 K |
| **Cavity intrinsic $Q$ factor at operating gradient** | $5 \cdot 10^{10}$ |
| **Peak resonant frequency detuning due to microphonic noise** | 6 Hz |
| **$Q_{ext}$ of FPC** | $3.5 \cdot 10^7$ |
| **Peak RF power per cavity** | 30 kW |
| **Total heat load at 1.9 K** | 2 kW |
| **Maximum HOM power per cavity** | 7.8 kW |

Table 3-4: Parameters of the main SRF linac.

The beam power loss will be compensated by a separate linac operating at 844 MHz, second harmonic of the main RF frequency. The energy loss compensation cavities are installed in the middle of the main linac. Energy spread caused by the curvature of the RF waveform causes spin depolarization of the electrons. To minimize this effect, an energy spread compensation linac is required. This SRF linac will operate at fifth harmonic of the main linac frequency, 2.1 GHz.

The main SRF linac will utilize five-cell 422 MHz cavities, scaled versions of the 704 MHz BNL3 cavity developed for high current linac applications [90,91,92]. Each cavity will be housed in an individual cryounit, a series of which will form one long cryomodule [106]. The cavities will operate at an accelerating gradient of 18.4 MV/m. The cavities will be powered from individual high power RF amplifiers. At 6 Hz peak detuning due to microphonic noise, the required peak RF power will be approximately 30 kW per cavity. High beam current and multi-turn operation of the ERL imposes stringent requirements on damping of higher order modes (HOMs) in the cavities to avoid beam breakup (BBU) instabilities. The HOM power is reaching 7.8 kW per cavity at a beam current of 50 mA and 8 ERL passes.

One of the primary design choices for an SRF linac is its frequency. There are several considerations affecting this choice for a high current multipass ERL: bunch structure, bunch length, energy spread, beam breakup instability threshold, SRF losses, RF power efficiency, cost and complexity considerations. In the eRHIC case, most of these considerations point toward lower frequency [107]. Here we briefly consider some of the effects:

- A continuous train of electron bunches can lead to accumulation of ions and the fast ion instability. To clear the accumulated ions a gap of about 0.95 µs (of the same duration as an abort gap for RHIC beams) is introduced. Such a gap in the electron beam induces a transient voltage on the ERL cavities. As the transient voltage fraction is proportional to the frequency squared, there is a clear advantage of using lower frequency cavities.

- Lower frequency allows us to proportionally increase the bunch length (the limitation on the bunch length is depolarization due to RF wave curvature). This, in turn, reduces various wake field effects. For example, the linac loss factor is approximately proportional to the cavity frequency squared.

- Transverse beam break up (BBU) is the dominant effect limiting the beam current in ERLs. The instability threshold current is inversely proportional to the frequencies of higher order modes. Also, the number of HOMs is reduced, as fewer low-frequency cavities are required to build the linac. This consideration is of special importance for a multi-pass ERL.

- During the last few years, new advances in the SRF technology demonstrated the possibility of preparing cavities with very low, of the order of one nanoOhm, residual resistivity. This is especially beneficial at low frequencies, where RF losses in the SRF cavities are dominated by residual losses as the BCS component becomes very small.

Additional considerations include better RF power efficiency at lower frequencies, possible lower sensitivity to environmental (microphonic) noise and some cost advantages related to reduced complexity of the system and smaller number of components in the linac. As at very low frequencies (~300 MHz and below) the size of elliptical cavities becomes inconveniently large, we have chosen a frequency of 422 MHz for eRHIC, which is the $45^{th}$ harmonic of the 9.38 MHz RHIC bunch repetition frequency. This RF frequency also would allow to operate at 14.1 MHz bunch repetition rate providing an opportunity for possible future luminosity upgrade.

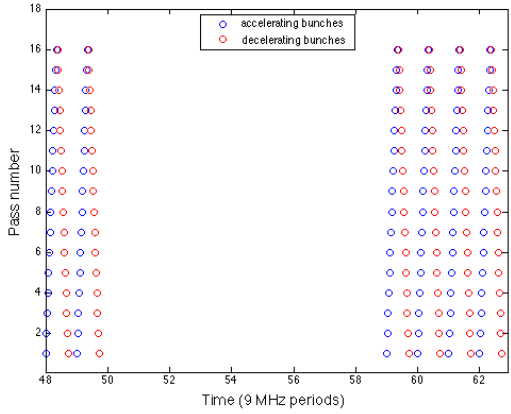

Figure 3-9: Bunch pattern in the eRHIC FFAG ERL with 16 passes, showing the ion clearing gap.

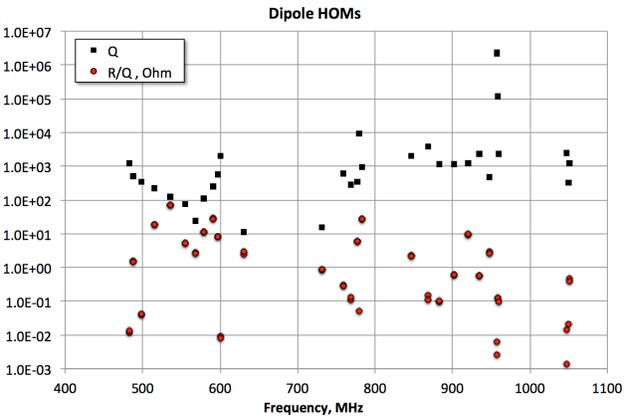

Figure 3-10: Quality factors and *R/Q*'s of the dipole HOMs.

In order to avoid bunches from different ERL passes piling on top of each other in the linac and to avoid uneven voltage transients, the circumference of the FFAG should be chosen appropriately. For eRHIC we have chosen it to be one RF wavelength longer than the circumference of RHIC. In this case all bunches travel in groups of *N* (*N* is the number of ERL passes) accelerating bunches separated by one RF period, followed by *N* decelerating bunches with the same bunch separation. This bunch pattern, shown in Figure 3-9, will produce a regular voltage transient and will ensure that the accelerating and decelerating bunches in the same pass have the same energy. At 422 MHz the transient is small, of the order of $10^{-3}$ in relative magnitude.

As we mentioned above, the dominant effect, which which limits the beam current in ERLs, is the transverse BBU. To evaluate this effect for eRHIC, we have scaled the previously developed BNL3 cavity shape [90,91,92] to 422 MHz and calculated parameters of several lowest dipole HOM pass bands. The results are shown in Figure 3-10. This data set was used for BBU simulations, which predicted a threshold beam current higher than 50 mA even with no HOM frequency spread (see

Table 3-8).

### *Linac Optics*

The goal of the linac optics is to minimize the beta function in the linac for all passes. In the eRHIC design, it was preferred to exclude quadrupoles from the linac to minimize the total length of the linac and leave more space for the spreader-combiner sections. When quadrupoles are excluded, the only free parameters are the initial optical functions at injection energy of the lowest energy pass. The optical functions of consecutive passes are connected by this rule:

$$\beta_n(s = L) = \beta_{n+1}(s = 0); \alpha_n(s = L) = -\alpha_{n+1}(s = 0)$$

After optimization of the initial optical functions, the beta function of the linac through 16 accelerating passes is shown in Figure 3-11, and the optics of the decelerating passes are the mirror image of the same figure.

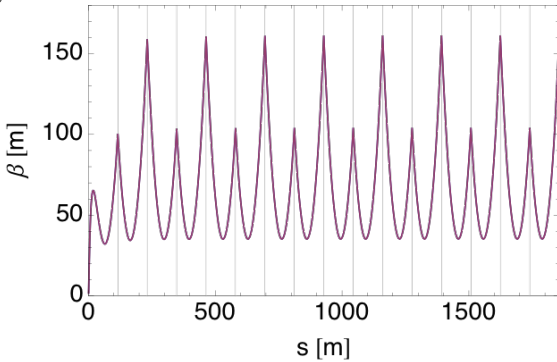

Figure 3-11: The beta function in the linac for 16 passes. The horizontal and vertical optics are identical. The grid lines separate the optics of each pass.

## 3.3.3 FFAG Lattice

We have selected a doublet lattice for both the low energy range (1.344-5.3 GeV) and higher energy range (6.622-21.164 GeV) with identical length magnets placed above each other. The NS-FFAG for eRHIC is made of:
- Identical cells in the arcs that follow the shape of the RHIC tunnel arcs.
- Spreaders/combiners matching the amplitude and dispersion functions of each energy in the arcs to the 1.322 GeV linac.
- Straight sections where the NS-FFAG cells are without any bending, with matching sections between these and the arcs.
- Two overpasses of the NS-FFAG cells for all electron beams with lower than colliding energy to avoid the detectors.
- Extraction beam line with magnetic septums.

Lattice optimization with respect to synchrotron radiation loss was accomplished by choosing the magnet gradients and transverse offsets to produce the least amount of radiation. Resulting parameters of basic cell elements for both FFAG beamlines are specified in
Table 3-5. The orbits in both low and high-energy rings, magnified 100x, are shown in Figure 3-12.

| Element    | Length (m) | Gradient (T/m) | Offset (mm) |
|------------|------------|----------------|-------------|
| All Drifts | 0.3        |                |             |
| BD (Low)   | 1.300      | 3.5            | -13.25      |
| QF (Low)   | 1.437      | -3.5           | 13.25       |
| BD (High)  | 1.300      | 17.0           | -8.09       |
| QF (High)  | 1.437      | -17.0          | 8.09        |

Table 3-5: Lattice specification for both low- and high-energy eRHIC FFAG rings. The lattice in both cases is a doublet BD-drift-QF-drift. The (rectangular) quadrupoles are placed such that the orientation of the quadrupole is parallel to the midpoint of the alignment arc and the magnetic center-line is on average offset in X from the alignment arc by the value shown.

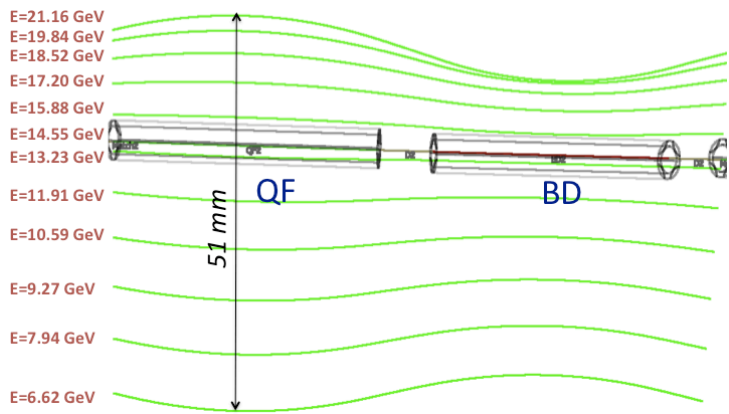

Figure 3-12: Orbits in large energy FFAG arc cell.

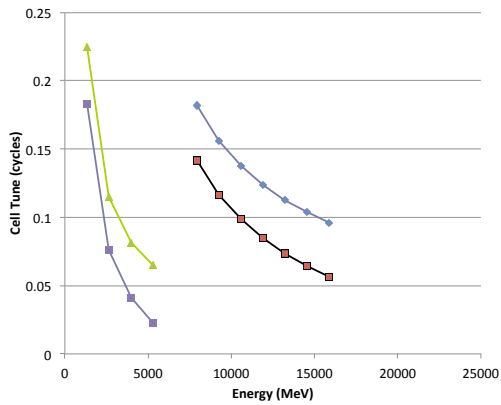
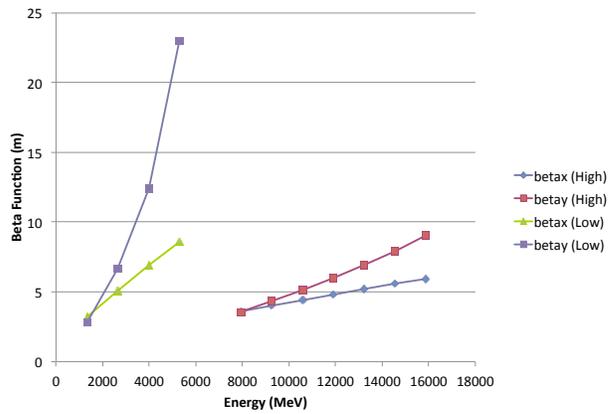

Figure 3-13: Tune (left plot) and betatron function (right plot) dependence on energy in the two basic low and high-energy NS-FFAG cells.

The tune and betatron function dependence on energy for the basic cell is shown in Figure 3-13. The range of stability lies between 0.5 and 0.0 as shown.

The path length dependence on energy is a parabolic function as shown for low and high energy rings in Figure 3-14. The minimum of the parabola depends on the relationship between the bending angles of the focusing and defocusing combined function magnets.

The synchrotron radiation power loss at the recirculation energies is shown in Figure 3-15. One can note that the energy dependence of synchrotron radiation losses in the optimized FFAG lattice is quite different from typical $E^4$ dependence in synchrotrons.

The straight cells of the FFAGs are just the arc cells with zero X-offset in the quadrupoles (so there is no dipole field) and zero curvature of the alignment path. This preserves the focusing structure and beta function behavior, leaving only the orbit offsets to be matched. Due to the large range of different energies in the FFAGs, this is most easily achieved adiabatically. Here, the transition cells have quadrupole offsets and angles that are both multiplied by a scaling factor u(s), which decreases smoothly from u=1 in the arc to u=0 in the straight. The transition from arc to straight section can be accomplished using 17 matching cells, or about 40 meters of beamline.

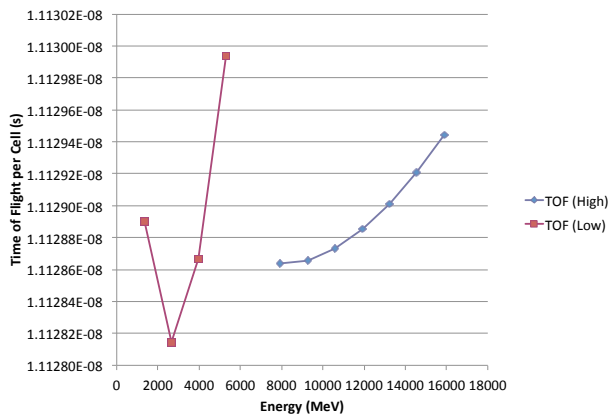

Figure 3-14: Time of flight for the arc cells of the low- and high-energy rings.

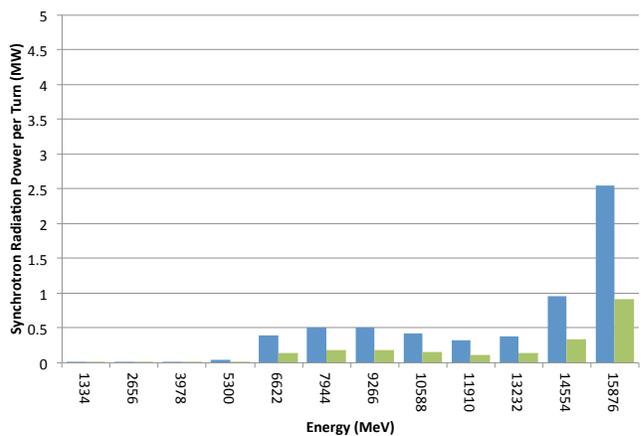

Figure 3-15: Synchrotron radiation power per turn produced in the FFAG arcs for currents of 10mA (blue) and 3.6 mA (green).

### 3.3.4 Permanent Magnet Design

The basic-cell lattice design requires fixed alternating gradient magnetic fields in the arc regions, where electron beams that have different energies will have different orbit excursions towards both sides. Due to the synchrotron radiation issues, the arc magnet design must provide free paths on both sides to accommodate special vacuum chambers for radiated energy absorption. These requirements demand that the arc magnet must be a Collins-type [108] quadrupole. Several designs of permanent magnet have been considered. These magnets can be built by using the rare-earth metal $Sm_2Co_{17}$ and low-carbon magnetic steel (1006), in order to eliminate the cost of major power supplies and cable connections in the arc region. The design magnet parameters are shown in Table 3-5.

Figure 3-16 shows one of possible magnet designs and the corresponding flux plot. The light blue area represents the permanent magnetic material blocks, whose magnetization direction is indicated by arrows. Figure 3-17 shows that a field gradient 25 T/m was reached, and that the gradient error is less than 1E-3.

Compared with electromagnets, the usual permanent magnets have disadvantages: (a) they have temperature effects because the remnant induction ($B_r$) is a function of temperature; (b) the magnetic fields (and gradients) will not be adjustable while the tuning is necessary since misalignments always exist and often vary.

Our corresponding solutions are: (a) use passive temperature compensators (a binary Ni-Fe alloy) as Fermilab implemented in their antiproton Recycler [109], to stabilize the magnetic fields and gradients during ambient temperature drifts; (b) design correction coils, mounted in space between the magnet and the vacuum chambers, powered by small current supplies, to adjust the orbit positions and field gradients. For every magnet we plan to provide a steering correction coil (either vertical or horizontal) and a correction coil for field gradient adjustments. These correction coils are current-dominated and their field quality is well preserved. Since the horizontal dipole field correction coil would require watercooling one might consider using a separate corrector magnet instead of it.

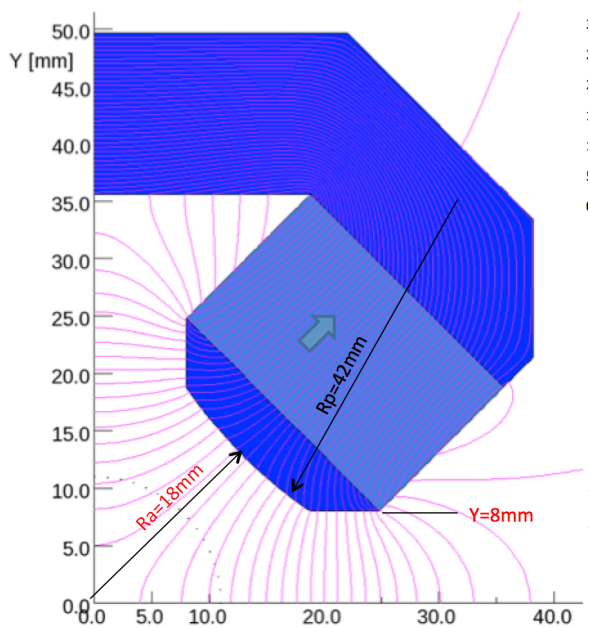

Figure 3-16: A quarter of magnet cross-section and magnetic flux plot. Light blue area is permanent magnet material

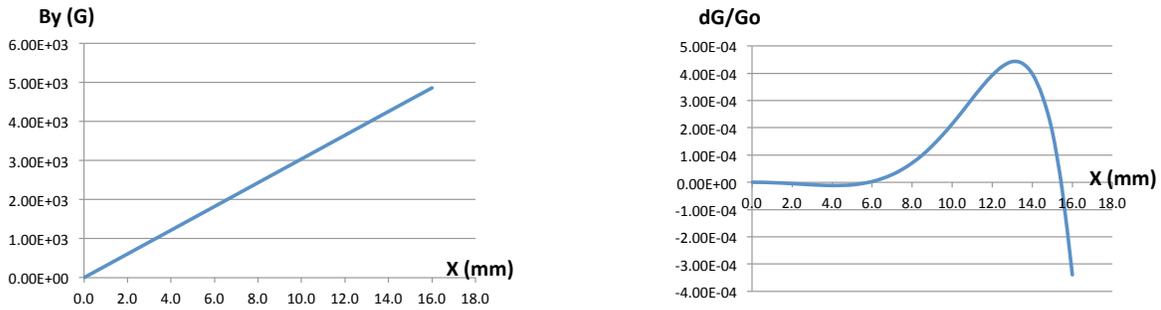

Figure 3-17: Field plot By vs. X and gradient variation plot vs. X.

### 3.3.5 Splitter and Combiner

The main function of the Splitter is to transport and optically match the beam bunches from the exit of the ERL to the entrance of the FFAG arc. The function of the Combiner is the same as that of the Splitter but in the opposite order (from the FFAG arc exit to the ERL entrance). Since the Splitter/Combiner places the beam bunches into separate beam lines it can also be used for beam diagnostics and for correction of important lattice parameters, like the betatron phase advances, the isochronicity ($R_{56}$) and the path length of electron trajectories. The acceleration of electrons to 21.2 GeV requires 16 passes of electron beam through the ERL. Therefore each Splitter/Combiner consists of 16 separate lines. (One additional beam injection line is also required for injecting the 10 MeV electrons into the ERL). The Splitter/Combiner layout has been designed to minimize the path-length difference introduced by its beam lines. The layout is shown in . In eRHIC there are two FFAG rings that accommodate the beam bunches of different energies. The path-length differences between the lines corresponding to the highest and lowest energies of each FFAG ring are shown in

Table 3-6.

| e-RHIC | HE-Ring [cm] | LE-Ring [cm] |
|---|---|---|
| 21 GeV | 12.1 | 24.6 |
| 15 GeV | 7.1 | 14.4 |

Table 3-6: The path-length difference between the high and low energy bunches of the High and Low energy FFAG rings for the 15 GeV and 21 GeV eRHIC operation mode.

The main constrain of beam optics is the matching of the beam parameters at the exit/entrance of the arcs. Figure 3-19 is an example of the beta and dispersion functions for the 15.9 GeV line. The blue filled rectangles are the main dipoles for the layout of the line and the unfilled rectangles are the quadrupoles.
The layout of the Splitter/Combiner introduces a path length increase, which is larger for the low energy lines. This path difference is minimized (see

Table 3-6) by properly laying out the lines. To compensate for the remaining path-length differences horizontal and vertical chicanes are introduced in the high-energy lines. In Figure 3-19 the red and green rectangles are dipoles forming chicanes for path compensation.

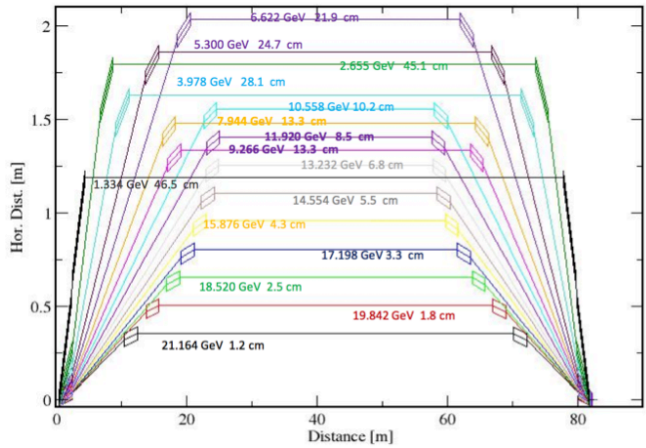
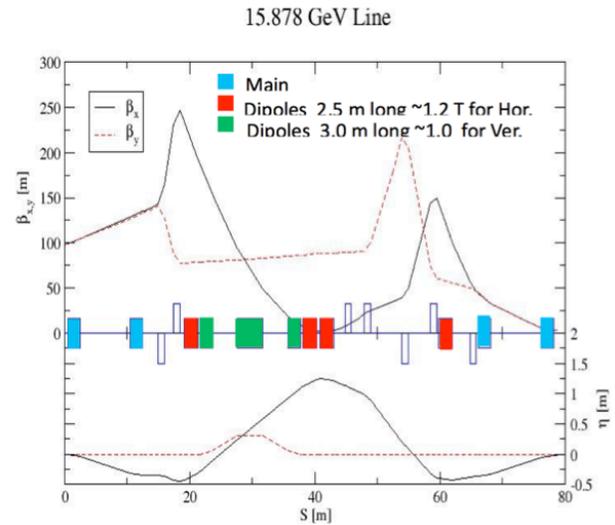

Figure 3-18: Layout of the Splitter/Combiner. The energies and the path-increase of each line are shown.

Figure 3-19: Horizontal, vertical beta and dispersion functions for the 15.9 GeV line. The Horizontal (red rectangles) and vertical chicanes (green rectangles) add 12 cm to the path of the beam.

### 3.3.6 Beam Orbit Measurement and Correction

The filling scheme, i.e. bunch fill pattern, has a gap in which a single bunch, well separated temporally from other bunches, will be placed and which will be used for diagnostic purposes. The beam's transverse positions will be measured using large-aperture button-type beam position monitors (BPMs) with conventional signal processing. Simulations with design beam parameters (~3·10$^{10}$ electrons/bunch, 4 mm rms bunch length) and a storage-ring style BPM design with 22 mm vertical aperture, four buttons of 6 mm diameter and 12 mm separation between each pair of buttons show ample signal response (~100 Volts, peak-to-peak) suitable for subsequent sampling of the signal using conventional signal processing (such as the Libera Brilliance Single Pass processor from Instrumentation Technologies). The nonlinear response for far off-axis beams was also mapped over a span of +/-15 mm range and found suitable for reconstruction of absolute beam positions. The pickup electrode geometry will be further optimized for precision trajectory measurements based on the total span of the orbits in each FFAG.

In addition to beam position monitors located in the FFAG proper (with 1 BPM every 2 cells), sets of BPMs will be located in the spreaders and combiners, the straight sections, the detector bypass lines and in the energy recovery linac.

Bunch-by-bunch measurements using fast time-based gating of emitted synchrotron radiation in the FFAG cells and alternative BPM designs are also being developed for intra-train beam position monitoring.

The unique feature of the orbits in the eRHIC FFAG design is that multiple accelerating and decelerating bunches pass through the same magnets with different horizontal offsets for beams of various energies, except in the spreaders and combiners where the beams are in separate vacuum chambers. In the FFAG beamlines, the beams of different energies respond differently to dipole correctors due to the energy-dependent tune. As a result, correction of one orbit will not improve other orbits passing through the same lattice. Therefore, dipole errors must be locally compensated to correct multiple orbits simultaneously.

The correction algorithm is based on that for a transfer line for which one needs to solve the linear equations $\Delta Y = Y_0 - Y = R * \theta$, where $Y_0$ is the target orbit, $Y$ is the measured orbit, $R$ is the response matrix, and $\theta$ is the correction strength. This can be extended for a multi-pass correction as $(\Delta Y_1, \Delta Y_2 ... \Delta Y_m)^T = (R_1, R_2 ... R_m)^T * \theta$ ,

where *m* is the number of passes. During the commissioning stage, beam may get lost at any point of the machine. In that case, the left side of the previous equation should be the measured orbit, which is a combination of any number of complete passes and a segment of one pass, and the response on the right hand side should change accordingly as well.

Orbit correction was simulated with reasonable estimates for the random alignment errors, gradient errors, angle errors in all the magnets, initial orbit errors and random BPM measurement errors. The simulated initial orbits (relative to the corresponding design orbit) and the final orbits with correction are shown in Figure 3-20. There are totally 9 accelerating passes whose path length in the arcs is 2136.5 m each. In the simulation, the beam could be thread through the machine further with corrections applied to the existing orbit. The orbit errors at the end of every pass are corrected by the correctors in the spreaders and combiners (assuming not perfectly) so that the orbit of the next pass starts with some preset initial errors. With multiple passes, the local errors can be found and corrected better as the number of measurements increases. The final orbit rms deviations of all passes was reduced from the mm scale to ~50 μm.

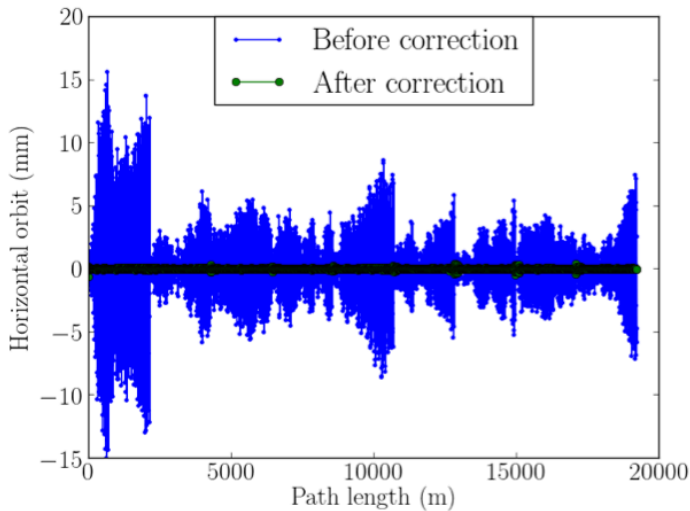

Figure 3-20: The orbits of 9 accelerating passes with various errors (blue), and the orbits after beam being thread through the accelerator and correction being applied simultaneously on all passes (green).

### 3.3.7 Electron-Hadron Frequency Synchronization

Since the hadron beams of eRHIC are not ultra-relativistic, at the fixed accelerator ring circumference the revolution frequency of the hadron beam depends noticeably on its energy. In order to have the hadron and electron repetition rates matched in the wide range of hadron energies the machine design has to incorporate a capability of varying the circumference of either hadron or electron rings. In eRHIC the hadron circumference control will be realized by radial shifts of the hadron closed orbit in hadron ring arcs. The radial orbit offsets of ±1.3 cm would provide up to 16 cm hadron circumference variation range allowing the electron-hadron synchronization in the energy range 100-250 GeV/u.

To make the synchronization at lower hadron energies the harmonic switching method is used. Switching of the ERL RF harmonic number (the ratio of the RF frequency to the revolution frequency) down by one unit allows operating with hadron energies 43-46 GeV. And when switching to even lower RF harmonics some of lower proton energies can be accessed. As shown in Figure 3-21 and Figure 3-22 switching the RF harmonic by one unit down implies that the RF frequency of the main linac cavities must be reduced by the revolution frequency of electrons (~78 kHz), while the circumference lengthening is reset to the maximum delay (16cm).

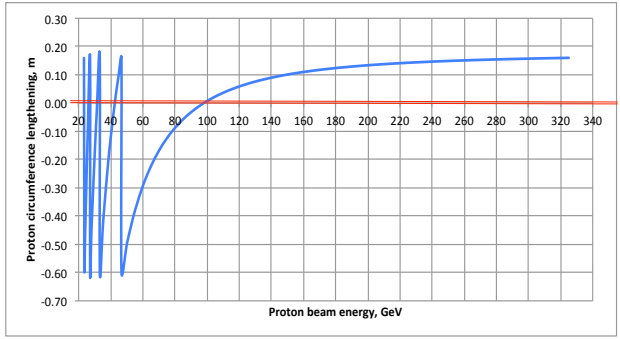
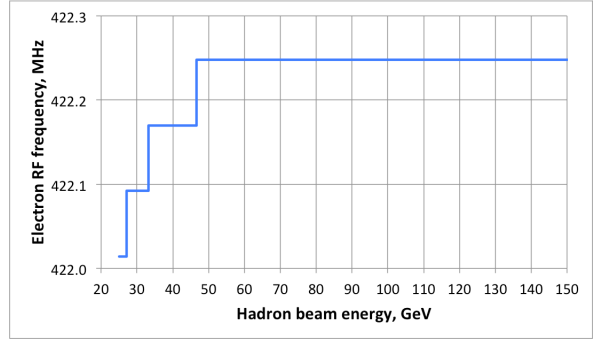

Figure 3-21: Required path-lengthening to be produced by the hadron delay line versus proton energy. Since the actual delay line can only increase the path length, compared with the original hadron circumference, only the energies corresponding to the positive path lengthening are accessible for the machine operation.

Figure 3-22: Main linac RF frequency dependence on the hadron energy.

### 3.3.8 Collective effects

Various collective effects were studied and three effects have been recognized as most important: the energy losses and energy spread due to collective effects, multi-pass beam breakup instability due to high order modes of SRF cavities, and the fast beam-ion instability.

*Energy losses and energy spread*

The following effects are investigated for potential energy losses and resulting energy spread: coherent synchrotron radiation (CSR), longitudinal resistive wall impedance, the higher order modes (HOM) of the SRF cavities, wall roughness of the beam pipe and synchrotron radiation.

|  | CSR | Machine impedances | Wall roughness | Synchrotron Radiation | Total |
|---|---|---|---|---|---|
| **Energy loss, MeV** | Suppressed | 2.4 | Negligible | 221 | 223 |
|  |  | 1.2 |  | 540 | 541 |
| **Full energy spread, MeV** | Suppressed | 3.8 | Negligible | 2.8 | ~5 |
|  |  | 2 |  | 6.7 | ~7 |
| **Power loss, MW** | Suppressed | 0.024 | Negligible | 2.2 | 2.2 |
|  |  | 0.004 |  | 2.0 | 2.0 |

Table 3-7: energy losses and energy spreads due to various collective effects with the top electron energy of 15.9 GeV (top) and 21.2 GeV (bottom).

Table 3-7 summarizes our estimations for the current design. As shown in the table, we expect that the energy loss due to CSR will be fully suppressed by the shielding effects of the vacuum chamber. Furthermore, the wall roughness of the extruded aluminum vacuum chamber can be reduced to sub-micron level[1] and its contribution to the energy spread is estimated to be negligible compared with other effects. The total

---

[1] We measured 0.2 μm rms surface height variation from a sample aluminum beam pipe provided by ANL.

power loss is about 2.4 MW, which will be compensated by a dedicated system with second harmonic RF cavities. The full energy spread of the electron beam at its last pass through the linac is comparable or larger than its final energy going to the beam dump. The possible techniques to reduce this energy spread are under exploration.

### *Multi-pass beam breakup*

Multi-pass beam breakup (BBU) is the major limiting factor of the average current in ERL [110]. The BBU threshold for eRHIC is calculated by using the BBU code GBBU [111]. The higher order mode frequencies and the corresponding R/Q can be found in Figure 3-10. In the simulation, the HOM frequency spread is considered from no spread to 1% rms frequency spread. For non-zero frequency spread, 50 random seeds are used to get reasonable statistics. With $10^{-3}$ rms errors, the threshold is 137±14 mA. This is well beyond the planned current for 21.2 GeV (3.6 mA) and even the maximum planned current (10 mA) at 9.3 GeV. For 9.3 GeV case, the pass number reduces from 16 to 7. The precise calculation for this case is needed in future. However, it is expected that the pass number reduction will yield even higher threshold.

| $\Delta f/f$ (rms) | Current Threshold (mA) | Standard Error (mA) |
|---|---|---|
| 0 | 53 | N/A |
| 5e-4 | 95 | 7 |
| 1e-3 | 137 | 14 |
| 3e-2 | 225 | 22 |
| 1e-2 | 329 | 37 |

Table 3-8: Current threshold of beam breakup of 21 GeV 16-pass ERL.

### *Fast beam-ion instability*

The fast beam-ion instability is caused by electron beams resonantly interacting with ions generated from ionizing the residual gas molecules. The instability is most pronounced when the ions are trapped in the beam passage by the periodic focusing force provided by the beam. In our current analysis, the ion is assumed to be $CO^+$ with 1 nTorr pressure.

Depending on which pass the electrons are traveling on, the exponential growth rate as estimated from linear theory[2] is 5~10 μs for the 15.9 GeV top energy operation and 10~20 μs for the 21.2 GeV top energy operation. The exponential growth is expected to saturate when the transverse oscillation amplitude of the ion centroids is comparable to the electron beam size. Since the ion oscillation amplitude is ~100 times larger than that of the electrons, the exponential growth of the electron coherent oscillation amplitude is expected to saturate at ~1% of the rms electron beam size.

A weak-strong code has been used to simulate the fast beam-ion instability in the two FFAG rings, which takes into account the non-linear space charge forces of the electron bunches and simultaneously simulates electron bunches from all energy passes. The simulation agrees with the theoretical estimation in the linear space charge limits and, in the absence of a gap between bunch trains, shows significantly slower but persistent growth with the non-linear space charge force being adopted. However, no growth of the coherent electron oscillation is observed from the simulation with a 560 ns gap introduced between two adjacent bunch trains.

---

[2] Theoretical estimates assume that ions generated by electrons with certain energy are trapped within their passage and hence do not interact with electrons with different energies as the trajectories from different energy passes are horizontally separated. However, the numerical simulation does not make this assumption.

## 3.3.9 Beam-Beam Effects

Beam-beam effects present one of the major restrictions in achieving higher luminosities. eRHIC adopts the linac-ring scheme to remove the beam-beam effect limit of the electron beam and aims for higher luminosity than a traditional ring-ring scheme. There are several challenging effects in the linac-ring scheme, including the electron disruption effect, the pinch effect, the ion-beam kink instability and the ion beam heating due to electron beam noise.

Electron disruption effect rises due to the large beam-beam parameter of the electron beam proposed in eRHIC. The strong nonlinear beam interaction field will distort the electron beam distribution and the large linear tune shift leads to significant mismatch between the designed optics and the electron beam distribution. The effect was studied in detail in [112]. Figure 3-23 shows the beam distribution after the collision and the electron beam size and emittance evolution in the opposing ion beam. The emittance growth and beam size blow-up due to the electron beam disruption effect are in acceptable range and will not affect the beam transport and energy recovery process in the beam decelerating stage.

The pinch effect describes the electron beam size shrinking in the interaction region due to the focusing beam-beam force, as shown in Figure 3-23. This effect will naturally boost the luminosity. For the design parameters, the pinch effect will boost the luminosity by factor 1.33. However, this effect also enhances the local beam-beam force to the opposing ion beam, which needs careful dynamics aperture study (Figure 3-29).

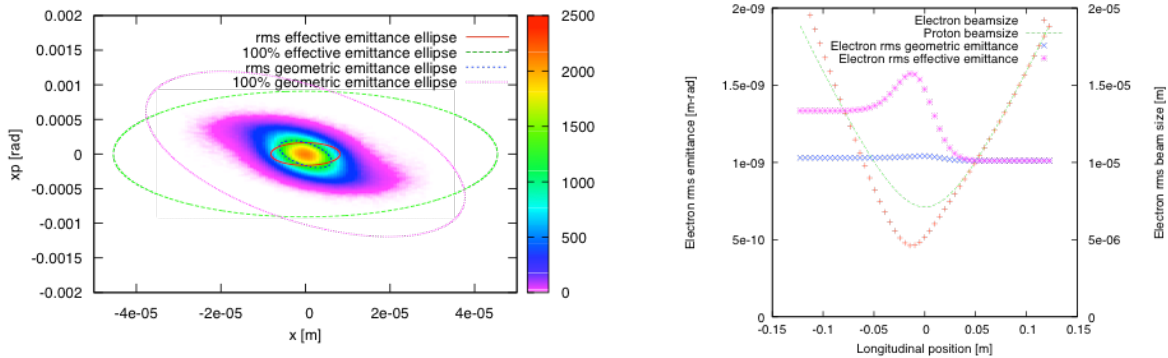

Figure 3-23: Left, electron beam distribution after the collision in transverse phase space ($x$-$p_x$); right, the electron beam parameter evolution in the opposing ion beam, e-beam travels from right to left.

For the ion beam, the largest challenge is the kink instability. The instability arises due to the effective wake field of the beam-beam interaction with the electron beam. The electron beam is affected by the head of the ion beam and passes the imperfection of the head portion to its tail. References [113] and [114] describe the instability in detail. The work in [114] predicts the threshold of the instability with two theoretical models (two-particle model and multi-particle circulant matrix model), as shown in Figure 3-23.
The eRHIC parameter exceeds the threshold; therefore a fast (few thousand turns) deterioration of the ion beam is expected if no countermeasure is implemented. Simulations also indicated that the current chromaticity in RHIC cannot suppress the instability.

To suppress instability two variants of feedback system have been studied (Figure 3-25). In reference [114], an innovative feedback system is presented as an effective countermeasure. In this feedback system, one electron bunch will be slightly steered transversely based on the feedback information of the previous electron bunches after collision. These electron bunches interact with the same ion bunch. The feedback system can successfully suppress the kink instability in a cost effective way, since there is no RHIC modification required.

An alternative traditional feedback system for the kink instability is also studied in [115]. It consists of a pickup, a kicker and the broadband amplifier between them. For the eRHIC parameters, the minimum bandwidth is determined as 50 MHz to 300 MHz from the simulation result.

The noise carried by the fresh electron beam may heat up the ion beam due to the beam-beam interaction. The random electron beam offset at the IP causes dipole-like errors for the ion beam, while the beam-size and intensity variation at the IP lead to quadrupole-like errors. The effects of both errors can be evaluated either theoretically or in simulation. The simulation shows that one-micron electron beam position offset at the IP causes an ion beam emittance growth of 20% per hour. The expected cooling time is much shorter than the emittance growth time. The same cooling time also allows the quad error of 0.1% (e-beam intensity or the beam size variation).

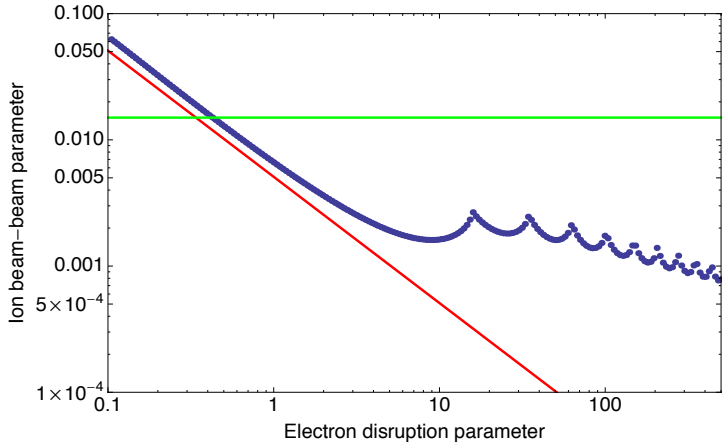

Figure 3-24: The threshold of kink instability, with choice of the synchrotron tune 0.004. The Blue dots denote the threshold calculated from macro-particles circulant matrix method. The red line represents the simple threshold form from simple two-particle model. The green line corresponds to the constant beam-beam parameter of 0.015.

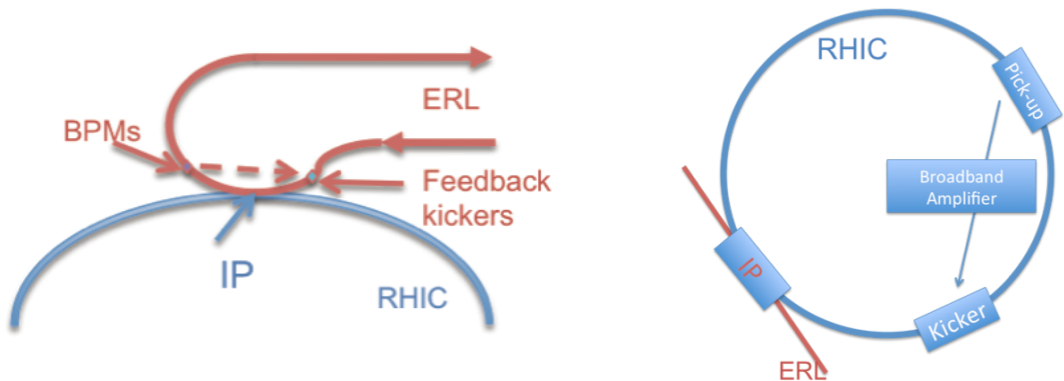

Figure 3-25: Left, dedicate feedback system of the electron accelerator to mitigate the kink instability; right, the pickup-kicker feedback system in RHIC for mitigating the kink instability.

### 3.3.10 Beam Polarization

The polarized electron beam is produced from the Gatling gun, with a polarization of ~85-90%, and the task is to preserve this high polarization through the acceleration cycle up to the collision points. The eRHIC experiments call for longitudinal polarization. With the cost saving intent in mind eRHIC avoids lengthy spin rotator insertions near the interaction regions. Instead, the beam polarization vector is allowed to rotate in the horizontal plane around the vertical guiding field during the beam re-circulations. The spin precession rate is directly proportional to the electron energy. With the accelerating gain of the main ERL chosen to be 1.322 GeV the orientation of beam polarization can be made longitudinal in both eRHIC experimental detectors, at IR6 and IR8.

The main depolarization effect is related with the spin decoherence due to the beam energy spread. The rms energy spread of ~0.001 is produced by the RF waveform of the ERL accelerating voltage. The effect of this energy spread on the beam polarization is shown in Figure 3-26.

To eliminate the spin de-coherence the energy spread compensating cavities, operating at 5th harmonic of the ERL RF frequency, have to be added to the main ERL. The parameters of the compensating cavity system have been selected to achieve a polarization remaining at 80% level for 4 mm rms bunch length up to an electron top energy of 21.2 GeV.

Another possible depolarization may come from the stochastic changes of the particle energy caused by the process of emission of the synchrotron radiation quanta. Since the spin precession rate is defined by the particle energy, the spontaneous changes of the particle energy lead to the diffusion of the spin rotation angle. Figure 3-27 shows the depolarization in terms of the rms spread of the spin angle caused by this spin diffusion. The corresponding resulting polarization loss is negligible at 15.9 GeV, and is only 2% at 21.2 GeV.

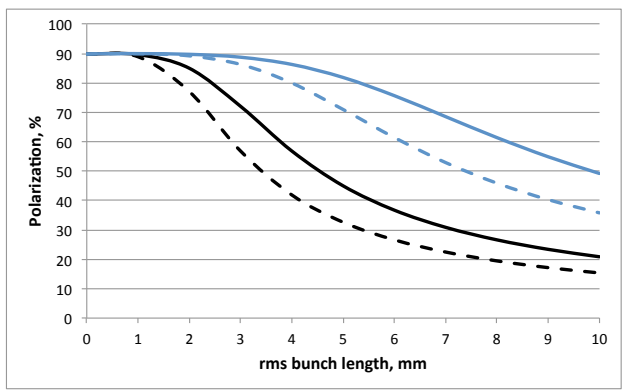

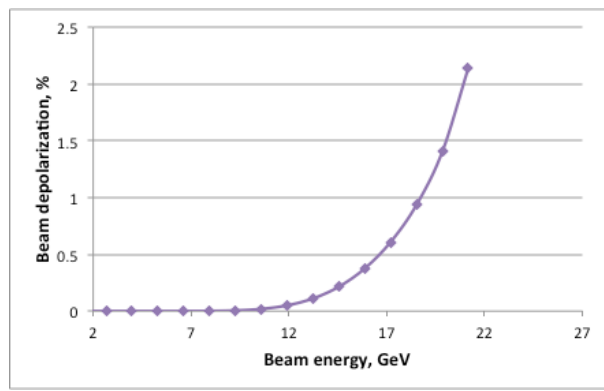

Figure 3-26: Average beam polarization versus the bunch length for 15.9 (solid lines) and 21.2 GeV (dashed lines) beam energies, as a result of the spin decoherence caused by the beam energy spread. Black lines show the depolarization caused by the main ERL RF waveform. Blue lines show the combined effect of the main ERL waveform and the energy spread compensation system.

Figure 3-27: Average beam depolarization due to the stochastic synchrotron radiation process.

### 3.3.11 Interaction Region Design

*Interaction Region Overview*

The main features of eRHIC interaction regions (Figure 3-28) are:
- Low $\beta^* = 5$ cm
- 10 mrad crossing angle and the crab-crossing scheme
- Magnets of hadron IR focusing triplets are large aperture superconducting magnets
- First magnet (the hadron quadrupole) is located at 4.5 m from the collision point, outside the detector.
- Detector components for registration of neutral and charged particles are placed near the forward hadron beamline.
- Arranged free-field electron pass through the hadron triplet magnets
- Gentle bending of the electrons to avoid the synchrotron radiation impact on the detector

The experimental requirements for the detection of forward propagating products of the collisions impact the IR design significantly. In the outgoing hadron beam direction, the IR magnets have to have enough aperture to pass the forward neutrons and forward scattered protons with a typical angle spread on the scale of ±10 mrad. In the outgoing electron beam direction arrangements have to be done to tag the scattered electrons with small scattering angles (25-35 mrad).

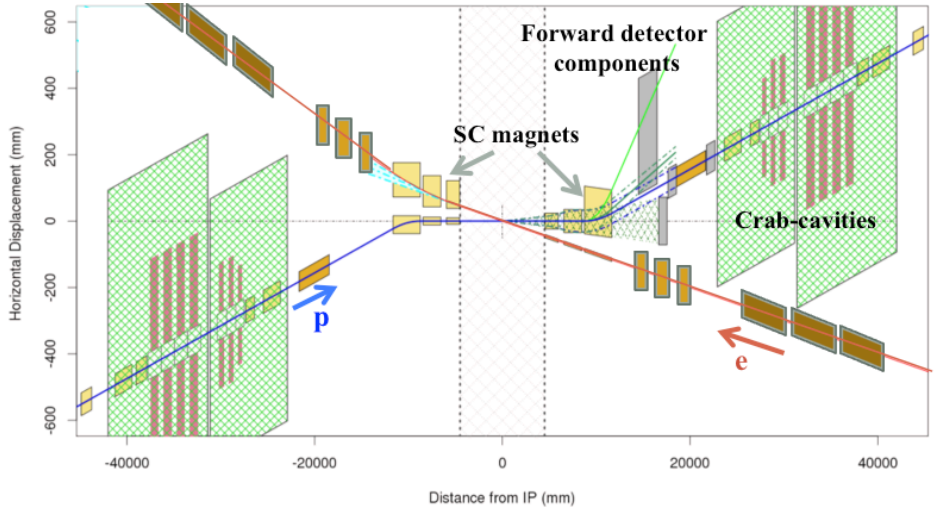

Figure 3-28: eRHIC interaction region layout. (the view from above)

### *Hadron IR Beamline*

$\beta^*$ = 5 cm is required for the high design luminosity. In the IR lattice design this small $\beta^*$ is realized in two steps. First, the interaction region quadrupoles are designed to provide a strong focusing which allows to achieve $\beta^*$=10 cm. Then, the squeeze from 10 cm to 5 cm is realized by introducing betatron waves in both planes, using the Achromatic Telescope Squeezing technique [116]. The eRHIC hadron lattice has a phase difference of 90° per cell in the arcs. The betatron wave is created by changing the quadrupole gradients ($\Delta G$= 7% with respect to the regular arc quadruple gradients) in two quadrupole pairs at the beginning of the arc before the IP.

24 families of sextupoles in the 90° degree lattice are able to correct the first and higher orders of chromaticities in the eRHIC lattice. The sextupole strength can be optimized also to minimize the lower order resonance driving terms. The resulting dynamic aperture (for the IR lattice variant with $\beta_{max}$ ~2200) obtained in the presence of the machine errors as well as beam-beam interactions is shown in Figure 3-29. Machine errors include 0.2% quadrupole and sextupole field errors and 100 microns magnet misalignments. At the momentum spread of the cooled hadron beam of ~2·10$^{-4}$, the sufficient dynamic aperture of 10$\sigma$ has been demonstrated. Further improvement may be expected from careful choice of the machine working point.

The main features of the IR superconducting magnets, forming the hadron IR triplet, include the large aperture, needed to pass through the forward momentum collision products, and near field-free region arranged for the electron beam passage through the magnets. Figure 3-30 shows how the electron passage is arranged through the magnet area between the coils of hadron IR magnet. The coil is splitted into separate inner- and outer- coil structures, and extended low field "sweet spot" is provided. Then, a combination of passive shielding and a relatively weak field-corrector coil affords a low field path for the electrons.

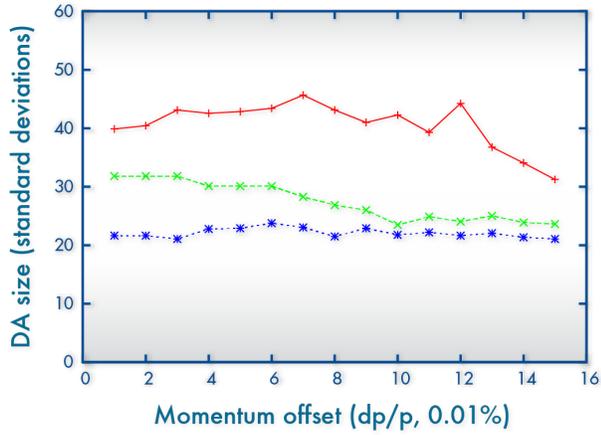

Figure 3-29: A plot of the optimized off-momentum dynamic aperture for eRHIC. The top curve (red, +) is the bare lattice, the middle curve (green, x) is with a beam-beam parameter of 0.015, and the bottom curve (blue, *) is with beam-beam and gradient errors.

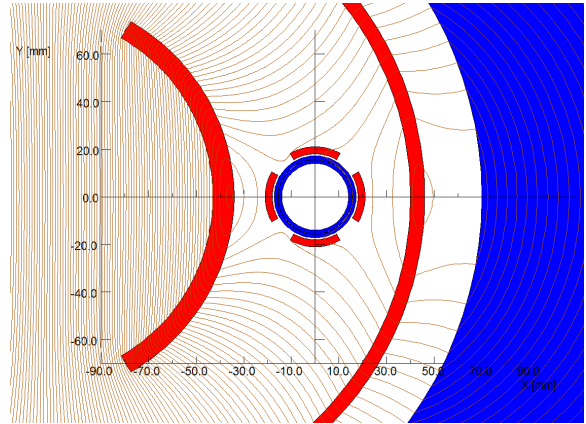

Figure 3-30: The electron pipe goes through "sweat spot", a weak field area of hadron IR magnet, which is arranged by SC coil configuration.

### *Electron Beamline*

The extraction of the top energy electron bunches from the high energy FFAG ring will be realized using a special pattern of dipole correctors which increases resonantly the orbit amplitude at particular energy, which then extracted through a thin septum magnet [117]. The exact details of the extraction scheme are under development. The beam is extracted into the individual beamline, which brings the electrons to the experimental detector along its axis and focuses the beam to small $\beta^*=5$ cm at the collision point. The beamlines upstream and downstream of the detector have a similar magnet and lattice structure.

The top energy beamline consists of two parts, determined by their functions. The first part, the vertical shift beamline transports the top energy electron beam over the hadron ring magnet line and down to the level of the detector center axis. This beamline is ~55m long and the bending is done with relatively strong magnetic field (0.081 T at 21.2 GeV).

The second part of the top energy beamline, the IR beamline, that is ~60m long, provides the final weak bending to put the electron beam exactly on the detector axis. The focusing magnets, including the final focusing triplet, provide $\beta^*=5$ cm at the collision point. This beamline contains the bending magnets with the field from 105 to 16 Gs at 21.2 GeV beam energy. Using the 16 Gs dipole magnets for the final bending produces a very low intensity soft synchrotron radiation, which does not create problems at the detector. The optical functions of the IR beamline are shown in Figure 3-31.

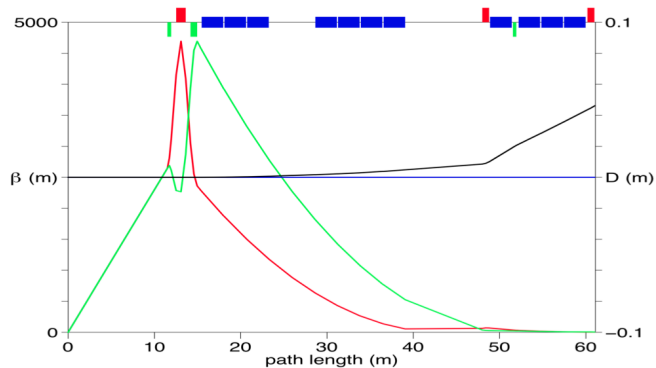

Figure 3-31: The horizontal (red) and vertical (green) beta-functions, and the horizontal (blue) and vertical (black) dispersion functions of the electron IR beamline. The collision point is located at 0 of the horizontal axis.

Since there are no strong bending magnets within 60 m from eRHIC detector, there are no strong synchrotron radiation sources near the experimental detector. The forward radiation coming from the upstream hard bend is completely masked and no hard radiation passes through the detector. Only soft bending is present in the vicinity of the detector. The forward radiation from the upstream soft bend passes through the detector but cannot penetrate through the beam pipe. The secondary backward radiation induced by the forward radiation generated in downstream bends can be mostly masked from entering the detector area.

## *Crab-Crossing*

Since the interaction region employs 10 mrad crossing angle between electron and hadron beams, the crab-crossing scheme is required to avoid more than an order of magnitude of luminosity loss. With the crab-crossing both electron and hadron beams come to the collision point rotated by 5 mrad in the horizontal plane. The beam rotation is realized by crab cavities. The crab cavities are placed on both sides of the interaction region area to ensure that the beam rotation does not propagate to the outside of the interaction region.

Two possible arrangements of hadron crab cavities have been considered. At one arrangement shown in Figure 3-28 the crab cavities are placed on each side of an interaction region. In this case the beam trajectory distortion produced by crab cavities remains local at each of two of eRHIC interaction regions. Another possible arrangement creates the trajectory distortion, which propagates through both eRHIC interaction regions as well as through the arc between them. Obviously the latter arrangement uses half as many cavities as the former, but requires arc lattice modification and, possibly, modifications of RHIC arc sextupole families.

The sinusoidal form of the crab-cavity voltage leads to the transverse deviation of particle at the head and tail of the bunch from the perfect linear x-s correlation shape. To exclude possible harmful effects of the beam-beam interactions the linear profile of x-s correlation should be within 1/10 of the transverse beam size. To satisfy this criterion with reasonable crab-cavity voltage the system involving higher harmonic cavities has been suggested. Table 3-9 lists the main parameters of the eRHIC hadron crab cavity system.

The main cavity design is based on a quarter wave (QW) coaxial resonator. The QW shape provides a very compact design, absence of lower and same order modes, and large separation of the fundamental and first higher order mode. The harmonic crab cavities will be of a similar design.

The crab cavities for electrons will operate at 676 MHz, as the electron bunches are short. A 1.9 MV voltage is required. Preliminary consideration for the eRHIC crab cavities can be found elsewhere [118,119].

| Crab-cavities | Number of cavities | RF frequency | Cavity voltages |
|---|---|---|---|
| **Main cavity** | 4 | 225 MHz | 6.2 MV |
| **2nd harmonic cavity** | 1 | 450 MHz | 2.8 MV |
| **3rd harmonic cavity** | 1 | 676 MHz | 0.76 MV |

Table 3-9: Parameters of the hadron crab cavities. Number of cavities is listed for one side of the interaction region. Cavity voltages are based on $R_{12}$ element of the horizontal transfer matrix between the crab cavity and the collision point equal to 16.7m.

# 3.4 Accelerator R&D Activities

eRHIC accelerator R&D activities have been underway for several years. Most of them have been funded through the Brookhaven Laboratory LDRD program. The aim of the R&D activities is to verify that major components of machine design based on advanced accelerator technologies can reach the performance required by the design. In the coming years the eRHIC R&D activities will continue with a goal to confirm all major design points before 2018.

## *BNL Gatling Gun*

The high average current polarized electron gun presented a number of state of the art challenges that are being addressed by the BNL Gatling Gun development project. The project has addressed a number of areas of specific interest, which are:

- Establish a full scale R&D Gatling Gun system with limited resources
- Large extreme vacuum chambers, systems and procedures
- High quantum efficiency GaAs cathode preparation capability
- The electron funneling mechanism.
- High voltage systems

The R&D Gatling Gun system design has been developed to begin operation with a minimum of two cathodes and can be expanded to demonstrate all essential aspects of gun operation. Figure 3-32 and Figure 3-33 show the components of the R&D Gatling system.

BNL funded two LDRD's for Gatling gun related developments one for the laser system to deliver short beam pulses to the gun and the other to develop the extreme vacuum system components and establish high quantum efficiency GaAs photocathode expertise in the Collider Accelerator Department and demonstrate proof of principle Gun operation. Phase one will develop the basic Gatling gun system components that produce a minimum of two beams that demonstrate the funneling principle with at least two beams and show how the operation of one cathode may affect the performance of another cathode. Cathode development began using bulk GaAs with a polarization of approximately 50%. Once basic Gatling gun operation is routinely established superlattice GaAs will be incorporated with polarization to exceed 80%.

A major advancement was the development of large XHV chambers required for large-scale cathode preparation and to accommodate the full 20 cathode array for gun operation. The first chamber constructed was the largest used in the Gatling gun system. The cathode preparation chamber named "Grand Central" is a spherical multiport chamber 0.9 meters in diameter. BNL worked closely the MDC company to develop the manufacturing, bake out and testing procedures to minimize surface outgassing. This became the largest know XHV chamber reaching $< 8 \times 10^{-12}$ Torr and demonstrated the large component feasibility that is necessary for the Gatling Gun concept. The second chamber was the Gatling Gun's main chamber that was developed by the Transfer Engineering Co. It reached the low $10^{-12}$ Torr range. This chamber demonstrated the feasibility of very large up to a 27-inch wire seal flange to achieve XHV conditions.

The cathode preparation system called "the tree" is used to apply Cesium and Oxygen to condition the cathode assemblies designed for use in the Gatling Gun. Quantum efficiencies of 8 percent have been achieved. Multiple Cathode preparation trees will be used on the Grand Central chamber to where reconditioned cathodes will be stored prior to reuse in the Gatling Gun.

The Components of the Gatling Gun system have been produced and are being assembled and prepared for phase one system testing. The first tests of multi-cathode gun operation are expected at the end of 2014.

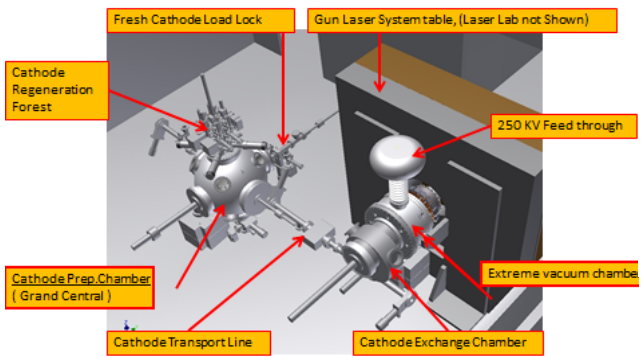
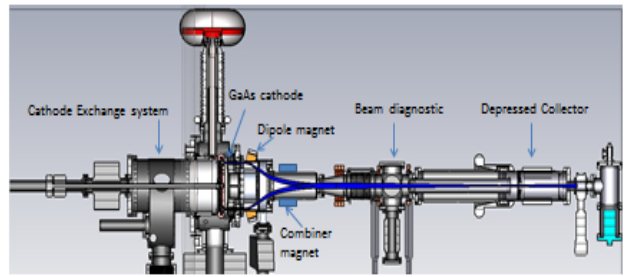

Figure 3-32: The Gatling Gun system composed of a polarized electron source, a cathode preparation system, and a photocathode drive laser system.

Figure 3-33: The source has a large XHV central chamber that supports the high voltage cathode shroud and an array of GaAs photocathodes, a cathode manipulation system, the electron funneling mechanism composed of first bend dipole magnets and combiner magnet, a beam diagnostic section and the depressed collector beam stop.

### *High Beam Current SRF Cavities*

One of the 704 MHz cavities fabricated for the high current application will serve as a vehicle for further eRHIC related R&D. In particular, over the next two years we plan to explore different options of cavity treatment in order to improve its intrinsic quality factor. Another R&D effort is dedicated to developing an efficient HOM coupler design. As mentioned above, there will be six such couplers attached to the cavity. We plan to investigate several options of the coupler design, select one, fabricate niobium prototypes and test them on the BNL3 cavity. To bring the HOM power from the cryogenic environment to an RF load outside the cryomodule, we will develop and a wideband RF window and a low thermal loss cable. These efforts will span three years.

As the eRHIC SRF frequency has changed from 704 MHz to 422 MHz, there will be R&D necessary to develop a multi-cell cavity at the new frequency. While the baseline geometry is simply a scaled version of the BNL3 cavity, an optimization will have to be performed to determine the new cavity shape as well as the number of cells per cavity. A prototype will have to be fabricated and tested. An estimated time for this R&D is three years.

Finally, we plan to develop an SRF cryounit based on the concept presented in [106. A single-cavity cryounit with end caps will be designed, built and tested. This effort will span three years, but its start will be offset by approximately one year with respect to the 422 MHz cavity R&D.

### *CEC Proof of Principle Experiment*

The Coherent Electron Cooling PoP experiment will be conducted in the IP2 of RHIC tunnel. The experiment layout is shown in Figure 3-34.

The electron beam will be generated by the 112 MHz superconducting RF gun. Two 500 MHz normal conducting cavities will provide energy chirp and the ballistic compression of the electron beam while it travels to the 704 MHz accelerating cavity, which boosts electrons to the final energy of 22 MeV. A dogleg structure merges electron bunches with the gold ions stored in the RHIC "yellow" ring. In the modulator section the gold ions imprint their position onto the electron beam thus creating the modulation of the electron density. The density modulation is then amplified in the FEL like structure, which also provides travel space for ions. The latter ones are moving forward or backwards with respect to center of the ion bunch depending on their energy. The amplified charge modulation in the electron bunch provides accelerating or decelerating field for the ions in the kicker section. With proper phasing one can set cooling or anti-cooling of the ion bunch. The goal is to observe the cooling of the ion bunch by measuring its longitudinal profile and/or spectral content of the signal from a wall current monitor and compare it with theoretical predictions and numerical simulations. The experiment equipment installation began in the 2013. During RHIC Runs 14 and 15 the gun structure (112 MHz and 500 MHz RF system and photocathode) will be commissioned. During the summer shutdown of 2015 the remaining equipment will be installed, including the accelerating cavity, beam transport, FEL system, and high power beam dump. The CeC Proof-of-Principle experiment will be performed during RHIC Run-16 and -17.

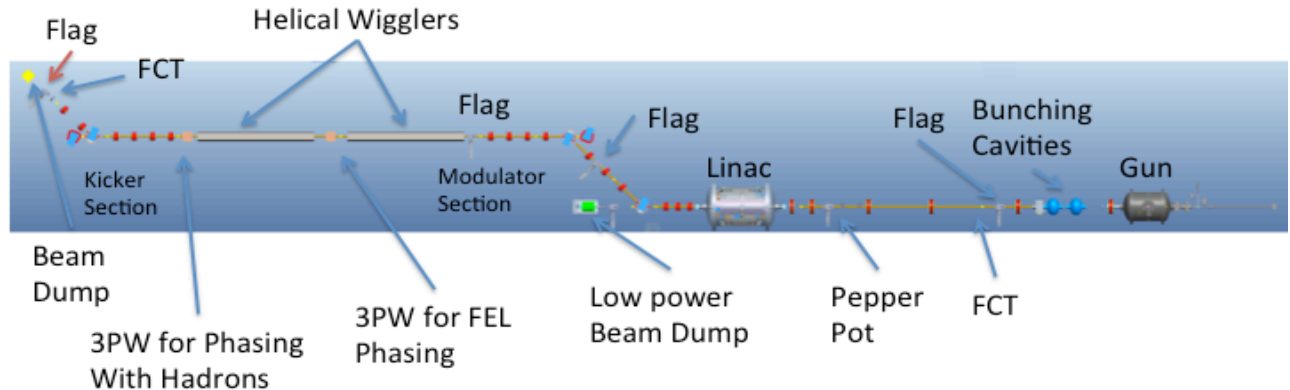

Figure 3-34: CeC PoP Experiment.

### Beam-Beam Effects

The beam-beam effects in this linac-ring scheme collider are very different from the traditional ring-ring type collider. In Sec. 0, the special effects and the proposed countermeasure are described. The results of the study come from either theoretical study or simulation. Therefore the opportunity for testing the understanding is essential to ensure the proposed luminosity in eRHIC.

The in-constructing 21.8 MeV electron beam for the CEC PoP experiment can be used also for this beam-beam test purpose. Through collision with the opposing ion beam in the blue ring with proper energy and optics, a similar beam-beam parameter of electron beam to that of the eRHIC can be achieved. All the special effects and the countermeasures can be tested, including

- Electron disruption effect and the pinch effect
- The kink instability of the ion beam and its countermeasures
- The noise heating effect on the ion beam

In addition, with low energy ion beam in the blue ring, the cross talk between the ion beam space charge effect and the beam-beam effect can be also studied. Such study can establish the understanding of the maximum space charge tune shift allowed in electron-ion collision.

### Crab Cavities

At present, there is a worldwide R&D effort to develop compact crab cavity for the HiLumi LHC upgrade. BNL is actively involved in this R&D and we have developed a Double Quarter Wave Crab Cavity (DQWCC) with strong HOM damping [120,121,122] A proof-of-principle DQWCC cavity was successfully tested in 2013 at BNL [123]. Further plans include fabrication of two prototype cavities in 2014; vertically testing them at BNL in 2015; design, fabrication and assembly a two-cavity cryomodule in 2014-2016; testing it in SM18 and SPS at CERN in 2016-2017. This R&D is synergetic with eRHIC and the DQWCC design can be easily scaled to frequencies of the eRHIC crab cavities for hadrons. The R&D of the eRHIC crab cavities for hadron beams can proceed quickly as soon as funds are available. However, because three different frequencies are involved, the development time will be about three years. Developing eRHIC crab cavities for electrons will require dedicated efforts and the estimated time needed for this is about four years.

# 4 eRHIC DETECTOR REQUIREMENTS AND DESIGN IDEAS

The physics program of an eRHIC imposes several challenges on the design of a detector, and more globally the extended interaction region, as it spans center-of-mass energies from 55 GeV to 141 GeV, different combinations of both beam energy and particle species, and several distinct physics processes. The various physics processes encompass inclusive measurements $ep/A \rightarrow e'+X$; semi-inclusive processes $ep/A \rightarrow e'+h+X$, which require detection of at least one hadron in coincidence with the scattered lepton; and exclusive processes $ep/A \rightarrow e'+N'A'+\gamma/h$, which require the detection of all particles in the reaction with high precision. The figures in section 4.1 illustrate the differences in particle kinematics of some representative examples of these reaction types, as well as differing beam energy combinations. The directions of the beams are defined as for HERA at DESY: the hadron beam is in the positive $z$-direction/pseudo-rapidity ($0^o$) and the lepton beam is in the negative $z$-direction/pseudorapidity ($180^o$).

## 4.1 Detector Requirements

All the different physics processes accessed at eRHIC require having the event and particle kinematics ($x$, $Q^2$, $y$, $W$, $p_t$, $z$) reconstructed with high precision. In order to access the full $x$-$Q^2$ plane at different center-of-mass energies, the detector must be able to reconstruct events over a wide span in $y$. This imposes certain requirement on both detector acceptance and resolution. At large $y$, where radiative corrections become large, as illustrated in Fig. 7.25 in [124] and the kinematics of the event is reconstructed from the scattered electron, there are two ways to address this: one is to calculate radiative corrections and correct for them; the other is to utilize the hadronic activity in the detector together with cuts on the invariant mass of the hadronic final state, which will reduce the impact of radiative corrections to a minimum.

At small lepton scattering angles or correspondingly small inelasticity radiative corrections are small, but the momentum and scattering angle resolution for the scattered lepton deteriorates. This problem is addressed by reconstructing the lepton kinematics purely from the hadronic final state using the Jacquet-Blondel method [125] or using a mixed method like the double angle method [126], which uses information from the scattered lepton and the hadronic final state. At HERA, these methods were successfully used down to $y$ of 0.005. The main reason this hadronic method renders better resolution at low $y$ follows from the equation $y_{JB} = (E - p_z^{had})/2E_e$, where $(E-p_z^{had})$ is the sum over the energy minus the longitudinal momentum of all hadronic final-state particles and $E_e$ is the electron beam energy. This quantity has no degradation of resolution for $y<0.1$ as compared to the electron method, where $y_e = 1-(1-cos\theta_e)E'_e/2E_e$. To allow for efficient unfolding of measured quantities, i.e. cross sections and asymmetries, for smearing effects due to detector resolutions and radiative events and retain the statistical power it is important to have a survival probability in each kinematic bin of ~80% or better.

Typically, one can reach for a given center-of-mass energy squared, roughly a decade of $Q^2$ at fixed $x$ when using only the electron method to determine the kinematics, and roughly two decades when including the hadronic or double angle method. If only using the electron method, one can increase the range in accessible $Q^2$ by lowering the center-of-mass energy. The coverage of each setting is given by the product of $y$ times $s$. As lower a $y_{min}$ that can be reached the fewer settings in $s$ are needed. However, this is an important consideration for any measurement, which needs to separate the cross-section components due to longitudinal and transverse photon polarization, i.e. the measurement of $F_L$ where one needs to have full $y$-coverage at all energies. Figure 4-1 (upper row) illustrates the dependence between $Q^2$ and the pseudo-rapidity of the scattered lepton. It is clearly shown that as higher the center-of-mass energy the more the lepton goes in the original electron beam direction, corresponding to negative pseudo-rapidity. A scattered lepton with $Q^2$ of 1 GeV$^2$ needs to be detected at a pseudo-rapidity of -3 to -4 increasing the lepton beam energy from 10 GeV to 20 GeV. Varying the hadron beam energy (Figure 4-1 lower row) has no influence on the scattered lepton pseudo-rapidity $Q^2$ correlation. Several eRHIC physics topics require going to low $x$ at low $Q^2$ such an eRHIC detector needs to have good electron identification and momentum/energy measurement at pseudorapidities < -2.

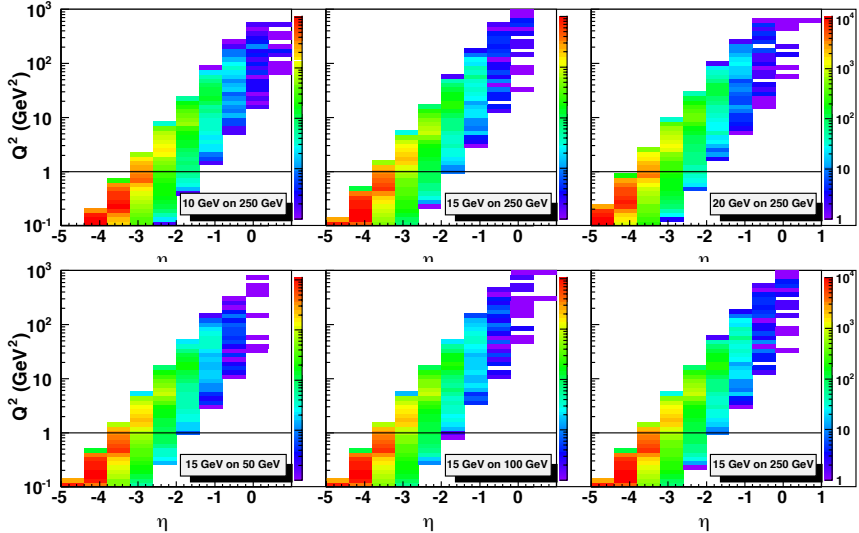

Figure 4-1: Q2 vs. pseudo-rapidity in the laboratory frame for the scattered lepton at different center-of-mass energies. The following cuts have been applied: 0.01<y<0.95

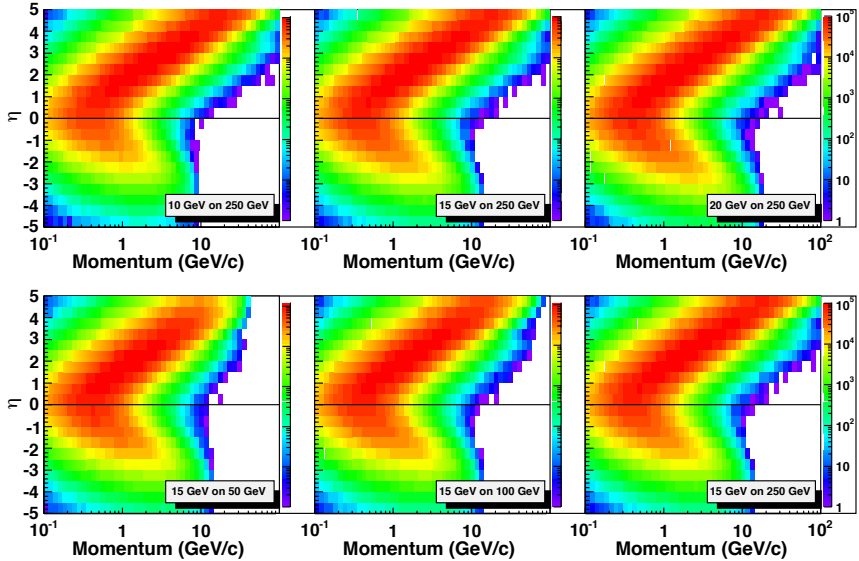

Figure 4-2: Momentum vs. pseudo-rapidity in the laboratory frame for pions from non-exclusive reactions at different center-of-mass energies. The following cuts have been applied: Q2 > 1 GeV2 0.01<y<0.95

Figure 4-2 shows the momentum versus pseudo-rapidity distributions in the laboratory frame for pions originating from semi-inclusive reactions for different lepton and proton beam energy combinations. For lower lepton energies, pions are scattered more in the forward (ion) direction. With increasing lepton beam energy, the hadrons increasingly populate the central region of the detector. At the highest lepton energies, hadrons are even largely produced going backward (i.e. in the lepton beam direction). For increasing hadron beam energies at fixed lepton beam energy the pseudo-rapidity distribution remains the same but the maximum hadron momentum increases at fixed pseudo-rapidity. The kinematic distributions for kaons and protons/anti-protons are essentially

identical to those of the pions. The distributions for semi-inclusive events in electron-nucleus collisions may be slightly altered due to nuclear modification effects, but the global features will remain.

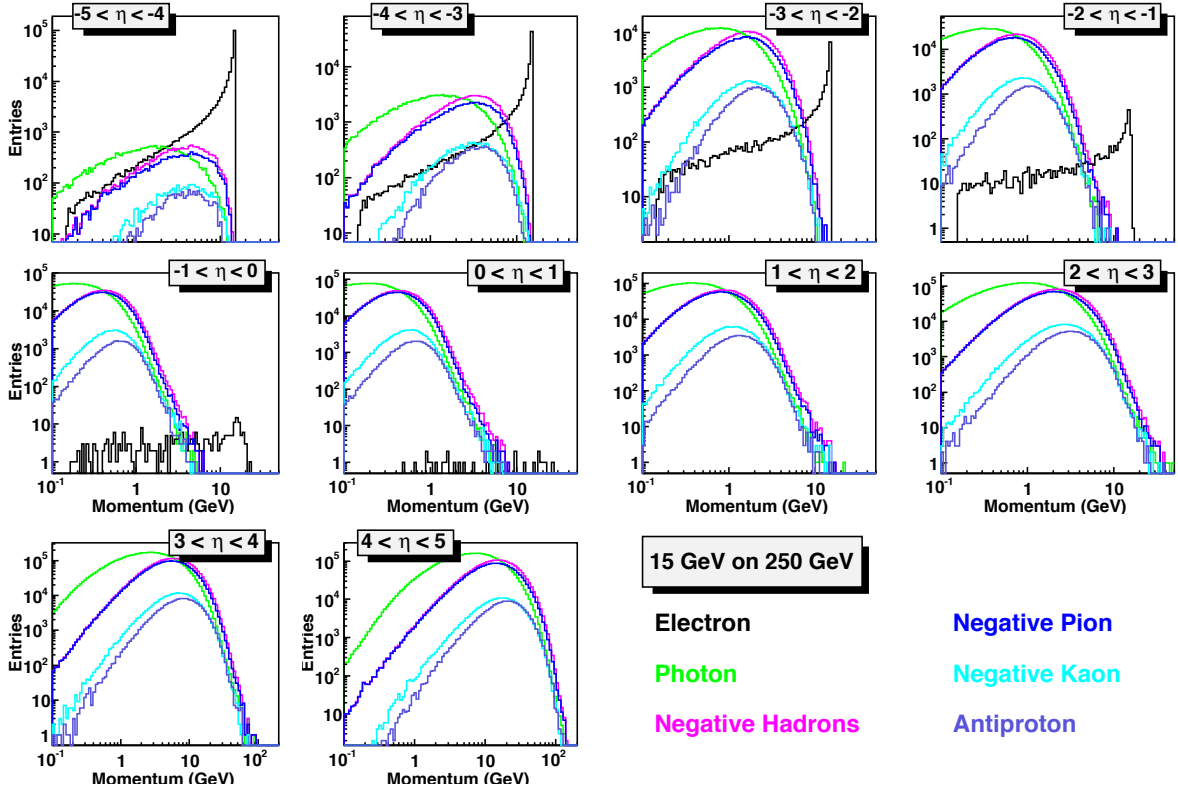

Figure 4-3: The momentum distribution for the scattered electrons (black), photons (greens), and negatively charged hadrons (magenta) for different pseudo-rapidity bins in the laboratory frame for beam energies of 15 GeV on 250 GeV. The distributions for negatively charged Pions (blue), Kaons (cyan) and antiprotons (violet) are shown as well. No kinematic cuts have been applied.

Figure 4-3 indicates the momentum/energy range of the scattered electron (black curve), photons (green), negative charged pions (blue) and kaons (cyan) as well as antiprotons (violet) and their sum (magenta) for a center-of-mass energy of 122 GeV as function of pseudo-rapidity. This plot provides on the one hand the needed information on the requirements for the scattered lepton identification as well as for the identifying pions, kaons and protons. For the entire pseudo-rapidity ($-5 < \eta < 5$) negative pions, kaons and antiprotons show the same momentum distributions, with negative pions having a factor ~3-5 higher multipliciy as negative kaons and antiprotons. In the central detector region ($-1 < \eta < 1$) the momenta are of typically 0.1 GeV/c to 4 GeV/c with a maximum of about 10 GeV/c. A combination of very high-resolution time-of-flight (ToF) detectors, a DIRC or a proximity focusing Aerogel RICH may be considered for particle identification in this region.

Hadrons with higher momenta go typically in the forward (ion) direction for low lepton beam energies, and in the backward direction for higher lepton beam energies. The most viable detector technology for this region of the detector is a Ring-Imaging Cerenkov (RICH) detector with dual-radiators. To achieve good pion-kaon-proton separation through a RICH detector an excellent momentum resolution is required for the momentum range the Cerenkov angle is still strongly changing. Having particle identification in the forward and backward direction for $1 < |\eta| < 3$ ensures that the $z$ and $p_t$ region critical for semi-inclusive and exclusive physics is covered, see Figure 4-4.

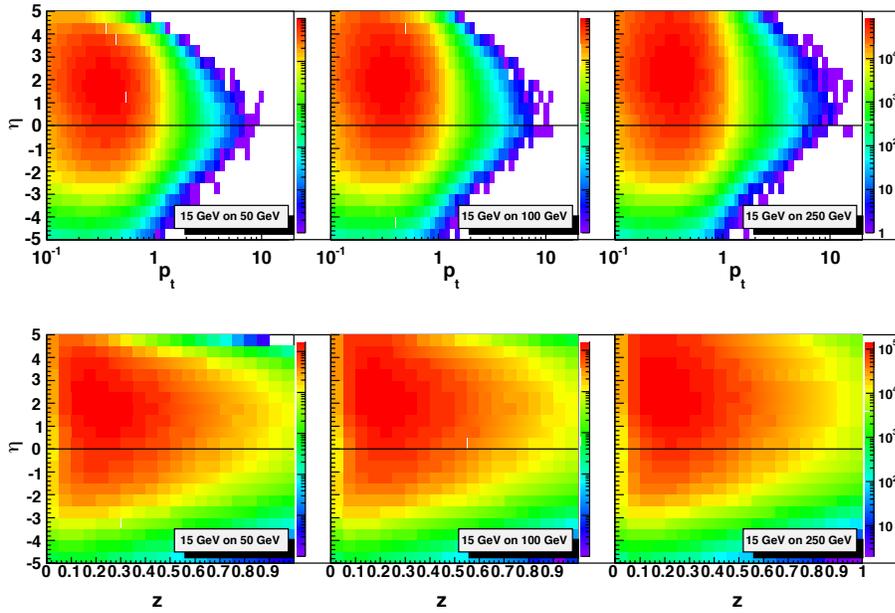

Figure 4-4: Transverse momentum $p_t$ and hadron momentum fraction z as fct. of pseudo-rapidity in the laboratory frame for pions from non-exclusive reactions at different center-of-mass energies. The following cuts have been applied: $Q^2 > 1$ GeV$^2$ 0.01<y<0.95. $p_t$ is calculated relative to the virtual photon.

To be able to measure identified hadron asymmetries as small as $10^{-4}$ that the hadrons are identified with a purity > 95% at preferably an efficiency of > 90%.

Events with $Q^2 < 10$ GeV$^2$ typically correspond to negative rapidities ($\eta < -2$) and $Q^2 > 10$ GeV$^2$ correspond to rapidities $-2 < \eta < 1$. Depending on the center-of-mass energy the rapidity distributions for hadrons (both charged and neutral) and the scattered lepton overlap and need to be disentangled. For $\eta < -3$ electron, photon and charged hadron rates vary from being comparable to a factor of 10 different. For the higher pseudo-rapidities electron rates are a factor of 100-1000 smaller than photon and charged hadron rates, and comparable at a 10 GeV/c total momentum. For very high $Q^2$-events a suppression factor of $10^5$ is needed. This adds another requirement to the detector: good electron identification. It is noted that the kinematic region in pseudo-rapidity over which hadrons and also photons need to be suppressed, typically by a factor of 10 - 1000, shifts to more negative pseudo-rapidity with increasing center-of-mass energy.

Measuring the ratio of the energy and momentum of the scattered lepton, typically gives a reduction factor of ~100 for hadrons. This requires the availability of both tracking detectors (to determine momentum) and electromagnetic calorimetry (to determine energy) over the same rapidity coverage. By combining information from these two detectors, one also immediately suppresses the misidentification of photons in the lepton sample by requiring that a track must point to the electromagnetic cluster. Having good tracking detectors with similar rapidity-coverage as electromagnetic calorimetry similarly aids in y-resolution at low y from the lepton method. The hadron suppression is further improved by adding a Hadron Calorimeter or a Cerenkov detector to the electromagnetic calorimetry or having tracking detectors, (e.g., a Time Projection Chamber) to provide good $dE/dx$. The resulting lepton purities should be > 99% with preferable a detection efficiency of > 90%.

There is specific interest in extracting structure functions with heavy quarks from semi-inclusive reactions for mesons, which contain charm or bottom quarks. To measure such structure functions as $F_2^C$, $F_L^C$, and $F_2^B$, it is sufficient to tag the charm and the bottom quark content via the detection of additional leptons (electrons, positrons, muons) in addition to the scattered (beam) lepton. The leptons from charmed mesons can be identified via a displaced vertex of the second lepton ($\tau$ ~150 μm). This can be achieved by integrating a high-resolution vertex detector into the detector design. For measurements of the charmed (bottom) fragmentation functions, or to study medium modifications of heavy quarks in the nuclear environment, at least one of the charmed (bottom) mesons must be completely reconstructed to have access to the kinematics of the parton. This requires, in addition to measuring the displaced vertex, good particle identification to reconstruct the meson via its hadronic decay products.

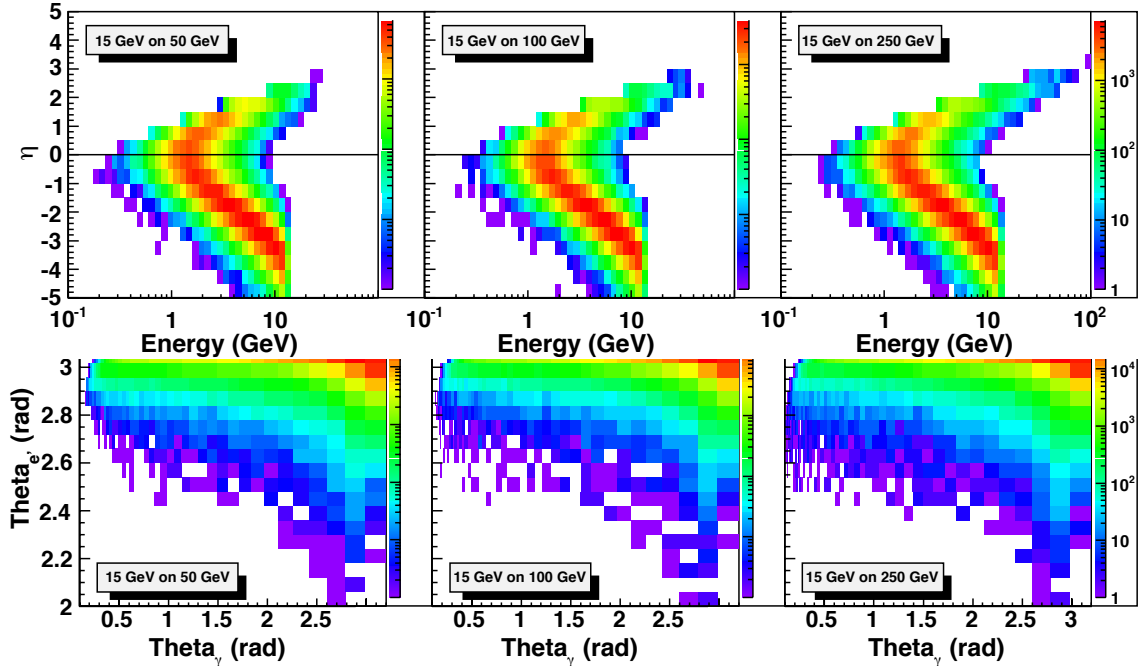

Figure 4-5: The energy vs. pseudo-rapidity in the laboratory frame for photons from DVCS (top) and the correlation between the scattering angle of the DVCS photon and the scattered lepton for three different center-of-mass energies. The following cuts have been applied: $Q^2 > 1$ GeV$^2$ 0.01<$y$<0.85 and -5< $\eta$ < 5.

Figure 4-5 shows the energy vs. rapidity distributions for photons from deeply virtual Compton scattering (DVCS), and the correlation between the scattering angle of the DVCS photon and the scattered lepton in the laboratory frame for different beam energy combinations. The general patterns follow the ones in Figure 4-2, but even at the low lepton beam energies the DVCS photons go more into the backward direction. To separate the DVCS events from their dominant background from Bethe-Heitler events it is important to measure the DVCS photon energy and the lepton momentum down to 1 GeV and to be able to resolve their scattering angle difference ($\theta_{e'}$-$\theta_\gamma$) down to below 1°. The most challenging constraints on the detector design for exclusive reactions compared to semi-inclusive reactions are, however, not given by the final state particle, but to ensure the exclusivity of the event. Exclusivity can be achieved by different methods. In electron-proton scattering by detecting all reactions products, especially the scattered protons going forward under extremely small scattering angles or requiring a rapidity gap between the hadron beam and produced pseudo-scaler/vector mesons and jets. To make the rapidity gap method highly efficient a detector with an acceptance to high pseudo-rapidities is needed. In lepton-nucleus scattering exclusivity can be ensured by the rapidity gap method or by vetoing the nuclear breakup by requiring no decay neutrons in the zero-degree calorimeter.

Figure 4-6 shows particle production rates for the 15 GeV on 250 GeV beam energy configuration, assuming an instantaneous luminosity of $10^{33}$cm$^{-2}$ s$^{-1}$. Events were simulated using PYTHIA-6, and the total cross section reported by PYTHIA was used to scale event counts to rates. No cuts, for example on event $Q^2$ or particle momentum, were applied. The $\eta$ range spans the expected acceptance of the main eRHIC detector. "Charged" particles refers to electrons, positrons, and charged pions and kaons, while "neutrals" refers to photons, neutrons and $K^0_L$.

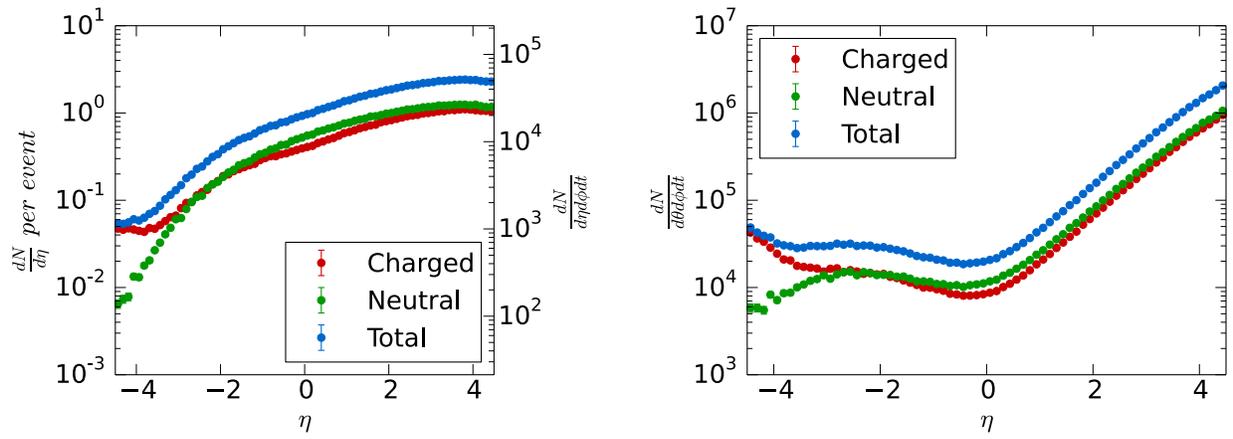

Figure 4-6: Particle production rates as a function of pseudo-rapidity at eRHIC for 15 GeV on 250 GeV e-p collisions and a luminosity of $10^{33}$cm$^{-2}$ s$^{-1}$. (a) mean numbers of particles per event (left axis) and particles per second per unit ($\eta$, $\varphi$) (right axis). (b) particles per second per unit ($\theta$, $\varphi$) i.e. the η-dependent flux at a distance of 1m from the interaction point.

## 4.2 Possible Detector Realizations

Three studies on a possible implementation for an eRHIC detector have been performed. Two studies are built on the existing RHIC detectors. Both the PHENIX and STAR collaborations have studied how the sPHENIX and STAR detectors would have to be upgraded/modified to fulfill the performance requirements as laid out by the eRHIC physics program. The third study is based on a "green field" design for an eRHIC detector, which is completely optimized to the physics requirements and the change in particle kinematics resulting from varying the center of mass energies from 55 GeV to 140 GeV. Details about all three studies are described in the following.

### 4.2.1 A Model Detector

A model for a detector implementation is shown in Figure 4-7, this detector concept closely follows the physics outlined in the EIC White Paper [1] and in section 4.1 of this document.

The compact tracker, located symmetrically with respect to the IP, consists of: a MAPS silicon barrel vertex detector and a set of forward/backward disks; a 2m long TPC with a gas volume outer radius of 0.8m and several GEM stations, all placed into a ~3T solenoid field. The TPC is specifically chosen as the main tracking element because of its small overall material budget, minimizing the rate of photon conversions on detector components, which is required in particular for the DVCS measurements. Besides this, the TPC should provide good charged PID in the momentum range up to a few GeV/c at central rapidities. Other detector options for the main tracker, such as a set of cylindrical micromegas planes are considered as well [127].

The vertex detector, covering the central rapidity range $-1 < \eta < 1$, is composed of the ALICE tracker upgrade elements [128]. It has 4 layers of high-resolution MAPS sensors with a 20 µm pixel size and an effective thickness of only ~0.3% radiation length per layer. As shown in the left panel of Figure 4-8, such a setup allows it to achieve a momentum resolution better than 3% for scattered electrons and secondary charged hadrons for momenta up to a few dozens of GeV/c in the pseudorapidity range $-3 < \eta < 3$. The right panel of Figure 4-8 demonstrates that, for a compact forward tracker design, it is critical to maintain a high detector space resolution. A 20 µm MAPS pixel size, the same as for the vertex detector, is anticipated at present.

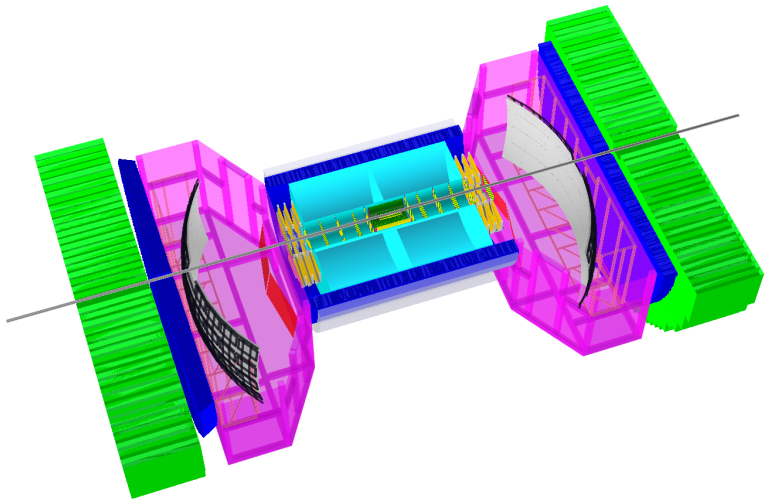

Figure 4-7: The eRHIC model detector implementation (BeAST = Brookhaven eA Solenoidal Tracker) with tracker, calorimeter and RICH components implemented in the EicRoot GEANT simulation framework [129].

As shown below in Figure 4-14, a momentum resolution of <3% should be sufficient for RICH-based hadron PID at forward rapidities ($1 < \eta < 3$), where the bulk of hadrons from semi-inclusive DIS reactions are expected to be located (see Figure 4-2 in section 4.1). At central rapidities ($-1 < \eta < 1$) the projected momentum resolution is certainly sufficient for the time-of-flight (ToF) based PID, which may be used for the particle momenta below 2-3 GeV/c. This topic, as well as technology choices for the ToF measurement and/or an option of proximity RICH installation in order to extend the PID range to ~ 5 GeV/s, are awaiting a more detailed R&D studies. In the electron-going direction, for the pseudo-rapidity range $-3 < \eta < -1$, the projected ~2-3% momentum resolution must suffice for the E/p-based lepton-hadron separation, making use of the perfect energy resolution of the backward crystal electromagnetic calorimeter (Figure 4-15).

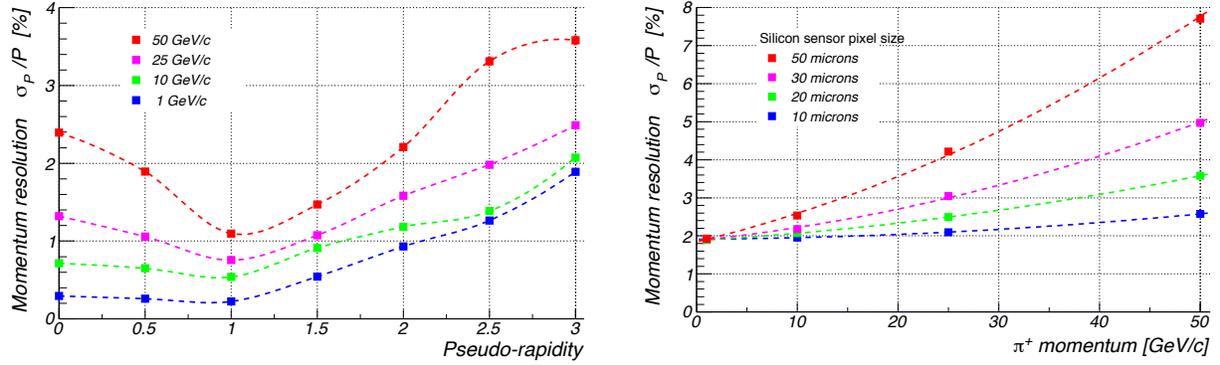

Figure 4-8: Left panel: expected momentum resolution of the baseline eRHIC detector as a function of pseudorapidity. Right panel: forward tracker momentum resolution at η = 3 vs secondary hadron momentum for various values of MAPS forward tracker pixel size.

Simulations also show that the MAPS-based vertex detector will allow for measurement of secondary decay vertices with an accuracy on order of 10-20 μm, which should be sufficient to identify events with charmed and bottom mesons (see section 4.1).

The detector will be equipped with a set of electromagnetic calorimeters, hermetically covering a pseudorapidity range of at least $-4 < \eta < 4$. The calorimeter technology choice is driven by the fact that a moderately high-energy resolution, on order of ~2-3% /$\sqrt{E}$, is needed only at backward (electron-going) rapidities (see section 4.1). Therefore in the present design the backward endcap calorimeter for the $-4 < \eta < -1$ range is composed of PWO crystals at room temperature, with the basic performance parameters taken from the very extensive PANDA R&D studies [130]. The calorimeter is located ~2700 mm away from the IP. The crystal length corresponds to ~22.5 $X_0$, and both the crystal shape and grouping follow the ideas of the PANDA and CMS [131] calorimeter designs. Both projective rectilinear and non-projective geometries are implemented in the simulation. A reasonably small crystal front facet size of 24x24 mm$^2$ is sufficient to achieve a cluster space resolution, corresponding to an angular resolution of an order of a few milliradians, which safely satisfies the requirements imposed by the DVCS event analysis [24].

For the barrel and forward endcap electromagnetic calorimeters, covering a pseudo-rapidity range of $-1 < \eta < 4$, a noticeably worse energy resolution suffices. In order to save costs, at present it is planned to use the STAR upgrade R&D building blocks of tungsten powder scintillating fiber sampling calorimeter towers, with a design goal of *~12%/$\sqrt{E}$* energy resolution [132]. The forward endcap calorimeter will be located at ~2700 mm from the IP in hadron-going direction. The barrel calorimeter will have an average installation radius of ~900 mm and be composed of slightly tapered towers, in order to avoid gaps in the azimuthal direction. Both calorimeter types will have a non-projective geometry and tower length corresponding to ~23 $X_0$. The typical anticipated energy resolutions for these two calorimeter types are shown in Figure 4-9.

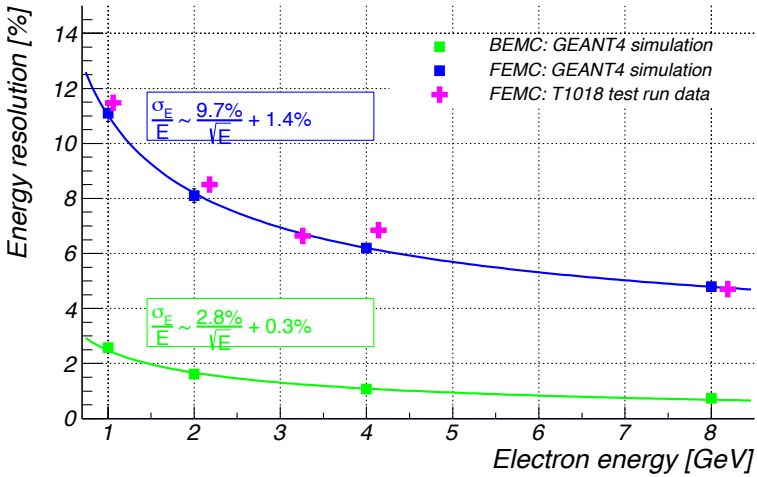

Figure 4-9: Expected energy resolution of crystal (BEMC) and sampling (FEMC) eRHIC baseline detector calorimeter types as simulated in GEANT. Realistic digitization and reconstruction parameters from PANDA [130] and STAR [132] R&D are taken.

It should be noted that although the forward and backward trackers are not assumed to provide a good charged particle momentum resolution for pseudorapidities $|\eta|>3$, they will still cover the same acceptance as the endcap electromagnetic calorimeters in this region, and therefore facilitate charged/neutral particle separation, as well as provide pointing resolution useful for calorimeter cluster reconstruction.

At least at the very backward rapidities ($\eta < -3$), where tracker momentum resolution is not sufficient to yield a reliable lepton-hadron separation based on E/p ratio, a hadronic calorimeter, installed behind the electromagnetic one, will be used for these purposes. Both forward and backward hadronic calorimeters are of a sandwich lead scintillator plate sampling type, based on the extensive EIC Calorimeter Consortium R&D. The anticipated hadron energy resolution for these calorimeters, being combined with electromagnetic calorimeter response, is expected to be of an order of ~40-45%/$\sqrt{E}$ (see Figure 4-10), which was confirmed in the recent test run at FNAL.

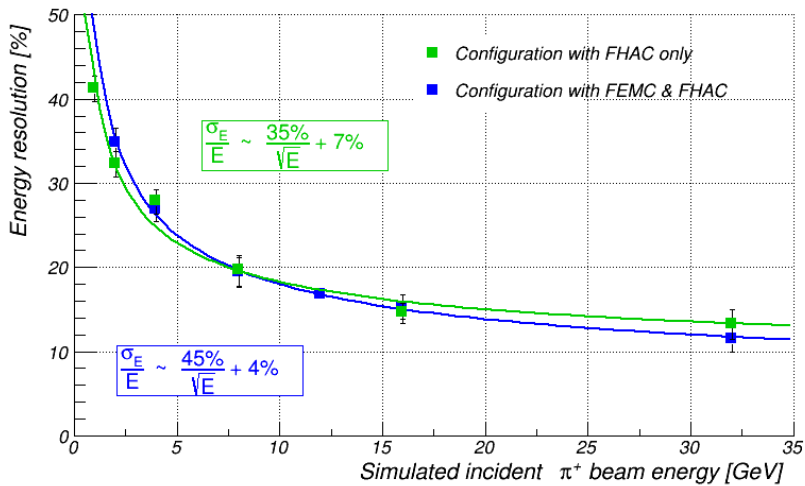

Figure 4-10: Simulated energy resolution of the forward calorimeter system for pions. Shown is the response from the lead-plastic hadronic sandwich calorimeter alone, as well as when the response from the electromagnetic tungsten powder scintillating fiber calorimeter installed in front of it, is added with a proper weight. The numbers are consistent with the results of T1018 test beam at FNAL in February 2014, as well as with [133].

The results of the momentum and energy resolutions obtained from EicRoot simulations of the tracking system and electromagnetic calorimeters were implemented in a fast-smearing generator. In addition, anticipated hadronic calorimeter performance of 38%/$\sqrt{E}$ was used in the forward direction alone. The effect of particle identification on kinematic reconstruction is negligible and is not included in the smearing generator. PYTHIA events, generated for a 15 GeV electron beam colliding with a 250 GeV proton beam, were passed through this smearing generator, and the event kinematics recalculated using the smeared momenta and en-ergies. Figure 4-11 shows the results of detector smearing on event kinematics calculated using the electron method.

As expected, due to the excellent resolution in both momentum and electron energy, *y*, *x* and $Q^2$ are exceedingly well reconstructed. Event purity is excellent at moderate-to-large *y* (typically > 90%) even with a relatively fine *x-Q2* binning of five bins per decade in *x* and four per decade in $Q^2$. However the quality of kinematic reconstruction does degrade at low *y* (corresponding to large *x* and low $Q^2$), as explained in section 4.1. This can be seen in the significant reduction in event purity in this region of the *x-$Q^2$* plane.

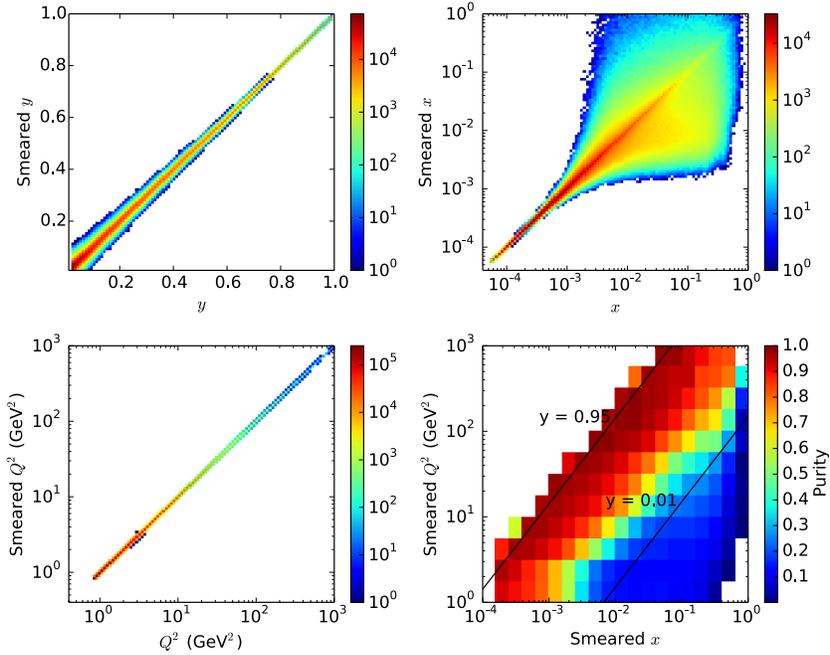

Figure 4-11: The correlation between smeared and true *y, x* and $Q^2$ (top to bottom left), and the resulting bin-by-bin event purity in the *x-$Q^2$* plane (bottom right), reconstructed using the electron method. Purity is defined as (Ngen - Nout) / (Ngen - Nout + Nin), where Ngen, out, in are the number of events generated in a bin, smeared out of it, and smeared into it from other bins, respectively.

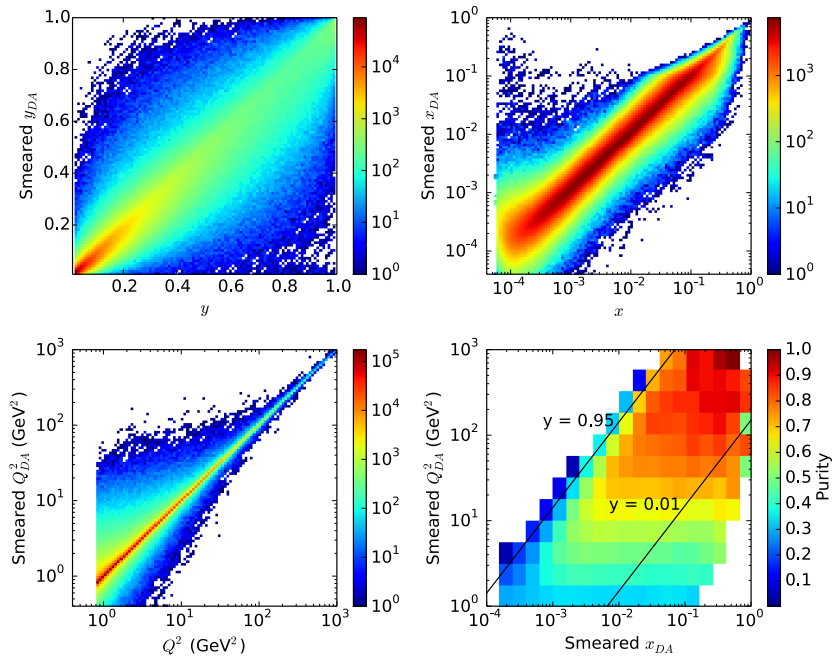

Figure 4-12: The correlation between smeared and true *y, x* and $Q^2$ (top, bottom left), and the resulting bin-by-bin event purity in the *x-$Q^2$* plane (bottom right), reconstructed using the DA method

As also explained in section 4.1 it is possible to overcome this degradation of resolution at low *y* by using hadronic information. Figure 4-12 shows the results of kinematic reconstruction using the "double-angle" (DA) method. This utilizes information from both the electron and the hadronic final state in the calculation of kinematic variables, and does not suffer the same degradation as the electron method at low *y*. This means it can be used in place of the electron method in this region. However, it is important to note that the DA method does not appear suitable as a general replacement for the electron method, as the resultant event purity is not as good as that attainable with the pure electron method at moderate-to-high *y*.

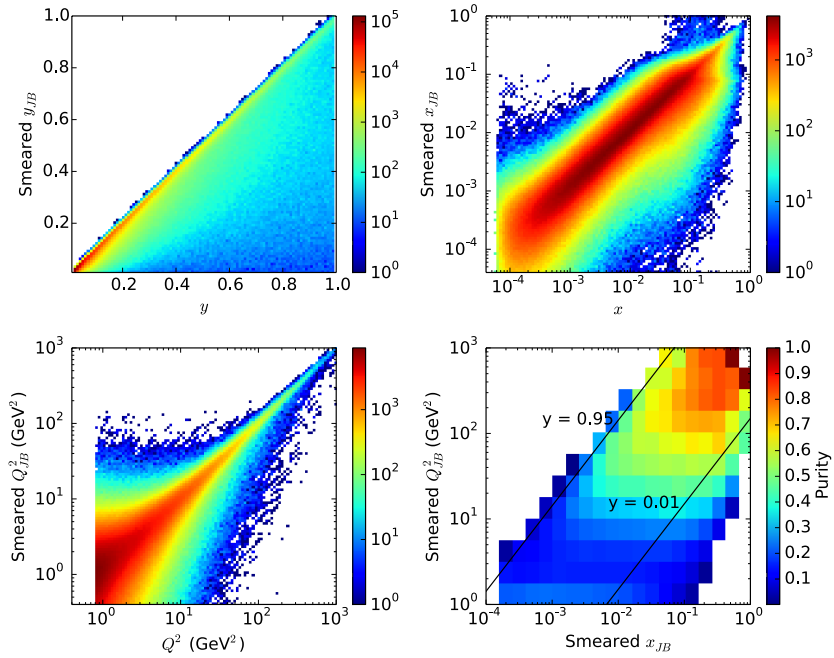

Figure 4-13: The correlation between smeared and true y, x and Q2 (top, bottom left), and the resulting bin-by-bin event purity in the x-Q2 plane (bottom right), reconstructed using the JB method.

Finally Figure 4-13 shows the resolution attainable with the Jacquet-Blondel (JB) method. This is a purely hadronic method of kinematic calculation, meaning it can be used in the absence of a measured scattered lepton. A drawback of this method is that it suffers from very poor resolution at low *y*. However, for charged current (CC) events, in which the scattered lepton is a neutrino, it is the only means of kinematic calculation available. Fortunately, as the majority of the CC cross-section resides at large $Q^2 > 100$, the JB method can be very successfully applied to the analysis of these events [7].

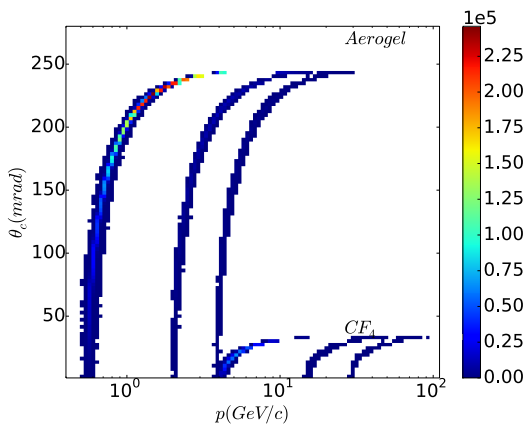

Figure 4-14: Hadron separation with BeAST momentum resolution, for aerogel and CF4 radiators. The high resolution allows clear separation of the pion (left), kaon (center) and proton (right) bands.

Identifying hadron species is key to meeting the physics aims of the SIDIS program. Figure 4-14 shows the separation of charged pions, kaons and protons by Cerenkov angle as a function of hadron momentum in the BeAST detector tracking acceptance. Aerogel (index of refraction, $n$ = 1.0304) and $CF_4$ ($n$ = 1.000558) are used as radiators. Momentum values are smeared according to the aforementioned prescription. The excellent momentum resolution allows a clear separation of species over a wide momentum range, which depends on the radiator material chosen. Use of a $CF_4$ radiator will permit hadron identification up to ~60 GeV/c. Note that these figures do not apply smearing due to uncertainties in the determination of the Cerenkov angle. Hence Figure 4-14 should be viewed as an upper limit on performance. The preliminary version of RICH detectors shown in Figure 4-7 is imported from the CbmRoot distribution [134].

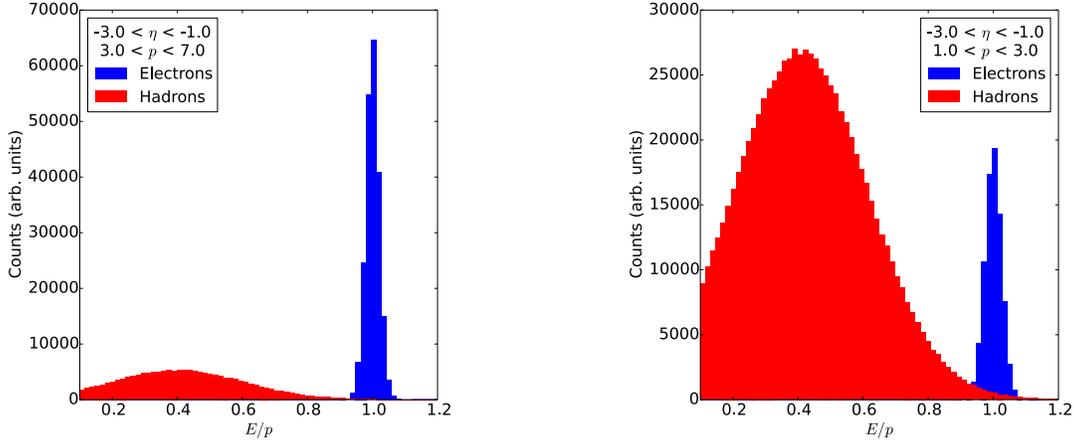

Figure 4-15: Electron-hadron separation by *E/p* in the region of the BeAST detector spanned by high-resolution tracking and electromagnetic calorimetry. In both momentum ranges the electron *E/p* distribution is sharply peaked around one, and is well separated from the broad hadron distribution, allowing clear separation of the two.

Figure 4-15 shows the ability of the BeAST detector to perform electron-hadron separation using tracking and electromagnetic calorimetry in the electron-going direction. Even in this direction, the particle yield is dominated at some momenta by hadrons. It is important to be able to separate the electron, on which we rely for kinematic calculations, from these hadrons. Momentum and electron energy smearing is as described above. The hadronic energy response is smeared to yield a Gaussian distribution, peaked at 40% of the hadron energy, with a sigma of 20% of the hadron energy. As can be seen in the figure, due to the excellent resolution in both momentum and energy for the electrons, the electron E/p distribution is sharply peaked around one, and very well separated from the hadron distribution. This gives us confidence that the BeAST design will be able to perform very effective electron-hadron separation in this region.

## 4.2.2 ePHENIX

The PHENIX Collaboration has proposed to build an eRHIC detector, here referred to as ePHENIX, upon sPHENIX [135], which is designed to further advance the study of cold and hot nuclear matter in nuclear collisions, with its main emphasis on jet measurements. A full engineering rendering of the ePHENIX detector – showing how ePHENIX builds upon sPHENIX – is shown in Figure 4-16[3].

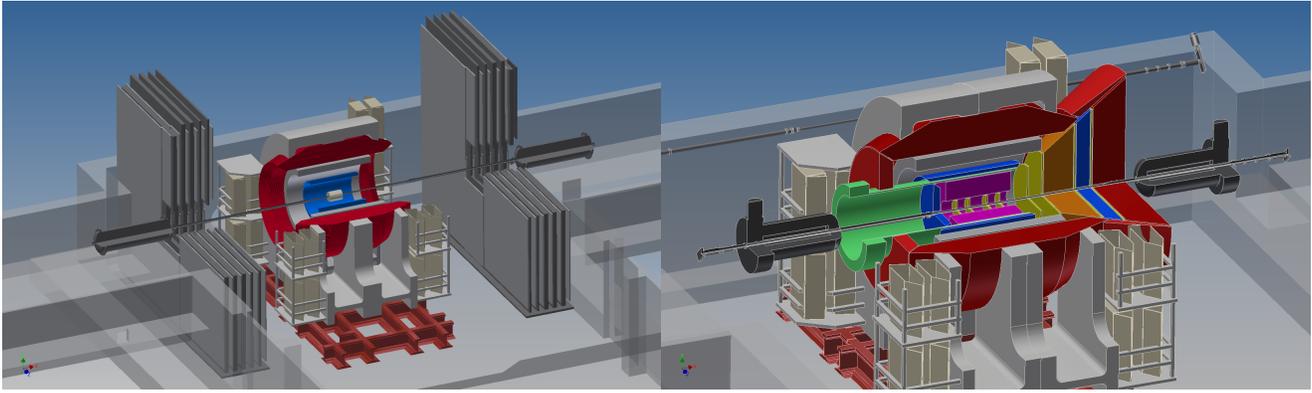

Figure 4-16: The evolution of the sPHENIX detector, with its focus on jets and hard probes in heavy-ion collisions, into ePHENIX, with additional capabilities supporting its focus on e+p and e+A collisions. (left) The sPHENIX detector in the existing PHENIX experimental hall. (right) The ePHENIX detector, in the same hall, showing the reuse of the superconducting solenoid and the electromagnetic and hadronic calorimeter system. The eRHIC focusing quadrupoles, each located 4.5 m from the interaction point, and the height of the beam pipe above the concrete floor, set the dominant physical constraints on the allowable dimensions of ePHENIX.

In addition to fully utilizing the sPHENIX superconducting solenoid and barrel calorimetry, ePHENIX adds new detectors in the barrel, electron-going and hadron-going directions [136], Figure 4-17. In the electron-going direction, a crystal calorimeter is added for electron identification and precision resolution. A compact time projection chamber, augmented by additional forward and backward angle GEM detectors, provides full tracking coverage. In the hadron-going direction, behind the tracking is electromagnetic and hadronic calorimetry. Critical particle identification capabilities are incorporated via a barrel DIRC, and in the hadron-going direction, a gas RICH and an aerogel RICH.

The sPHENIX detector concept reuses the BaBar superconducting solenoid to provide a 1.5 Tesla longitudinal tracking magnetic field. Its field is shaped in the forward directions with an updated yoke design in the ePHENIX detector. The BaBar solenoid has higher current density at both ends and its length of ±1.9m provides a long path for magnetic bending. It is therefore expected to give good analyzing power even for high momentum charged tracks in the hadron-going direction. The ePHENIX tracking system utilizes a combination of GEM and TPC trackers to cover the pseudorapidity range of $-3 < \eta < 4$. The momentum resolution for the full device is summarized in Figure 4-18 It will be provided by TPC position resolution of $r\Delta\varphi$=300μm with 65 readout rows, and GEM resolutions of 100μm and 50μm for outer and inner tracking regions, with minimal material along the particle path.

The ePHENIX detector will have full electromagnetic calorimeter coverage over $-4<\eta<4$. The sPHENIX barrel electromagnetic calorimeters will be used in ePHENIX, covering $-1<\eta<1$ with an energy resolution of *~12%/√E*. In addition, crystal (with energy resolution of *~1.5%/√E*) and lead-scintillator electromagnetic calorimeter (with energy resolution of *~12%/√E*) are planned for the electron-going and hadron-going direc-tion, respectively. With such coverage, it will provide excellent photon and electron measurements in exclusive processes and determination of DIS kinematics with scattered electron measurements. For the latter, the emphasis is put on the electron-going direction covered by high precision crystal calorimeter and giving access to lower x kinematics.

---

[3] The sPHENIX design is continuously developing with careful consideration of its future serving as a basis for an EIC detector, ePHENIX; the presented here sPHENIX design represent its status from February 2014.

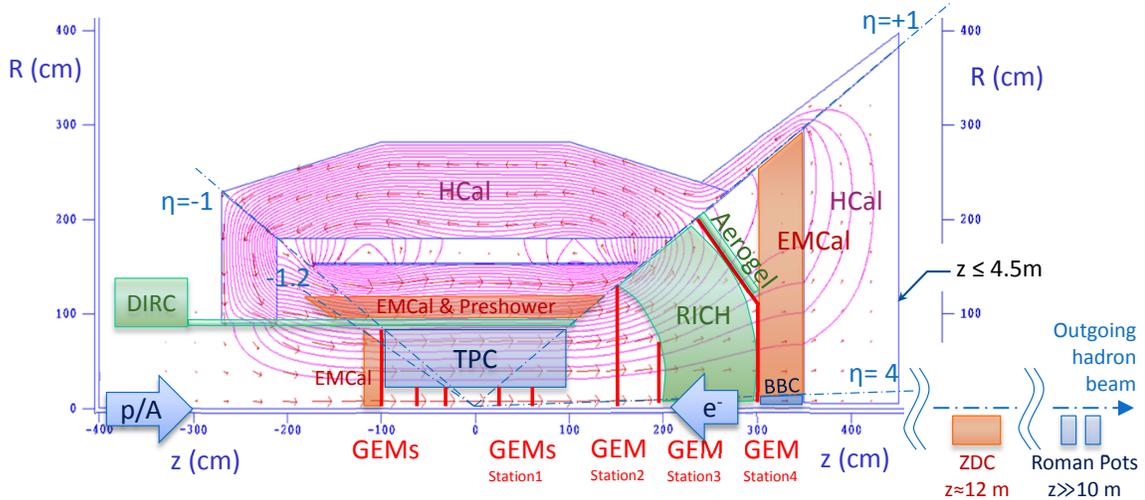

Figure 4-17: A cross section through the top-half of the ePHENIX detector concept, showing the location of the superconducting solenoid, the barrel calorimeter system, the EMCal in the electron-going direction and the system of tracking, particle identification detectors and calorimeters in the hadron-going direction. Forward detectors are also shown along the outgoing hadron beamline. The magenta curves are contour lines of magnetic field potential as determined using the 2D magnetic field solver, POISSON.

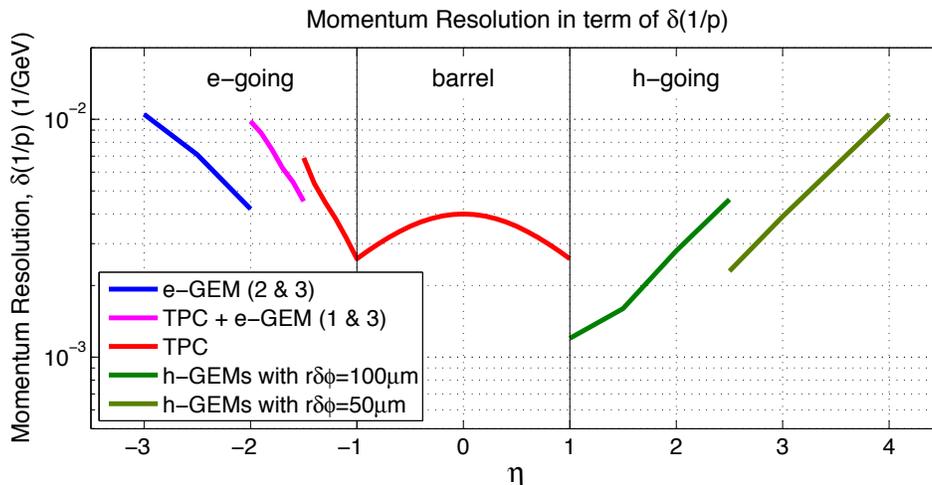

Figure 4-18: Momentum resolution over the full pseudorapidity coverage of the planned tracking system in the high momentum limit. Multiple scattering contribution to the relative momentum resolution (not shown on the plot) was studied with GEANT4 simulation, and found to vary from below 1% at low pseudorapidity to ~3% at $|\eta|=3$.

The different response of the EMCal to hadrons and electrons, along with a direct comparison of energy deposited in the EMCal and momentum measured in the tracking system (i.e., $E/p$ matching) provides a significant suppression of hadronic background in DIS scattered electron measurements: from a factor of 20–30 at momenta near 1 GeV/c to a factor of greater than 100 for momenta above 3 GeV/c. Further enhanced electron identification is expected from the use of the transverse shower profile. These will provide high purity for DIS scattered electron measurements at momenta

>2–3 GeV/c when colliding 10–15 GeV electron beam with 250 GeV proton beam. This only marginally limits the (x, $Q^2$) space probed in our measurements (effectively limiting $y<0.8$ at low $Q^2$).

The precision for the DIS kinematics reconstruction is defined by detector acceptance and resolutions. While scattered electron measurements provide good precision for $Q^2$ determination for almost the entire kinematical space, the precision for $y$ (and hence for $x$) deteriorates as $1/y$. Oppositely, measurements of hadron activity (Jacquet-Blondel method) will provide good measurement for $y$ and poorer for $Q^2$, particularly at lower $Q^2$, see Figure 4-19-middle. Combining these two methods is expected to give good resolutions for the whole $(x,Q^2)$ space, see Figure 4-19-right.

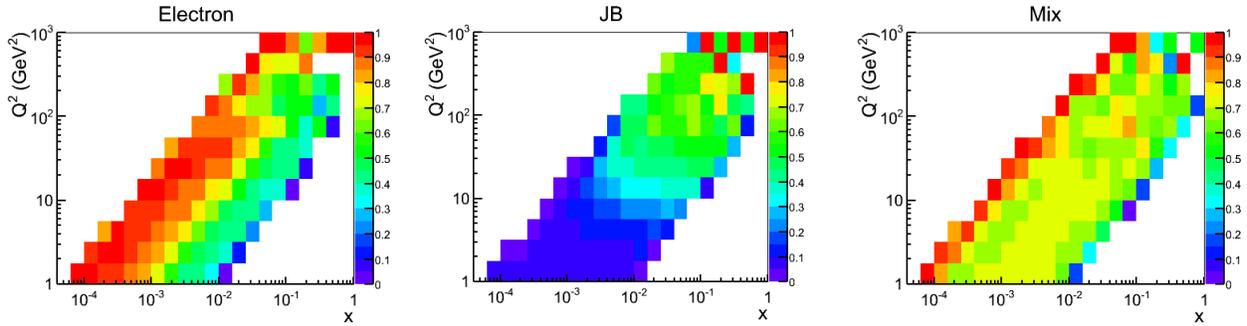

Figure 4-19: For 15 GeV × 250 GeV beam energy configuration, event purity in (x,$Q^2$) bins, defined by the likelihood of an event to remain in its true (x,$Q^2$) bin after resolutions smearing; left – for electron method, middle – for Jacquet-Blondel method, and right – for "Mixed" method, when $Q^2$ is defined from electron method, y is defined from Jacquet-Blondel method, and x=$Q^2$/(sy).

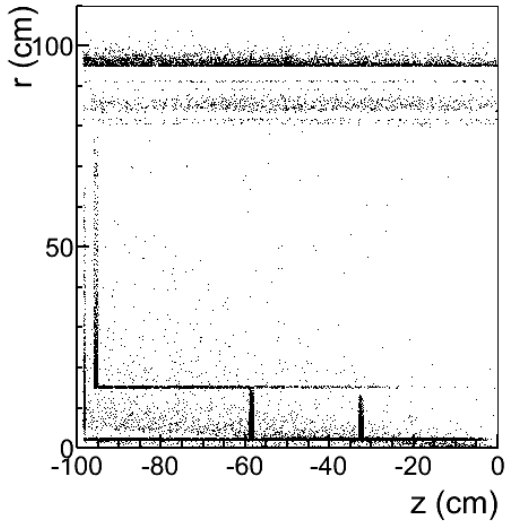

Figure 4-20: Obtained from ePHENIX GEANT4 simulation, the distribution of the vertices (*r,z*) for Bremsstrahlung photon radiation from the scattered electron detected in backward EMCal located at *z*=-100cm; collision point corresponds to (*r,z*)=(0,0). The main sources are beam pipe at *r*=2cm, GEM stations at *z*=-32, -58 and -98cm, TPC frame (*r*=15-80cm and *z*=-95cm), and DIRC and barrel EMCal behind the TPC at *r*>85cm (see also ePHENIX schematic view in Figure 4-15).

Bremsstrahlung photon radiation and photon conversion in the material between the collision point and EMCal will affect the scattered electron reconstruction. Figure 4-20 shows the main sources for Bremsstrahlung photon radiation from the scattered electron detected in crystal EMCal (in electron-going direction), the 1mm thick beryllium beam pipe being the dominant contributor. Its effect on the $(x,Q^2)$ resolution with electron method was evaluated with ePHENIX GEANT4 simulation and found to be negligible for $y<0.5$, and leading to 3-7% decrease of true event purity in $(x,Q^2)$ bins corresponding $y$=0.5-0.95. Using the same simulation framework it was found that the contribution of electrons from photon conversion is negligible (<1%) at momenta >3-5 GeV/c when colliding a 10-15 GeV electron beam with a 250 GeV proton beam. At lower momenta the electron-positron pairs will be well identified by our tracking system in the magnetic field and additionally suppressed by *E/p* matching cut. As was summarized in the beginning of the section, ePHENIX will have three PID systems: (1) a DIRC covering $|\eta|$ <1 providing π-K separation below 3.5–4 GeV/c (depending on purity and efficiency requirements), (2) an proximity-focused aerogel-based RICH covering $1 < \eta < 2$ providing π-K (K-p) separation below 6 (10) GeV/c and (3) a gas-based RICH covering $1 < \eta < 4$ providing π-K separation for $3 < p < 50$ GeV/c and K − p separation for $15 < p < 60$ GeV/c (depending on the balance between efficiency and purity chosen). These three detectors cover a majority of the kinematics of interest at eRHIC.

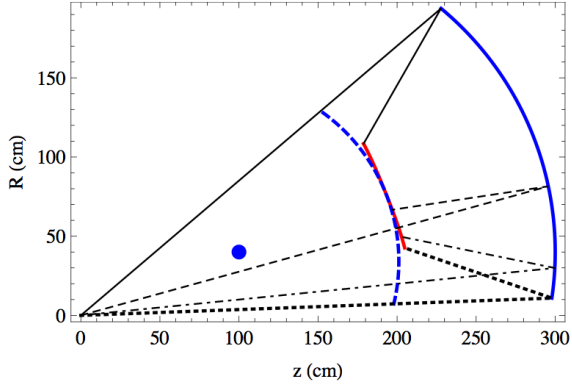

Figure 4-21: The cross-section of the gas-based RICH detector in the *r-z* plane that crosses the mirror center. The interaction point is centered at (0, 0). The geometric center of the mirror is shown as the blue dot at (*r, z*) = (40 cm, 100 cm). The mirror and RICH entrance window are shown by the solid and dashed blue curves, respectively. Several example tracks and the central axis of their Cerenkov light cone are illustrated by the black lines. The Cerenkov photons are reflected by the mirror to the focal plane, shown in red.

Figure 4-21 shows a design of the gas-based RICH detector. Due to the limited space constraints required by the IR design, it is not possible to reflect the light outside of the acceptance, and so any photon readout must be in the path of the particles. The gas-based RICH uses $CF_4$ as a Cerenkov radiator, with the Cerenkov photons focused to an approximately flat focal plane using spherical mirrors of 2 m radius. The photon detector consists of CsI-coated GEM detectors placed at the focal plane (red line in Figure 4-21). This design is currently funded as an EIC R&D project. Figure 4-22 illustrates the ePHENIX PID capabilities in the most challenging very forward direction at $\eta$=4. The combined information from tracking system and energy deposit in HCal helps to improve momentum resolution for high momentum tracks particularly at higher rapidities, where momentum resolution from tracking degrades. It is notable that the limitation on the mass resolution comes from the estimated 2.5% radius resolution per photon for the RICH from the EIC R&D RICH group. Our calculation includes also the smearing effect from the residual magnetic field in RICH volume. The effect is minimized by filed design that ensures that field component is mostly parallel to the track inside RICH. Another source for Cerenkov ring smearing are tracks originating from an off-center vertex, leading to focal plane offset from nominal position. The effect was found sub-dominant in the proposed RICH design, for the vertex distributed with Gaussian $\sigma_Z$=40cm around nominal collision point at *z*=0.

ePHENIX EMCal and tracking coverage up to $\eta$=4 along with HCal coverage up to $\eta$=5 will also provide excellent capabilities for detecting diffractive events through the rapidity gap method.

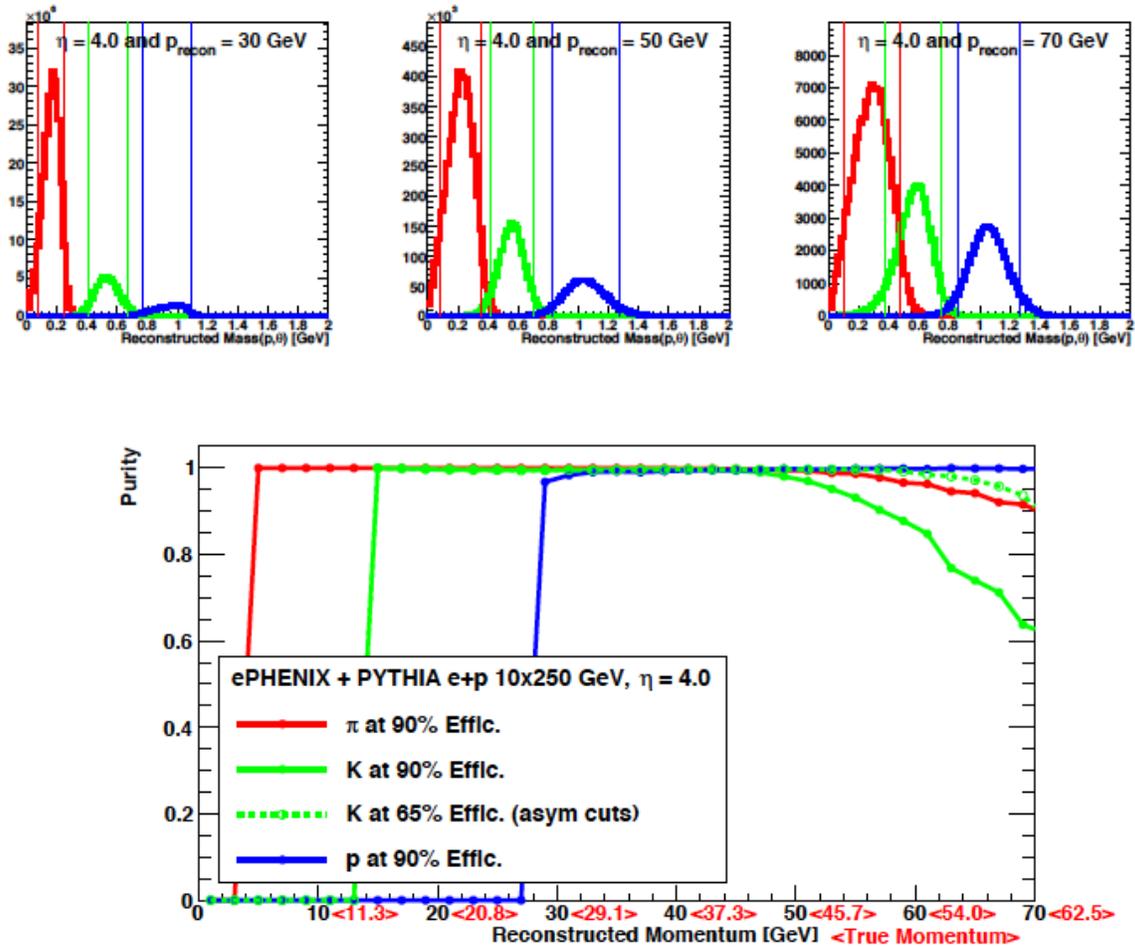

Figure 4-22: Top: Reconstructed mass distribution via m(p,θCrk) at η=4 for reconstructed momenta 30 GeV/c (left), 50 GeV/c (middle) and 70 GeV/c (right), for pions (red), kaons (green) and protons (blue), with the parent momentum and particle abundances from the PYTHIA generator. Vertical lines indicate the symmetric mass cuts corresponding to 90% efficiency. Note that particle true momentum is on the average smaller than reconstructed momentum (see bottom plot). Bottom: π, K, p purities at pseudorapidity 4.0 as a function of reconstructed momentum, based on symmetric cut on reconstructed mass corresponding to 90% efficiency (solid lines), and asymmetric cut with stricter selection on the kaons with efficiency 65% (dashed line); Also indicated in angle brackets are the values of the average true momentum at each reconstructed momentum, which are different due to momentum smearing and sharply falling momentum spectra.

### 4.2.3 eSTAR

The STAR collaboration has proposed a path to evolve the existing STAR detector [137] to an initial-stage eRHIC detector, here referred to as eSTAR. In this plan an optimized suite of detector upgrades will maintain and extend the existing low-mass mid-central rapidity tracking and particle-identification capabilities towards more forward rapidities in both the electron and hadron going beam directions. This plan is described in [138], which contains also a capability assessment for key measurements of the eRHIC science program.

Figure 4-23 shows a side-view of the baseline eSTAR detector layout. This baseline plan consists of three essential upgrade projects, namely endcap Time-of-Flight walls located between the TPC and the magnet pole-tips on the East and West sides of the interaction region (ETOF and WTOF, covering the regions $1<|\eta|<2$ in pseudo-rapidity), a GEM-based Transition Radiation Detector (GTRD) between the TPC and ETOF in the forward electron direction, covering $-2<\eta<-1$, and a Crystal ElectroMagnetic Calorimeter with pre-shower (CEMC, covering $-4 < \eta < -2$). Furthermore, eSTAR will rely on a replacement upgrade of the Inner Sectors of the existing Time-Projection-Chamber prior to a completion of the RHIC Beam-Energy Scan program with $A + A$ collisions and on a subsequent upgrade in the form of a new Forward Calorimeter System (FCS) with associated Forward Tracking System (FTS) on the West side of STAR. The upgrade sequence will enable STAR to complete the high-priority experimental programs with $A + A$, polarized $p + p$, and $p + A$ collisions outlined in the STAR decadal plan [139]. In the side-view presented here, the FCS is closer to the interaction region than in the decadal plan and eSTAR LoI [138] to ensure compatibility with the eRHIC interaction region design. At the time of the writing of this document, a structural analysis of the support and floor remains to be completed.

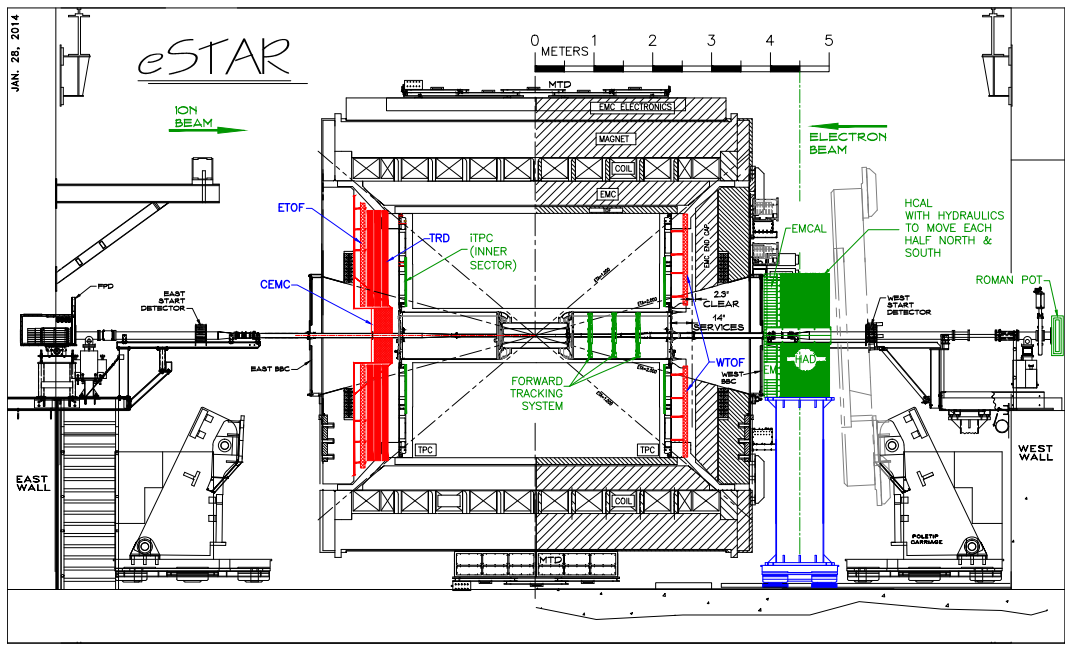

Figure 4-23: The eSTAR layout with the proposed upgrades of iTPC, Forward Calorimetry System (FCS), the Forward Tracking System (FTS), Endcap TOF (E/W TOF), BSO Crystal Calorimeter (CEMC), GEM based TRD. In this configuration, the electron beam is from right to left (eastward) while hadron beam from left to right (westward).

The existing STAR mid-rapidity acceptance and particle identification capabilities, paired with the FCS and FTS aimed at high (total) energies, form its key strengths into the eRHIC era. The STAR mid-rapidity region with the upgraded TPC, MTD, existing BEMC, and TOF is relatively well matched to the demands of inclusive and semi-inclusive deep-inelastic scattering measurements at hard scales $Q^2 > 10$ GeV$^2$. The extension of this coverage to smaller $x$ and $Q^2$ requires forward instrumentation in the electron going direction, in particular to identify and measure the forward scattered electron with good efficiency, purity, and resolution. Coverage over the region $-4 < \eta < -1$ (on the east end of STAR, opposite to the EEMC) is provided with the CEMC, a BSO crystal calorimeter, the GTRD and ETOF. Together, they expand the $Q^2$ range of inclusive and semi-inclusive measurements accessible to STAR to cover most of the conventional deep-inelastic regime, $Q^2 > 1$ GeV$^2$, including the region of low-$x$ that is of particular scientific interest.

STAR has demonstrated hadron rejection capability at a level better than $10^5$ at $p_T = 2$ GeV at mid-rapidity. The proposed upgrade of iTPC further improves the hadron rejection by more than an order of magnitude at mid-rapidity. In addition, GTRD and ETOF upgrades are proposed to achieve the necessary electron identification in the pseudo-rapidity range ($-2 < \eta < -1$) in the DIS kinematics of the scattered electrons essential to the eSTAR physics program. Going even more forward in the electron scattered direction, the requirements for hadron rejection become less stringent. However, the requirement of precise measurement of electron kinematics and the need to reject photon conversion and misidentification as an electron become increasingly demanding. Although the $h/e$ ratio is about 1000 at mid-rapidity, hadron contamination is negligible and concentrated in a limited momentum range at very forward rapidity. However, the photon becomes the major source of background. We have proposed to leave only the beam pipe and its necessary support structure along this direction, and to install a crystal calorimeter and preshower to precisely determine the electron energy and angle with minimum material along the electron path. In this rapidity range, the crystal calorimeter material along the beam direction amounts to about 25 radiation length and less than one interaction length. An adequate hadron rejection factor is likely to be achievable by a combination of pre-shower hit position, energy deposition and shower lateral distribution in CEMC and shower leakage detected by existing Beam-Beam Counter (BBC) behind the CEMC.

| Coverage | Orientation | Tracking | EMC | HCAL | Resolution (momentum or energy) |
|---|---|---|---|---|---|
| $-4<\eta<-2$ | Electron Beam direction; EAST | | BSO | | $\sigma_E/E = 2\%/\sqrt{E} \oplus 0.75\%$ |
| $-2<\eta<-1$ | | iTPC+GTRD+ETOF | | | $\sigma_p/p = 1/(p_T/p_Z - 1/6) \times (0.45\% p_T \oplus 0.3\%)$ $\oplus (p_Z/p_T) \times 0.2\%/p/\beta$ |
| $-1<\eta<1$ | Mid-Rapidity | TPC+TOF | SMD+EMC | | $\sigma_E/E = 14\%/\sqrt{E} \oplus 2\%$ $\sigma_p/p = 0.45\% p_T \oplus 0.3\% \oplus 0.2\%/p/\beta$ |
| $1<\eta<1.7$ | Hadron Beam direction; WEST | iTPC+TOF | | | $\sigma_p/p = 1/(p_T/p_Z - 1/4) \times (0.45\% p_T \oplus 0.3\%)$ $\oplus (p_Z/p_T) \times 0.2\%/p/\beta$ |
| $1<\eta<2$ | | iTPC+FTS | SMD+EMC | | $\sigma_E/E = 16\%/\sqrt{E} \oplus 2\%$ |
| $2.5<\eta<5$ | | FTS | W-fiber EMC | HCAL | $\sigma_E/E = 12\%/\sqrt{E} \oplus 1.4\%$ $\sigma_E/E = 38\%/\sqrt{E} \oplus 3\%$ |

Table 4-1: eSTAR detector subsystems coverage and resolution.

Table 4-1 lists the detector subsystems in different pseudorapidity and their energy (momentum) resolutions. The resolutions of existing detector subsystem are obtained from the actual performance while the resolutions of proposed detectors are based on simulation and prototype test results. The performance of the existing detectors has been reported in multiple STAR publications, including for example Refs. [137,139]. An assessment of eSTAR resolution and event purity is shown in Figure 4-24 for the electron method. The corresponding results for the hadron method will be shown in a future update.

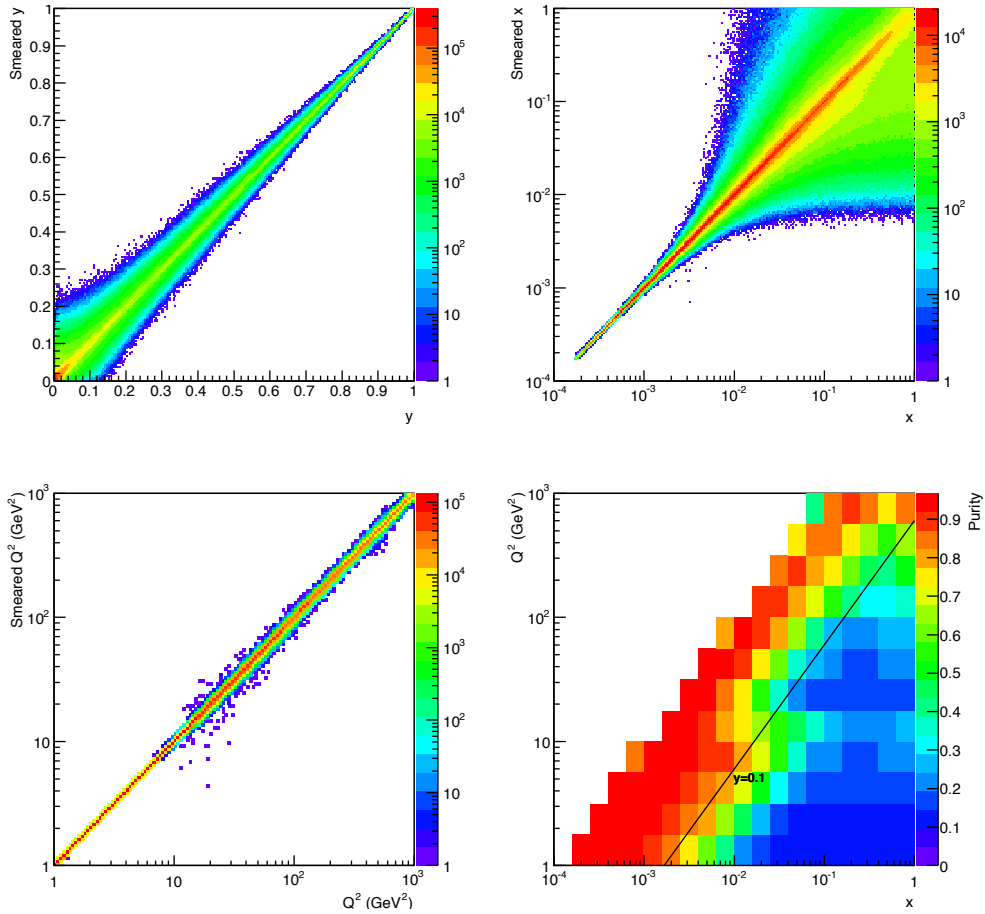

Figure 4-24: The correlation between smeared and true $y$, $x$, and $Q^2$ (top left and right, and bottom left) and event purity in the $(x,Q^2)$ plane (bottom right), as reconstructed using the electron method. Purity is defined as defined as (Ngen - Nout) / (Ngen - Nout + Nin), where Ngen, out, in are the number of events generated in a bin, smeared out of it, and smeared into it from other bins, respectively. The collision system is a 15 GeV electron beam and a 100 GeV hadron beam.

## 4.2.4 Zero-degree Calorimeter, Low angle hadron and lepton tagger designs

To achieve the physics program as described in earlier sections, it is extremely important to integrate the detector design into the interaction region design of the collider. Particularly challenging is the detection of forward-going scattered protons from exclusive reactions, as well as of decay neutrons from the breakup of heavy ions in non-diffractive reactions. In general, for exclusive reactions, one wishes to map the four-momentum transfer (or Mandelstam variable) $t$ of the hadronic system, and then obtain an image by a Fourier transform, for $t$ close to its kinematic limit up to about 1-2 GeV. One of the most challenging constraints for the interaction region design from exclusive reactions is the need to detect the full hadronic final state.

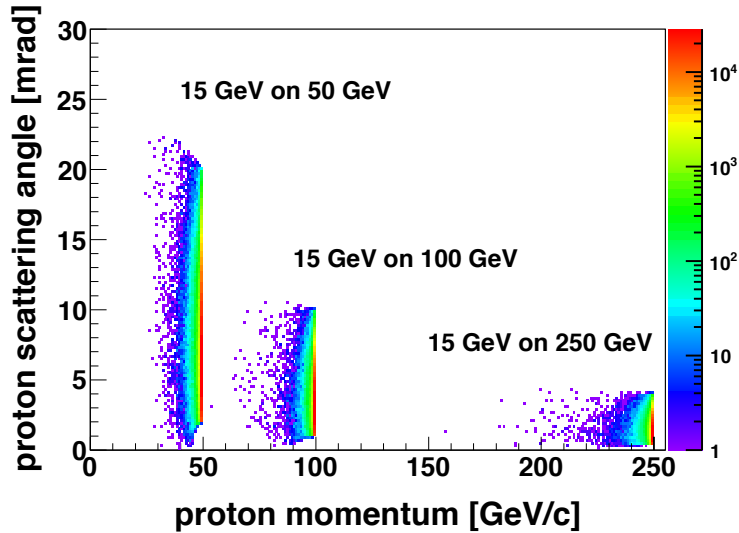

Figure 4-25: The scattered proton momentum vs. scattering angle in the laboratory frame for DVCS events with different beam energy combinations. The following cuts have been applied: 1 GeV$^2$ < $Q^2$ <100 GeV$^2$, 0.01 < $y$ < 0.85, $10^{-5}$<$x$<0.5 and 0.01 < $t$ < 1 GeV$^2$. The angle of the recoiling hadronic system is directly and inversely correlated with the proton energy. It thus decreases with increasing proton energy.

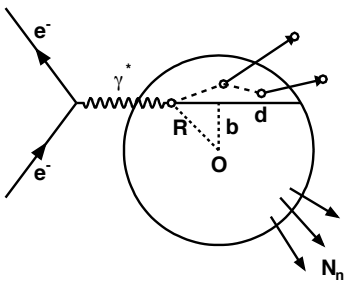 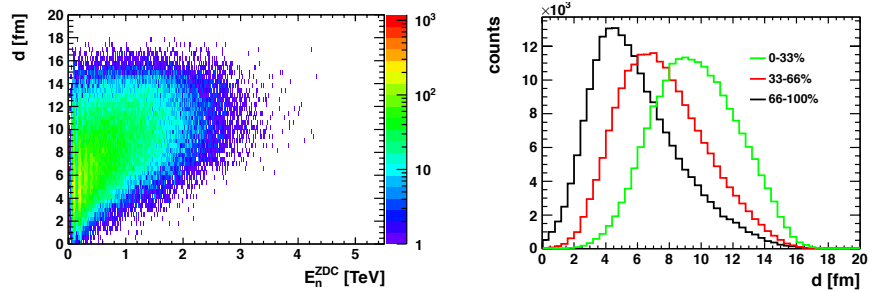

Figure 4-26: Relevant quantities to describe the collision geometry. $b$ represents the impact parameter. $R$ shows the spatial displacement of the interaction point to the center of the nucleus. $d$ is the traveling length, which defines the pro-jected virtual photon traveling length from the interaction point to the edge of the nuclear medium.

Figure 4-27: (left) Correlation between traveling length $d$ and energy deposition in the ZDC. All the forward neutrons can be detected in the ZDC. (right) Traveling length distribution in different forward neutron energy bins. The black line corresponds to peripheral collisions (66-100%), while the red and green lines correspond to the 33-66% and 0-33%, respectively.

Figure 4-25 shows the correlation between proton scattering angle and its momentum, and illustrates that the remaining baryonic states go in the very forward ion direction. Even at a proton energy of 50 GeV, the proton scattering angles only range to about 25 mrad. At proton energies of 250 GeV, this number is reduced to one/fifth. In all cases, the scattering angles are small. Because of this, the detection of these protons is extremely dependent on the exact interaction region design.

The only possible way ensuring exclusivity for lepton-nucleus collisions for heavy nuclei is to veto the nuclear break up. This is realized by requiring no decay-neutrons in the zero-degree calorimeter. In SIDIS, collision geometries in e+A (See Figure 4-26) can be determined by utilizing the ZDC. The number of forward neutrons produced and detected in the ZDC is expected to be sensitive to the path length of the parton and fragmentation of the colliding nucleon along the virtual photon direction in the nucleus. See Figure 4-27 for the correlation between the forward neutrons and the distance, details are described in [140].

The geometry information is an additional and useful gauge for investigating properties of partonic interactions in nuclei. While the impact parameter $b$ has a correlation with the number of the neutrons in the ZDC, the most "central" collision in e+A ($b\sim 0$) can be identified from the events with the highest neutron multiplicity since the longest path length of the nucleon fragmentation in the nucleus is expected to be at $b = 0$. This will be an effective use of selecting events with maximized nuclear effects in SIDIS e+A collisions such as for the di-hadron correlation studies. The eRHIC design features a 10 mrad crossing angle between the protons or heavy ions during collisions with electrons. This choice removes potential problems for the detector induced by synchrotron radiation. To obtain the eRHIC luminosities strong focusing close to the IR is required to have the smallest beam sizes at the interaction point. A small beam size is only possible if the beam emittance is also very small. The focusing triplets are 4.5 meters away from the interaction point (IP). To accomplish a small emittance for the ion and proton beams, the beams need to be cooled. The eRHIC interaction region design relies on the existence of small emittance beams with a longitudinal RMS of ~5 cm, resulting in a $\beta^*= 5$ cm, details about the IR layout can be found in Section 3.3.11. To ensure the previously described requirements from physics are met, four major requirements need to be fulfilled: high luminosity (> 100 times that of HERA); the ability to detect neutrons; measurement of the scattered proton from exclusive reactions (i.e. DVCS), and the detection of spectator protons from deuterium and He-3 breakup. The eRHIC IR design fulfills all these requirements for the outgoing proton beam direction. The apertures of the interaction region magnets allow detection of neutrons with a solid angle of ~4 mrad, as well as the scattered proton from exclusive reactions, i.e. DVCS, up to a solid angle of ~9 mrad. The detection of the scattered proton from exclusive reactions is realized by integrating several "Roman Pot" stations into the warm section of the IR. The electrons are transported to the interaction point through the heavy-ion/proton triplets, seeing zero magnetic fields.

There are many eRHIC physics topics beyond what was discussed in the EIC White paper [1], which benefit from tagging the scattered lepton at $Q^2$ values significantly below 1 GeV$^2$. Scattered leptons with a $Q^2 < 0.1$ GeV$^2$ cannot be detected in the man detectors, therefore as in HERA a special low-$Q^2$ tagger is needed. This requires the integration of a dipole in the outgoing lepton beam-line to separate the scattered leptons from the core of the beam. The scattered leptons will be detected in an electromagnetic calorimeter. Such a low-$Q^2$ tagger needs to be well integrated into the IR design and care needs to be taken separate the scattered leptons from leptons from the bremsstrahlung process, which due to its high cross section and the high eRHIC luminosity will be dominant.

To study all these physics driven requirements discussed above are fulfilled by the eRHIC IR design (see Section 3.3.11 for details) the current interaction region design has been implemented into the EicRoot simulation framework. The current design includes the electron beam magnets, the FFAG electron bypass, and the hadron beam magnets (see Figure 4-28). Magnetic fields have been implemented in the electron and hadron beams so that simulations can be done to study the acceptance of particles through the aperture of the magnet yokes, as well as to optimize detector geometry and placement.

In addition to the ongoing effort for the main detector system (see Section 4.2.1), recent simulations have been done to integrate and develop the detectors far down the beam lines to capture protons (a Roman Pot) and electrons (a low $Q^2$ tagger), which scatter at very small angles (<10 mrad). Particles that scatter at such a small angle will go outside the acceptance of the main detectors and so detector subsystems far down the beam line (>10m) are needed.

Development of a low $Q^2$ tagger is underway within the EicRoot framework. This is a simple, small, and inexpensive detector that will consist of two to three tracking layers followed by an electromagnetic calorimeter, which will allow reconstruction of the scattering angle and energy of the electrons of interest. This detector will be placed about 15m from the interaction point. On the other side of the IR, we have the Roman Pot detector to capture the protons that scatter at small angle, placed roughly at 18m from the interaction point. A simple design for this has also been implemented in EicRoot, allowing for detailed acceptance studies to optimize the IR design.

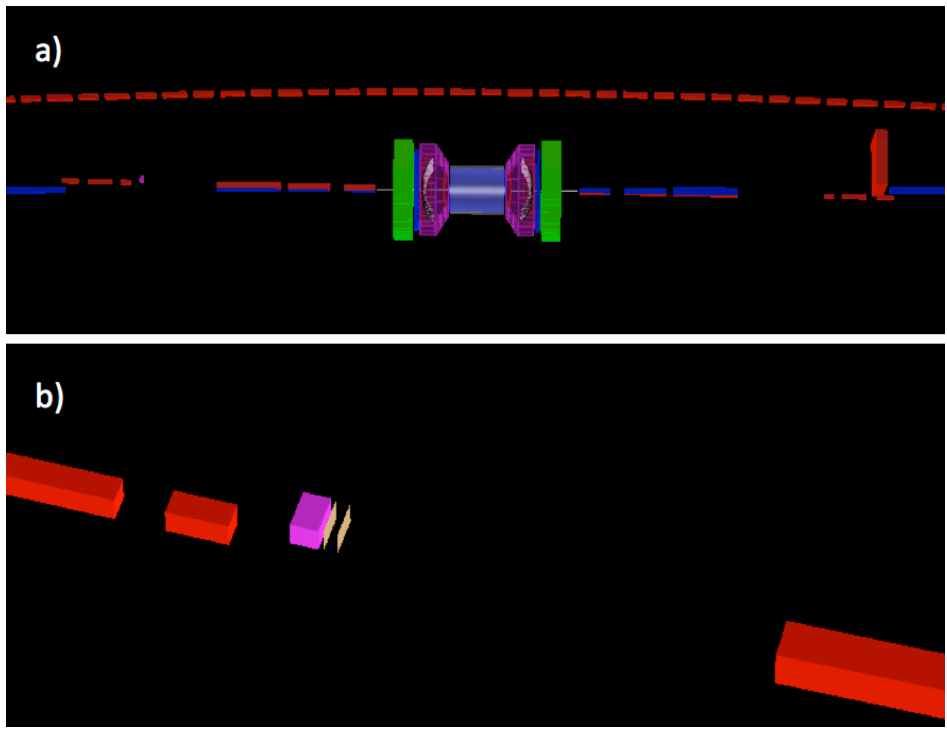

Figure 4-28: Panel a: An event display showing the IR setup in the EicRoot framework. The main detector is seen in the center of the figure. The electron beam (center red), FFAG bypass (top red), and hadron beam (blue) lines are all implemented in the simulation. The low $Q^2$-tagger is shown as the small detector on the left (at about 15m). The placement of the Roman Pot detector is shown by the red block on the right side of the figure (at 18m). Panel b: A zoomed in view of the low $Q^2$-tagger.

## 4.3 Luminosity and Polarization Measurements

The bremsstrahlung process ep → epγ was used successfully for the measurement of luminosity by the HERA collider experiments [141]. It is a precisely known QED cross-section, and has high rates, which allowed negligible statistical uncertainty. Different from HERA, where only the lepton beam was polarized, in eRHIC both the lepton and proton/light ion beams will be polarized. Then the bremsstrahlung rate is sensitive to the polarization dependent term a in the cross section: $\sigma_{brems} = \sigma_0(1 + aP_e P_h)$.

Thus, the polarization ($P_e, P_h$) and luminosity measurements are coupled, and the precision of the luminosity measurement is limited by the precision of the polarization measurement. This also limits the precision of the measurement of double spin asymmetries

$$A_{LL} = \frac{1}{P_e P_p}\left[\frac{N^{--/++} - RN^{+-/-+}}{N^{--/++} + RN^{+-/-+}}\right]$$

through the determination of the relative luminosity $R = L^{++/--}/L^{+-/-+}$.

The straightforward method of measuring bremsstrahlung uses a calorimeter at zero degrees in the lepton direction to count the resulting photons, the distribution of which is strongly peaked in the forward direction. The calorimeter is also exposed to the direct synchrotron radiation fan and must be shielded, degrading the energy resolution. At peak HERA luminosities, the photon calorimeters were hit by 1-2 photons per HERA bunch crossing, at which rate pileup effects were already significant. At eRHIC luminosity of $10^{33}$ cm$^{-2}$ s$^{-1}$ the mean number of photons per bunch crossing is over 20. The distributions are broad, with a mean proportional to the number of photons per bunch crossing. The counting of bremsstrahlung photons thus is effectively an energy measurement in the photon calorimeter, with all of the related systematic uncertainties (e.g. gain stability) of such a measurement.

An alternative method of counting bremsstrahlung photons, used with smashing success by the ZEUS collaboration at HERA, employs a pair spectrometer. A small fraction of photons are converted to e$^+$e$^-$ pairs in the vacuum chamber exit window. A dipole magnet splits the pairs and each particle hits a separate calorimeter adjacent to the unconverted photon direction. This has several advantages over a zero degree photon calorimeter:

1. The calorimeters are outside of the primary synchrotron radiation fan
2. The exit window conversion fraction reduces the overall rate
3. The spectrometer geometry imposes a low energy cutoff in the photon spectrum, which depends on the magnitude of the dipole field and the transverse location of the calorimeters

The variable parameters of the last two points (conversion fraction, dipole field and calorimeter locations) may be chosen to reduce the rate to less than or of order one e+/e- coincidence per bunch crossing even at nominal eRHIC luminosities. Thus counting of bremsstrahlung photons is simply counting of e+/e- coincidences in a pair spectrometer, with only small corrections for pileup effects.

Compton back-scattering is the established method to measure lepton beam polarization in e+p colliders. At HERA, there were two Compton back-scattering polarimeters [142]: one measuring the transverse polarization (TPOL) of the beam through a position asymmetry and one measuring the longitudinal polarization (LPOL) of the beam through an energy asymmetry in Compton back-scattered photons. The TPOL and LPOL systematic uncertainties of RUN-I were 3.5% and 1.6% and Run-II 1.9% and 2.0%, respectively. In spite of the expected high luminosity at eRHIC, these systematic uncertainties should be reduced to ~1%. Unlike the HERA electron synchrotron, each bunch in an eRHIC ERL would pass only once through the interaction region. This requires control and monitoring of bunch-to-bunch fluctuations of both intensity and polarization. For example, the Gatling gun polarized electron source has several cathodes, which may have significant variations among the cathodes. Also, the question arises at which point during the bunches single pass through the ERL to measure polarization. A list of significant challenges to polarization measurements include:

1. Fluctuations in polarization from cathode to cathode in the Gatling gun
2. Fluctuations in bunch current from cathode to cathode
3. Polarization losses from the polarized source to full energy
4. Polarization deterioration during collision
5. Possible polarization profile for the lepton bunches

The current and polarization variations among the cathodes can be straightforwardly addressed by performing both the luminosity and polarization measurements such that information for cathode separately can be extracted. These measurements need to be further divided among the approximately 120 RHIC hadron bunches to monitor fluctuations among bunches. To address the loss of polarization of a bunch passing through the ERL, an ideal solution is to measure the polarization as close to the IP as possible. Both the luminosity and the electron polarization detector design and integration into the machine lattice are part of the EIC R&D activities.

To measure the hadron beam polarization is very difficult as, contrary to the lepton case, there is no process that can be calculated from first principles. Therefore, a two-tier measurement is needed: one providing the absolute polarization, which has low statistical power and a high statistical power measurement, which measures the relative polarization. At RHIC [143], the single spin asymmetry $A_N$ of the elastically scattered polarized proton beam on a polarized hydrogen jet is used to determine the absolute polarization. This measurement provides the average polarization per fill and beam with a statistical uncertainty on the order of ~5% and a systematic uncertainty of 3.2%. High-statistics bunch-by-bunch relative polarization measurements are provided, measuring the single spin asymmetry $A_N$ for scattering the polarized proton beam of a carbon fiber target. To obtain absolute measurements, the pC-measurements are cross-normalized to the absolute polarization measurements from the hydrogen-jet polarimeter. The pC-measurements provide the polarization lifetime and the polarization profile per fill with high statistical precision. The achieved total systematic uncertainty for single spin asymmetries is 3.4%. The systematic uncertainties could be further reduced by monitoring continuously the molecular hydrogen contamination in the jet, improving the operational stability of the carbon fiber targets, and by developing methods to monitor the silicon detector energy calibration at the recoil carbon energy. All are under development for the polarized p+p program at RHIC.

The same measurement concept can be followed for a polarized He-3 beam [144]. The absolute polarization measurement would be provided by the single spin asymmetry $A_N$ of the elastically scattered polarized He-3 beam on a polarized He-3 target. It will be critical to ensure that the scattering was really elastically and that both the beam or target He-3 nucleus stayed intact. This puts additional requirements on the detection system for the scattered He-3 nuclei. To measure the polarization lifetime and profile for each fill as well as the bunch-by-bunch polarization the single spin asymmetry $A_N$ for scattering the polarized He-3 beam of a carbon fiber target can be utilized. The asymmetry is predicted to be a factor of 2 reduced compared to the pC case and has a steeper dependence on the kinetic energy of the scattered carbon nucleus, like for the absolute He-3 polarization measurement it is important to ensure the scattering occurred elastically.

# 5 IMPLEMENTATION: SCHEDULE & COST

The final RHIC campaign is envisioned to be carried out over roughly the next decade. Its purpose will be to complete critical measurements that take full advantage of the versatility of RHIC, covering nearly the entire QCD phase diagram in nucleus-nucleus collisions and exploiting the unique capabilities of RHIC as a polarized hadron collider. This final campaign will be one of definitive measurements with broad capability for new discovery, which, in combination with those at LHC, JLab, and elsewhere will set the stage for an Electron Ion Collider to take the next step toward a complete picture of the evolution of the structure of QCD matter and its properties. The BNL plan includes a smooth transition from the scientific program of RHIC to the first eRHIC experiments with minimum interruption of physics operation between the end of RHIC and the start of eRHIC.

The eRHIC concept is built with the goal of providing a facility capable of addressing the compelling science questions for the next frontier of QCD research at a cost that can realistically fit within the U.S. Nuclear Physics planning for new construction in the coming decade. The design for eRHIC is aiming at a highly cost effective facility. It would provide an electron-beam energy of up to 21.2 GeV, with e-nucleon luminosity of more than $10^{33}$cm$^{-2}$sec$^{-1}$ at 15.9 GeV and one high-luminosity intersection region equipped with crab crossing cavities and a second interaction region upgradable to low $\beta^*$ and available for a second detector.

Given the breadth and scope of the physics program for an Electron Ion Collider, as well as the size and diversity of the interested scientific community, we envision an experimental program with two general-purpose detectors. These can be extremely cost effective, taking full advantage of the existing infrastructure in the experimental halls presently occupied by the PHENIX and STAR detectors. As described in Section 4, the detailed requirements for detectors to carry out the program of an electron-ion collider have been the subject of considerable recent study, both in terms of the physics requirements leading to the "golden measurements" of Ref. [1] and in terms of current instrumentation technology.

At the request of the BNL management, both the PHENIX and STAR collaborations have developed conceptual plans, see Section 4.2.2 and 4.2.3, to assess the feasibility of meeting the requirements of an EIC scientific program, given the machine parameters specified for eRHIC, through upgrades of the existing RHIC detectors. In the case of PHENIX, the starting point is the sPHENIX configuration [135], assumed to be in place at the conclusion of the RHIC program, providing an open-geometry detector using the former BaBar solenoidal magnet. These exercises demonstrate that the large RHIC detectors can form a cost-effective base for detectors capable of initiating an experimental program at eRHIC. There is also a strong interest in the community for developing a purpose-built detector that could be mounted in either of the two presently used experimental halls; its design is described in Section 4.2.1. In practice, the initial detector configuration will be determined through a proposal-driven process, with the outcome very dependent on funding realities and the amount of non-DOE investment.

As noted in Section 3 in order to meet science-driven performance goals within realistic cost constraints the eRHIC design incorporates several challenging technology developments. Foremost among these are the high-energy multi-pass ERL, a high brightness, high current polarized electron source, and coherent electron cooling of the hadron beams. Each of these is being addressed by intense R&D efforts at BNL and elsewhere. In addition a community-wide generic R&D program for EIC detector technologies has been funded at BNL. Over the past three years this peer-reviewed program has made good progress in clarifying detector design issues and beginning to address the needs for instrumentation development that arise from the scientific requirements and machine environment of an EIC.

Based on the expected timelines for technical readiness and for the community and agency approval process for such a project, BNL is planning for a start of eRHIC construction in FY 2019 parallel to operating RHIC, with start of eRHIC operations in ~2025. Table 5-1 shows the current plan for the RHIC operating schedule prior to the final shutdown for the installation of eRHIC. Such a plan, of course, is reviewed annually based on funding projections and evolving physics priorities.

| Years | Beam Species and Energies | Science Goals | New Systems Commissioned |
|---|---|---|---|
| 2014 | 15 GeV Au+Au<br>200 GeV Au+Au<br>$^3$He+Au at 200 GeV | Heavy flavor flow, energy loss, thermalization, etc.<br>Quarkonium studies<br>QCD critical point search | Electron lenses<br>56 MHz SRF<br>STAR HFT<br>STAR MTD |
| 2015-16 | Pol. p+p at 200 GeV<br>p+Au, p+Si at 200 GeV<br>High statistics Au+Au<br>Pol. p+p at 510 GeV?<br>Au+Au at 62 GeV? | Extract $\eta/s(T)$ + constrain initial quantum fluctuations<br>More heavy flavor studies<br>Sphaleron tests<br>Transverse spin physics | PHENIX MPC-EX<br>Coherent e-cooling test |
| 2017 | No Run | | Low energy e-cooling upgrade |
| 2018-19 | 5-20 GeV Au+Au (BES-2) | Search for QCD critical point and onset of deconfinement | STAR ITPC upgrade<br>Partial commissioning of sPHENIX (in 2019) |
| 2020 | No Run | | Complete sPHENIX installation<br>STAR forward upgrades |
| 2021-22 | 200 GeV Au+Au with upgraded detectors<br>Pol. p+p, p+Au at 200 GeV | Jet, di-jet, γ-jet probes of parton transport and energy loss mechanism<br>Color screening for different quarkonia | sPHENIX |
| 2023-24 | No Runs | | Transition to eRHIC |

Table 5-1: A long-term view of the RHIC operations schedule leading to a transition to eRHIC.

As this table indicates, the plan for completing the RHIC science program includes gap years during which the RHIC beams and experiments are not operating. The first of these is in 2017 when the electron cooling system for low-energy ion beams will be installed, prior to the scheduled Beam energy Scan-II runs in 2018-2019. Another gap year is 2020, corresponding to the completion of the installation of sPHENIX and possible forward upgrades of STAR. In this plan, RHIC operations cease after FY 2022, with two gap years prior to start-up of eRHIC operations in 2025.

The years when RHIC is not operating provide extended shut-down periods needed for major equipment installation. They also provide opportunities for re-directing significant amounts of workforce and other resources from RHIC operations toward offsetting the cost of eRHIC construction, including new detector equipment. Optimizing the amount of this cost offsetting is a key element of the strategy for a cost-effective transition from RHIC to eRHIC.

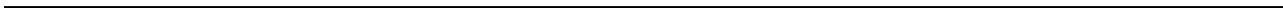

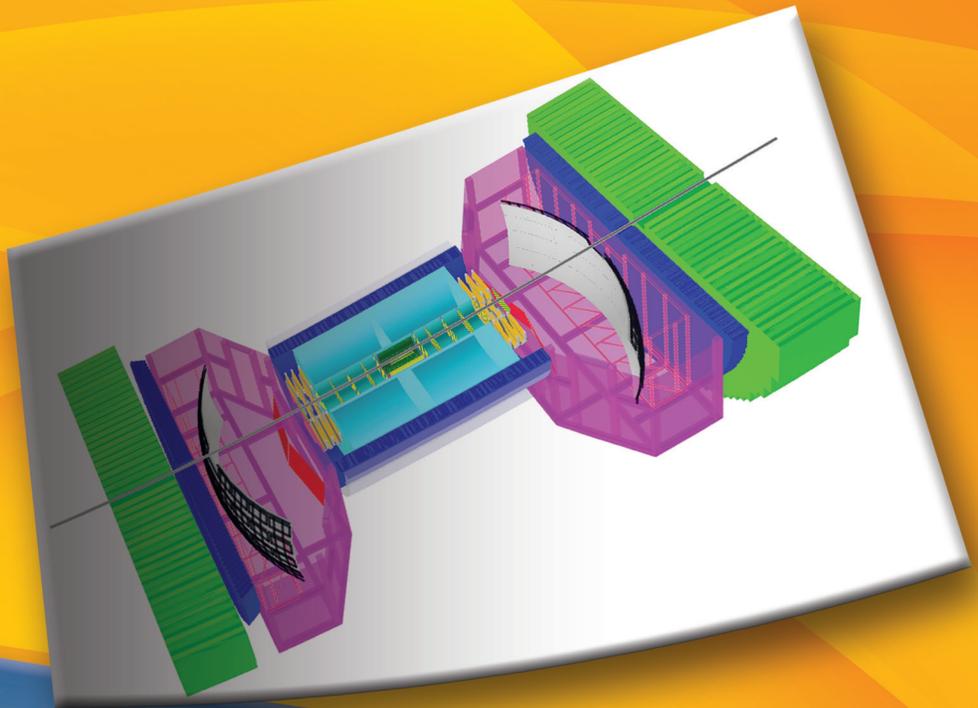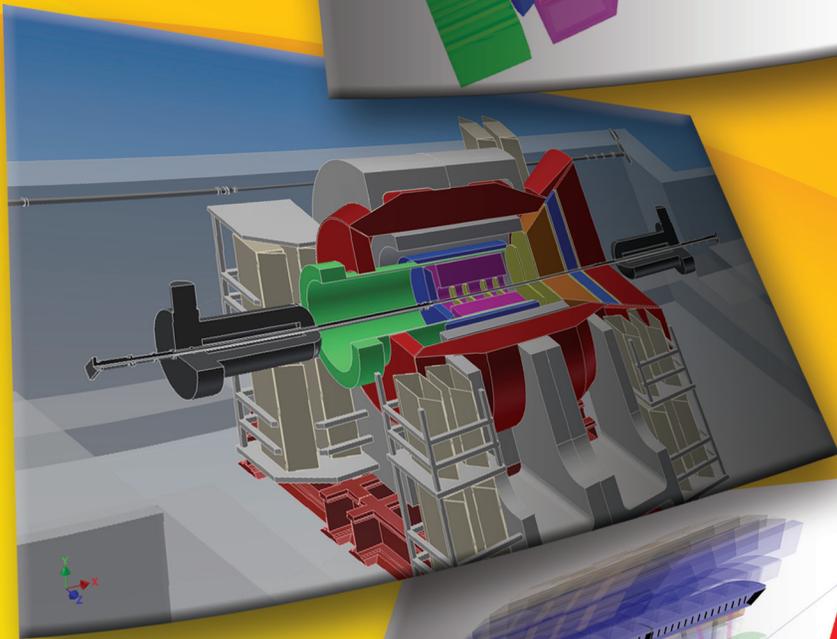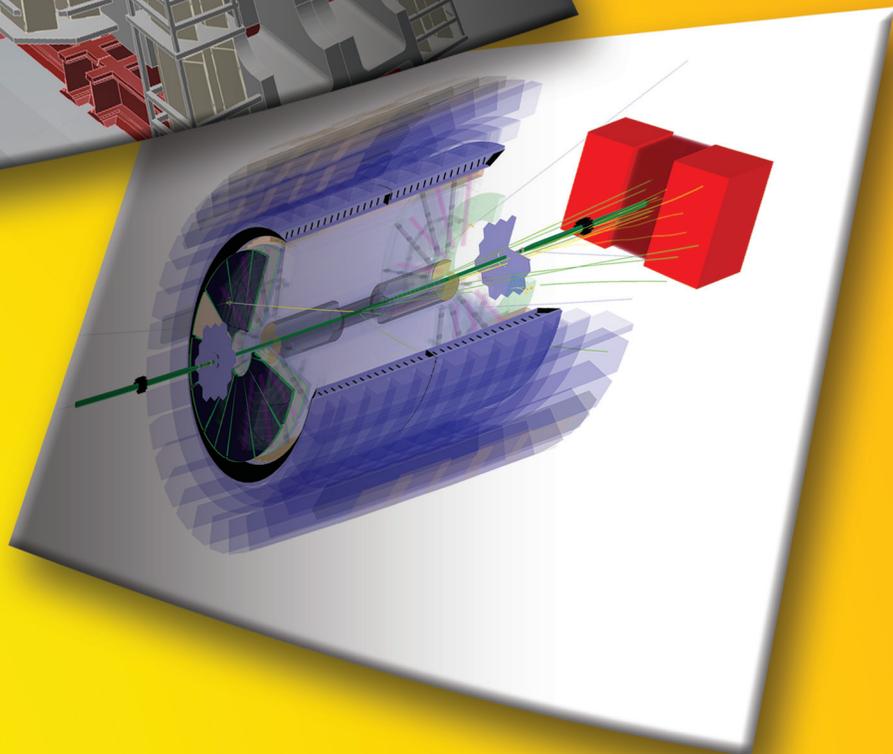